\documentclass[12pt]{article}

\usepackage{newtxtext,newtxmath}
\usepackage{amsmath}

\usepackage{graphicx}

\usepackage[letterpaper,margin=1in]{geometry}

\renewenvironment{abstract}
	{\quotation}
	{\endquotation}

\date{}

\makeatletter
\renewcommand{\fnum@figure}{\textbf{Figure \thefigure}}
\renewcommand{\fnum@table}{\textbf{Table \thetable}}
\newif\ifscience@tightcaption
\science@tightcaptionfalse
\newcommand{\tightcaptionspacing}{\science@tightcaptiontrue}
\newcommand{\standardcaptionspacing}{\science@tightcaptionfalse}
\newcommand{\sciencecaptionof}[2]{%
  \@makecaption{\csname fnum@#1\endcsname}{#2}%
}
\long\def\@makecaption#1#2{%
  \vskip\abovecaptionskip
  \begingroup
  \ifscience@tightcaption
    \renewcommand{\baselinestretch}{1}\selectfont
  \else
    \renewcommand{\baselinestretch}{1.25}\selectfont
  \fi
  \sbox\@tempboxa{#1: #2}%
  \ifdim \wd\@tempboxa >\hsize
    #1: #2\par
  \else
    \global\@minipagefalse
    \hb@xt@\hsize{\hfil\box\@tempboxa\hfil}%
  \fi
  \endgroup
  \vskip\belowcaptionskip}
\makeatother

\usepackage{scicite}

\usepackage[hyphens]{url}
\usepackage{booktabs}
\usepackage{tabularx}
\usepackage{array}
\usepackage{makecell}
\usepackage[table]{xcolor}

\def\scititle{
    TRACE: A Multi-Agent System for Autonomous Physical Reasoning for Seismology 
}
\def\scititlesupp{
    TRACE: A Multi-Agent System for Autonomous Physical Reasoning for Seismology 
}
\title{\vspace*{-3.0em}\bfseries \boldmath \scititle}

\author{
	Feng~Liu$^{1,2}$,
	Xin~Cui$^{3}$,
	Jian~Xu$^{4}$,
	Wang~Xinghao$^{2}$,
	Zijie~Guo$^{2}$,\\[1pt]
	Jiong~Wang$^{2}$,
	S.~Mostafa~Mousavi$^{1}$,
	Xinyu~Gu$^{2}$,
	Hao~Chen$^{2}$,
	Ben~Fei$^{2}$,\\[1pt]
	Lihua~Fang$^{5}$,
	Fenghua~Ling$^{2\ast}$,
	Zefeng~Li$^{3\ast}$,
	Lei~Bai$^{2\ast}$\and
    \small$^{1}$Department of Earth and Planetary Sciences, Harvard University, Cambridge, MA 02138, USA.\and
    \small$^{2}$Shanghai Artificial Intelligence Laboratory, Shanghai 200232, China.\and
    \small$^{3}$School of Earth and Space Sciences, University of Science and Technology of China, Hefei 230026, China.\and
    \small$^{4}$Department of Earth and Planetary Sciences, McGill University, Montreal, Quebec H3A 0E8, Canada.\and
    \small$^{5}$Institute of Earthquake Forecasting, China Earthquake Administration, Beijing 100036, China.\and
    \small$^\ast$Corresponding authors. Emails: lingfenghua@pjlab.org.cn; zefengli@ustc.edu.cn; bailei@pjlab.org.cn.
}

\begin{document} 

\maketitle

\vspace{-1.5em}
\begin{abstract} \bfseries \boldmath
    Modern seismic networks resolve earthquake sequences in unprecedented detail, yet explaining how large earthquakes emerge from evolving fault systems remains difficult. We introduce TRACE, a seismology-guided artificial intelligence agent that plans and executes workflows while preserving auditable evidence chains from observations to physical interpretation. We evaluated TRACE through 104 benchmark tasks and two complementary earthquake sequences. For the well-studied 2019 Ridgecrest sequence, TRACE constructed a high-resolution catalog from continuous waveforms and retrospectively recovered delayed cascading activation between the $M_w$ 6.4 and $M_w$ 7.1 earthquakes without a prescribed target interpretation. In the less-understood 2025--2026 Sanriku sequence off northeastern Japan, TRACE developed a testable interpretation of progressive destabilization within a segmented megathrust. Its synthesis linked coupled seismic--aseismic activation around the $M_J$ 6.9 sequence and subsequent persistent, spatially segmented shallow-interface activity to a megathrust patch that lay between regions of past large coseismic slip and later hosted the $M_J$ 7.7 rupture. These results open a path from seismic observations to testable physical insight.
\end{abstract}

    Large earthquakes arise from complex, evolving fault systems whose behavior is governed by tectonic loading, the lasting imprint of past ruptures, and dynamic stress interactions \cite{stein_1999_Role,scholz_2019_Mechanics,wirth_2022_Occurrence}. Since fault processes are hard to observe directly, our understanding of them relies critically on seismic and geodetic observations. Individual seismicity metrics, such as seismicity-rate changes, earthquake migration, and magnitude--frequency distributions, can provide valuable clues, yet they are rarely diagnostic when analyzed in isolation \cite{ogata_1988_Statistical,schorlemmer_2005_Variations,peng_2009_Migration,scholz_2015_Stress}. Similar signatures may arise from distinct processes, including aftershock relaxation, aseismic slip, stress transfer, fluid migration and structural heterogeneity \cite{dieterich_1994_Constitutive,king_1994_Static,shapiro_1997_Estimating,peng_2010_Integrated}. Joint analysis of the temporal evolution, spatial distribution and magnitude--frequency characteristics of seismicity can reduce this ambiguity and provide a more complete basis for physical interpretation. Complementary information on fault geometry, rupture history, interseismic coupling and structural segmentation can further constrain whether a proposed interpretation is physically plausible within the fault system \cite{moreno_2010_2010,wirth_2022_Occurrence}. The central challenge, therefore, is to integrate these complementary observations and fault-system constraints into physically grounded interpretations while preserving a transparent link between evidence and inference.

    Modern seismic networks provide increasingly dense, high-fidelity waveform records, and machine-learning methods have automated key stages of raw-data processing and earthquake-catalog construction \cite{zhu_2018_PhaseNet,mousavi_2020_Earthquake,zhu_2022_Earthquake,zhang_2022_LOCFLOW}. These advances have greatly improved earthquake catalogs, but most automated approaches address predefined processing tasks rather than open-ended mechanism questions. Moving from these observational products to physical interpretation still depends largely on expert judgment in analysis planning, validation of intermediate results and synthesis of complementary evidence, making the process difficult to reproduce or update systematically.

    Advances in AI-assisted scientific reasoning and large language models (LLMs) have shown that planning, code generation and tool use can support parts of scientific work \cite{wei_2022_Chainofthought,wang_2023_Scientific,boiko_2023_Autonomous,romera-paredes_2024_Mathematical}. Recent work has extended these capabilities to multi-agent discovery systems, co-scientist workflows and domain-specific autonomous agents \cite{ghareeb_2026_Multiagent,gottweis_2026_Accelerating,ferber_2026_Autonomous}. For seismological inference, however, LLM-generated reasoning cannot itself serve as scientific evidence; agent-supported interpretations must be constrained by physical assumptions, grounded in auditable data products and produced through reproducible analysis workflows.

    To address this need, we developed TRACE (Trans-perspective Reasoning and Automated Comprehensive Evaluator), a seismology-guided AI agent designed to construct auditable evidence chains from seismic observations to physical interpretation. TRACE converts open-ended mechanistic questions into executable analysis plans, evaluates them against seismological constraints, executes analyses with validated scientific tools and checks intermediate products for computational and physical consistency. It then synthesizes evidence across complementary analyses while retaining explicit links among data products, parameter choices, physical assumptions, intermediate results, alternative explanations and the resulting interpretation. We evaluated TRACE hierarchically through bounded computational tasks, multi-step workflows and applications to two earthquake sequences that posed complementary scientific questions.

    We applied TRACE to two earthquake sequences with complementary scientific roles. The well-studied 2019 Ridgecrest sequence provided a retrospective test of whether TRACE could reconstruct an established physical interpretation from continuous waveforms without a prescribed hypothesis \cite{ross_2019_Hierarchical,chen_2020_Cascading,huang_2020_Spatiotemporal,toda_2020_Long}. The less-understood 2025--2026 Sanriku sequence provided a discovery setting in which TRACE evaluated the evolution of shallow-interface seismicity before the April 2026 $M_J$ 7.7 earthquake and its relation to longer-term megathrust structure. Together, the two applications test whether an auditable agent can recover established earthquake-process interpretations and formulate testable hypotheses where the governing processes remain unresolved.

\section*{Auditable physical interpretation with TRACE}
    
    TRACE links seismic observations to physical interpretation through an auditable analysis workflow (Fig.~\ref{fig:fig1_workflow}a--c). Depending on data availability, it begins with continuous seismic waveforms or existing earthquake catalogs and phase-arrival records. Case-specific inputs, including station metadata, velocity models and fault geometry, constrain the analyses performed in each application. Each step of the workflow retains executable scripts, parameter choices, diagnostic records and intermediate products, allowing interpretive claims to be traced back to the observations and analyses that support them. TRACE then integrates these checked products into an evidence-supported physical interpretation while retaining viable alternatives and indicating how the interpretation could be tested.

    The workflow comprises three linked stages. Planning and Task Decomposition translates open-ended seismological questions into geophysically constrained hypotheses and executable analyses, with candidate pathways evaluated against explicit seismological constraints. Automated Execution and Diagnostics runs the resulting analyses, checks intermediate outputs for computational and physical consistency, and retains alternative workflows that pass validation. Scientific Integration and Inference synthesizes evidence from spatiotemporal patterns, seismicity-rate and magnitude--frequency distributions, waveform similarity, source characteristics and fault-system constraints. Using these evidence streams, TRACE compares candidate physical interpretations and identifies the interpretation best supported by the available evidence. Together, these stages preserve an auditable evidence chain from observations to physical interpretation.

    The technical architecture and evaluation framework are detailed in Materials and Methods \cite{methods} and Supplementary Figs.~\ref{fig:ed_trace_architecture} and \ref{fig:ed_trace_benchmark}. The evaluation framework assessed three levels of capability: bounded tool execution, multi-step workflow control, and integrated seismological inference. Levels 1 and 2 formed a cross-backbone benchmark of 74 basic and 30 multi-step tasks, whereas Level 3 evaluated end-to-end observation-to-interpretation reasoning in the complementary Ridgecrest validation and Sanriku discovery applications.

\section*{Delayed cascading activation between the 2019 Ridgecrest $M_w$ 6.4 and $M_w$ 7.1 earthquakes}
    
    The 2019 Ridgecrest earthquake sequence provides a validation setting for TRACE in a complex strike-slip fault system. The sequence included an $M_w$ 6.4 foreshock followed approximately 34 h later by the $M_w$ 7.1 mainshock within an interacting, approximately orthogonal fault network \cite{ross_2019_Hierarchical}. Starting from continuous waveforms and mapped surface-rupture traces \cite{kendrick_2019_Geologic}, TRACE analyzed the interval without a prescribed target interpretation. Published Ridgecrest interpretations were used only for retrospective comparison after the analysis was complete. TRACE constructed a high-resolution earthquake catalog and recovered delayed, fault-bound activation toward the segment that later hosted the $M_w$ 7.1 mainshock. Spatial timing, seismicity-rate evolution and $b$-value diagnostics supported delayed cascading activation.

    \subsection*{From waveforms to a traceable earthquake catalog}
        
        To resolve the spatiotemporal organization of seismicity between the two earthquakes, TRACE constructed a catalog with fine-scale relative hypocentral locations and a documented processing history. The workflow requested continuous waveforms from the Southern California Earthquake Data Center (SCEDC) for 4--26 July 2019 \cite{scedc_2013_Southern} and retained quality-controlled waveform records, phase-pick tables and an associated-event catalog with initial locations as intermediate data products. The resulting dataset comprised 853 station-day waveform records from 47 selected three-component stations. Using a PhaseNet probability threshold of 0.4, TRACE retained 1.45 million $P$- and 1.64 million $S$-phase picks \cite{zhu_2018_PhaseNet}. For association and initial location, GaMMA associated picks from the 30 stations nearest to the $M_w$ 6.4 epicenter into 140,505 initially located events \cite{zhu_2022_Earthquake}. Local magnitudes were then estimated from simulated Wood--Anderson waveforms, using the associated phase picks to define amplitude-measurement windows. Finally, HypoDD double-difference relocation refined the associated catalog using catalog-derived $P$- and $S$-phase differential times \cite{waldhauser_2000_DoubleDifference}, yielding a final TRACE catalog of 84,474 relocated earthquakes (Fig.~\ref{fig:fig2_ridgecrest_location}).

        During the interval between the $M_w$ 6.4 and $M_w$ 7.1 earthquakes, the relocated TRACE catalog contained 4,714 events. Its principal spatial and temporal distributions, together with the numbers of events with $M \ge 2$ and $M \ge 3$, were comparable to those in the USGS/ANSS ComCat catalog \cite{USGS_2017_Advanced} and the catalog of Ross et al. \cite{ross_2019_Hierarchical}. This cross-catalog consistency supported the use of the TRACE catalog for subsequent analyses of fault activation. To preserve auditability, TRACE retained data selections, processing parameters, intermediate picks, associated events and relocated hypocenters as provenance records. These records link the catalog and downstream diagnostics to the workflow that generated them. Detailed processing counts, parameter settings, relocation diagnostics and catalog comparisons are provided in Supplementary Note 3.

    \subsection*{Fault-organized evidence for delayed cascading activation}
        
        The relocated catalog resolved a systematic change in the spatial organization of seismicity during the 34-h interval (Fig.~\ref{fig:fig3_ridgecrest_analysis}). Time-window maps, proximity to mapped surface ruptures, and activation-onset diagnostics suggested that the seismic activity did not expand as a near-synchronous diffuse response around the $M_w$ 6.4 foreshock. Most earthquakes during this interval localized within 2 km of mapped surface-rupture traces, with grid-cell onset times indicating a delayed activation toward the $M_w$ 7.1 rupture zone (median 10.0 h) compared to the $M_w$ 6.4 zone (median 5.0 h). Early activity after the $M_w$ 6.4 foreshock followed the SW-NE-trending surface-rupture traces, whereas later activity organized along the NW-SE-trending fault family associated with the later mainshock. This delayed, fault-bound organization defined a NW-SE activation corridor across the interval, providing spatial-timing evidence against a simple regional aftershock response or a coherent migration front.

        Seismicity rate and magnitude--frequency diagnostics provided additional constraints on whether the corridor reflected transient aftershock decay or sustained activation of the fault system that later hosted the $M_w$ 7.1 rupture. Seismicity near the SW--NE surface-rupture traces declined, whereas activity on the NW--SE fault family persisted into the late part of the interval. TRACE then defined southern and northern portions of the NW--SE fault family for comparison (Supplementary Fig.~\ref{fig:supp_ridgecrest_RA3_rate_omori}). Their cumulative $M \ge 3$ counts and Omori--Utsu decay diagnostics showed that the northern portion, closer to the later mainshock source region, retained activity more persistently than the southern portion \cite{omori_1895_Aftershocks,ogata_1988_Statistical}. The $M$ 5.4 event served as an analytical separator between the early response dominated by the $M_w$ 6.4 foreshock and subsequent activation on the NW--SE fault family. For the $b$-value analysis, TRACE instead defined two 5 km comparison regions centered on the $M_w$ 6.4 and $M_w$ 7.1 reference locations (Fig.~\ref{fig:fig3_ridgecrest_analysis}e). After the separator, the $M_w$ 7.1 comparison region had a lower $b$-value ($b \approx 0.67$) than the $M_w$ 6.4 comparison region ($b \approx 0.96$) and the regional background ($b \approx 1.07$). This $b$-value contrast complemented the timing and seismicity-rate evidence as a proxy for spatial variations in stress state \cite{schorlemmer_2005_Variations,scholz_2015_Stress,nanjo_2020_Were}.

        Taken together, the delayed onset, organization along mapped surface ruptures, persistent activity late in the interval and lower $b$-value on the $M_w$ 7.1 side led TRACE to identify delayed cascading activation between the $M_w$ 6.4 foreshock and the $M_w$ 7.1 mainshock. Ross et al. resolved the interlocked orthogonal fault network using a template-matched seismicity catalog and geodetic and seismic source analyses, while Huang et al. used matched-filter detection and relocation to identify spatially variable pre-mainshock seismicity rates and a low-$b$ nucleation region \cite{ross_2019_Hierarchical,huang_2020_Spatiotemporal}. Complementary geodetic, waveform and Coulomb-stress analyses indicated that the $M_w$ 6.4 rupture promoted failure near the later hypocenter \cite{chen_2020_Cascading,toda_2020_Long}. TRACE integrated catalog construction, activation timing, fault organization, seismicity-rate evolution and $b$-value contrasts within a single auditable workflow. Its convergence with these prior studies provides retrospective validation of the observation-to-interpretation process through a documented catalog-based evidence chain. The seismicity evidence constrained the organization and timing of the cascade but did not uniquely identify the process responsible for the delay. Plausible contributions include Coulomb stress transfer, a seismicity response governed by rate-and-state friction, afterslip and other postseismic deformation \cite{king_1994_Static,dieterich_1994_Constitutive,yue_2021_2019}. Detailed diagnostics are provided in Supplementary Note 4.

\section*{Persistent shallow-interface activation before the 2026 Sanriku $M_J$ 7.7 rupture}

    Large megathrust earthquakes rupture strongly coupled patches embedded within heterogeneous plate interfaces \cite{moreno_2010_2010,wirth_2022_Occurrence}. Around these patches, seismic swarms, slow slip, tremor, very-low-frequency earthquakes and repeating earthquakes can respond to small stress perturbations and transfer stress to adjacent locked regions \cite{uchida_2016_Periodic,obara_2016_Connecting,nishikawa_2023_Review}. These signals provide diagnostic evidence for short-term interface activation, but no single observation uniquely separates aftershock relaxation, aseismic slip, stress transfer and structural heterogeneity \cite{dieterich_1994_Constitutive,king_1994_Static,peng_2010_Integrated}. A central problem is therefore to determine whether localized activation reflects a broader change in the state of a segmented megathrust capable of hosting a large rupture.

    The 2025--2026 Sanriku sequence placed this question in a recent, incompletely understood megathrust setting. In April 2026, an $M_J$ 7.7 earthquake ruptured the shallow plate interface off Sanriku, where recurrent $M$ 7--8 earthquakes occur on a margin that also hosts repeating earthquakes, slow slip, tremor and very-low-frequency earthquakes \cite{yamanaka_2004_Asperity,uchida_2003_Interplate,nishikawa_2023_Review}. We therefore used TRACE to reconstruct the seismicity evolution before the $M_J$ 7.7 earthquake and identify organized shallow-interface activation. TRACE began with the Japan Meteorological Agency (JMA) unified hypocenter catalog and associated arrival-time data spanning June 2025 to May 2026 \cite{jma_2026_SeismologicalBulletin}. It relocated 25,646 of 33,482 regional events using the double-difference method \cite{waldhauser_2000_DoubleDifference}. TRACE then compared the November 2025 $M_J$ 6.9 activity, the December 2025 regional $M$ 7.5 sequence and the April 2026 $M_J$ 7.7 earthquake. Among the candidate sequences, the $M_J$ 6.9 and $M_J$ 7.7 source regions formed the closest pair at comparable shallow depths, and $M \ge 4$ earthquakes were distributed across the intervening region (Figs.~\ref{fig:fig4_japan_mechanism}f and \ref{fig:fig5_japan_evidence}a). TRACE therefore defined a focused region containing both source regions and the shallow seismicity between them.

    \subsection*{Coupled seismic and aseismic activity around the $M_J$ 6.9 sequence}
        
        The November 2025 $M_J$ 6.9 sequence occupied a patch on the shallow interface near the northern edge of the 2011 $M_w$ 9.0 Tohoku-Oki high-slip domain \cite{iinuma_2012_Coseismic} and south of the 1994 $M_w$ 7.7 Sanriku-Oki rupture region \cite{nagai_2001_Comparative} (Fig.~\ref{fig:fig5_japan_evidence}a). The magnitude-time plot showed that the $M_J$ 6.9 event was preceded by local seismicity and followed by activity that did not resemble simple aftershock decay (Fig.~\ref{fig:fig5_japan_evidence}b). Over the $\sim$10 days before the $M_J$ 6.9 event, local seismicity formed a compact cluster and expanded across $\sim$16 km during the final 2 days. Using the 90th-percentile distance envelope as a bulk migration metric, TRACE estimated an advance of $\sim$4.3 km d$^{-1}$; the $b$-value estimate for this short window was $0.51$ for 201 events above completeness, and moderate earthquakes became increasingly frequent (Supplementary Fig.~\ref{fig:supp_sanriku_swarm_migration}b--f). During the first 0.5 days after the $M_J$ 6.9 event, seismicity reached distances of $\sim$75 km from the source region. The 90th-percentile distance envelope had an apparent slope of $\sim$16.2 km d$^{-1}$, and the $b$-value was $0.43$ for 308 events above completeness (Supplementary Fig.~\ref{fig:supp_sanriku_swarm_migration}). These distance-envelope diagnostics record rapid shallow-interface expansion at rates comparable to those reported for seismicity associated with slow slip in subduction settings \cite{obara_2016_Connecting,nishikawa_2023_Review}.

        Waveform similarity provided an independent constraint on repeated failure within the activated region. Cross-correlation screening of 20,301 candidate event pairs identified 231 high-confidence repeating pairs, grouped into 145 repeating-earthquake families (Fig.~\ref{fig:fig5_japan_evidence}d--f; Supplementary Fig.~\ref{fig:supp_sanriku_repeating_qc}). These families overlapped the $M_J$ 6.9 activated region and were concentrated in time around the $M_J$ 6.9 sequence \cite{nadeau_1999_Fault,uchida_2019_Repeating}. Independent S-net borehole strain observations indicate substantial aseismic moment release during this stage \cite{yoshida_2026_Evolution}. Together, the migration, repeating-earthquake and strain observations provide evidence consistent with coupled seismic and aseismic activity around the $M_J$ 6.9 sequence.

    \subsection*{Persistent shallow-interface activation after the $M_J$ 6.9 sequence}
        
        TRACE next evaluated the temporal evolution of seismicity after the $M_J$ 6.9 sequence (Fig.~\ref{fig:fig4_japan_mechanism}a-e; Supplementary Fig.~\ref{fig:supp_sanriku_system_definition}c--f). A long-term background comparison (Supplementary Note 5) identified anomalous activity across magnitude thresholds, with observed-to-expected count ratios of 7.70, 8.80 and 9.64 for $M \ge 3$, $M \ge 4$ and $M \ge 5$ earthquakes, respectively. TRACE then tracked the temporal evolution using five fixed analytical windows spanning background activity, the $M_J$ 6.9 activation sequence, the following interval, the final pre-$M_J$ 7.7 interval and the early $M_J$ 7.7 response. The total rate of relocated events increased from approximately 4.5 d$^{-1}$ during the background window to 162 d$^{-1}$ during the $M_J$ 6.9 sequence. It subsequently declined to approximately 13 d$^{-1}$ but remained above the background-window rate, before increasing slightly to approximately 16 d$^{-1}$ during the final pre-$M_J$ 7.7 interval (Fig.~\ref{fig:fig5_japan_evidence}c; Supplementary Note 5). These rate changes showed that seismicity across the focused shallow-interface region did not return to background after the $M_J$ 6.9 sequence and increased again before the $M_J$ 7.7 rupture.

        Magnitude--frequency statistics varied across the same analytical windows (Fig.~\ref{fig:fig5_japan_evidence}c; Supplementary Note 5). The $b$-value decreased from approximately 0.81 before the $M_J$ 6.9 sequence to 0.61 during it, partially recovered to 0.77 afterward, and fell again to 0.65 before the $M_J$ 7.7 rupture. These repeated low-$b$ intervals complemented the rate evidence as a proxy for temporal changes in stress state within the focused shallow-interface region \cite{schorlemmer_2005_Variations,scholz_2015_Stress,gulia_2019_Realtime}. Spatially, $M \ge 4$ earthquakes persisted in separated patches across the focused shallow-interface region (Fig.~\ref{fig:fig4_japan_mechanism}a--f). Together, the elevated rates, repeated low-$b$ intervals and spatially segmented $M \ge 4$ activity are consistent with a persistently perturbed shallow-interface state before the $M_J$ 7.7 rupture.

    \subsection*{Historical large-slip constraints on the $M_J$ 7.7 rupture patch}
        
        The seismicity-rate and $b$-value observations constrain the short-term evolution of the focused region but do not establish its longer-term capacity for large rupture. We therefore compared the activated region with historical coseismic-slip models. The November 2025 $M_J$ 6.9 earthquake lay at the southeastern end adjacent to the northern margin of the 2011 $M_w$ 9.0 Tohoku-Oki high-slip domain \cite{iinuma_2012_Coseismic}, whereas the April 2026 $M_J$ 7.7 earthquake occurred at the northwestern end, closer to the 1994 $M_w$ 7.7 Sanriku-Oki rupture and the margin of the 1968 $M_w$ 8.3 Tokachi-Oki rupture \cite{nagai_2001_Comparative} (Fig.~\ref{fig:fig5_japan_evidence}a). The $M_J$ 7.7 rupture occurred within this bounded patch rather than in the central part of either neighboring historical large-slip region (Fig.~\ref{fig:fig4_japan_mechanism}g). At a regional convergence rate of approximately 8 cm yr$^{-1}$ \cite{demets_2010_Geologically}, the 58 years since the 1968 earthquake correspond to approximately 4.6--4.8 m of relative plate motion. Under full locking, this value represents an upper bound on the accumulated slip deficit, but it shows that meter-scale loading of the patch was physically plausible before the $M_J$ 7.7 rupture.

        The preceding analyses provide complementary constraints on the short- and long-term state of the Sanriku megathrust. Persistently elevated seismicity and the renewed decrease in $b$-value indicate that the shallow interface remained perturbed after the $M_J$ 6.9 sequence, whereas historical rupture models place this activity within an intervening patch between previous large-slip domains. No continuous migration front connected the $M_J$ 6.9 and $M_J$ 7.7 source regions, arguing against a simple model of progressive rupture propagation. Instead, the observations support a model of progressive destabilization within a segmented megathrust: the $M_J$ 6.9 sequence marked a phase of coupled seismic–aseismic activation, shallow-interface activity remained elevated afterward, and the later $M_J$ 7.7 rupture occurred within this rupture-capable patch. Under this interpretation, long-term megathrust segmentation defined the structural setting for a large rupture, whereas the $M_J$ 6.9 sequence and subsequent activation marked a transient perturbation of that pre-existing system. The detailed stress-transfer processes governing this evolution remain unresolved.

\section*{Implications and limitations}
    
    TRACE provides an auditable framework for linking heterogeneous seismic observations to physical interpretation while retaining the analytical choices and assumptions underlying each interpretation. The Ridgecrest application retrospectively validated this observation-to-interpretation process by recovering delayed cascading activation without a prescribed target interpretation. Although prior exposure of the LLM backbones to published Ridgecrest studies during pre-training or fine-tuning cannot be excluded, no Ridgecrest publication or target interpretation was supplied during the analysis, and the reported interpretation had to be supported by the catalog and intermediate evidence generated within the workflow. By contrast, the Sanriku application extended TRACE from retrospective validation to scientific discovery, supporting a testable model of progressive destabilization within a segmented megathrust before the $M_J$ 7.7 rupture. Together, these cases show that TRACE can both recover established earthquake-process interpretations and develop evidence-grounded, testable interpretations where the governing processes remain unresolved.

    The Sanriku application links the months-long evolution of shallow-interface activation between the $M_J$ 6.9 and $M_J$ 7.7 earthquakes to the longer-term rupture segmentation of the megathrust. Historical rupture patterns bracketed an intervening patch where substantial slip deficit could accumulate, providing a long-term structural setting capable of hosting a large rupture. Within this patch, the $M_J$ 6.9 sequence marked a phase of coupled seismic--aseismic activation, after which seismicity remained elevated and spatially segmented rather than returning to background or propagating as a continuous migration front. The subsequent $M_J$ 7.7 rupture within the same patch is therefore consistent with progressive destabilization within a pre-existing rupture-capable patch. This interpretation extends previous work on interactions between seismic and aseismic slip by suggesting that the significance of short-term interface activation depends on the long-term rupture architecture in which it occurs \cite{uchida_2016_Periodic,obara_2016_Connecting,nishikawa_2023_Review}.

    This auditability is essential because downstream interpretations of any earthquake sequence are highly sensitive to upstream choices in catalog construction, velocity models, association algorithms and statistical windows, all of which require scientific justification. Auditability exposes, rather than eliminates, the dependence of physical interpretation on these choices and on expert judgment. TRACE systematically logs all processing decisions, validates intermediate products against empirical and physical constraints, and preserves competing workflows and alternative interpretations as distinct branches within the evidence record. These records make it possible to assess whether an interpretation is supported by complementary evidence or depends on a narrow set of procedural choices. They do not guarantee correctness, and expert review remains necessary to identify omitted alternatives, inappropriate analyses and propagated errors.

    The two applications began at different observational levels, with continuous waveforms for Ridgecrest and operationally curated JMA event and phase products for Sanriku. For Sanriku, TRACE also evaluated an alternative route based on deep-learning phase picks from continuous S-net waveforms. Human-supervised quality control revealed numerous implausible $S$ arrivals, and the resulting relocated catalog was judged less coherent than the JMA-based catalog. The JMA arrivals were therefore retained for relocation, while a separately assembled waveform dataset was used for repeating-earthquake cross-correlation. This decision illustrates how auditable intermediate products can prevent upstream processing errors from propagating into physical interpretation. TRACE can expose how observational inputs shape an interpretation, but it cannot compensate for limitations in network geometry, catalog completeness, waveform quality or velocity models. Applying TRACE in regions without comparable curated products would therefore require local validation of detection, association and location workflows, because processing models and parameter choices do not necessarily transfer across networks, noise conditions and station geometries.

    The Sanriku analysis constrains the fault-system evolution more clearly than the detailed stress-transfer processes underlying it. Earthquake-derived observations resolve the spatial and temporal evolution of shallow-interface activation, while repeating earthquakes are consistent with repeated failure of small asperities that may be reloaded by surrounding interface slip. Independent S-net borehole strain observations provide additional evidence for substantial aseismic activity during the $M_J$ 6.9 episode \cite{yoshida_2026_Evolution}. However, these observations do not resolve the amount or geometry of aseismic slip, or whether and how aseismic slip transferred stress to the later $M_J$ 7.7 rupture patch. Thus, the interpretation of progressive destabilization is supported at the fault-system level, whereas the relative contributions of aseismic slip, stress transfer and other postseismic processes remain unresolved.

    The complementary Ridgecrest and Sanriku applications suggest that auditable observation-to-interpretation reasoning may be transferable across diverse fault systems. Establishing how broadly this capability transfers will require evaluation against independent expert analyses under varying network geometry, catalog completeness, waveform quality and tectonic setting. More broadly, TRACE provides a way to transform increasingly rich seismic observations into reviewable, evidence-grounded and testable explanations of how fault-system evolution gives rise to large ruptures.


\begin{figure}
    \centering
    \includegraphics[width=0.99\linewidth]{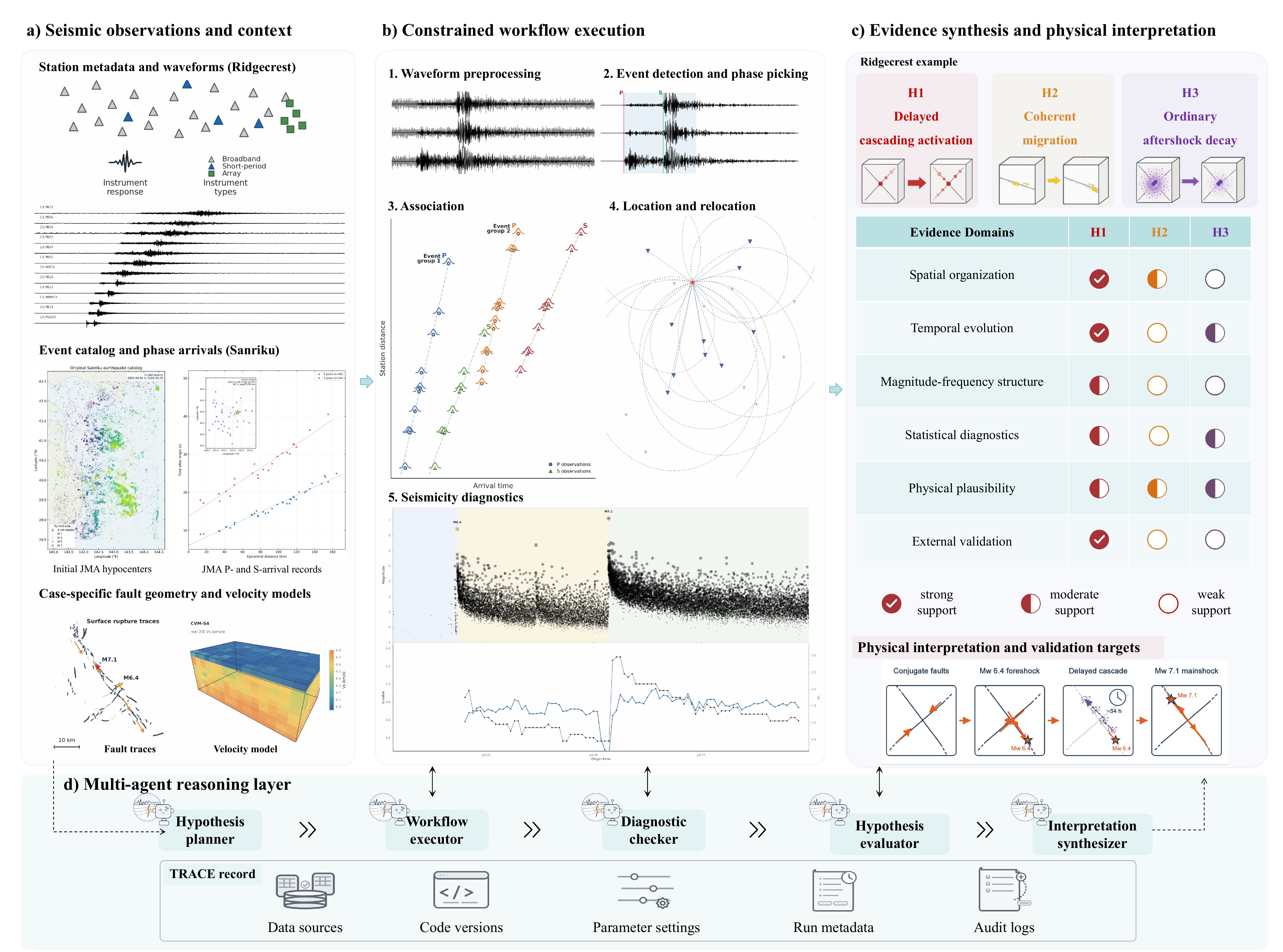}
    \caption{
    \textbf{TRACE framework for constrained workflow execution and physical interpretation.}
    \textbf{a}, Seismic observations and case-specific geophysical inputs available to TRACE, including station metadata, waveform records, event catalogs, phase-arrival records, fault geometry and velocity models. These inputs define the observational and physical boundary conditions for subsequent analysis.
    \textbf{b}, Constrained workflow execution. TRACE links waveform preprocessing, event detection and phase picking, association, location and relocation, and seismicity diagnostics through diagnostic checks of intermediate products. Depending on input availability, workflows may enter at different stages.
    \textbf{c}, Evidence synthesis and physical interpretation. The evidence matrix summarizes qualitative support for candidate interpretations derived from the scientific question and analysis outputs; the lower schematic shows the best-supported interpretation and targets for further validation.
    \textbf{d}, The coordinated agent layer supports hypothesis planning, workflow execution, diagnostic checking, interpretation evaluation and synthesis. The TRACE record retains data sources, code versions, parameter settings, run metadata and audit logs. Miniature panels illustrate representative inputs, workflow products and evidence domains; quantitative application results are shown in subsequent figures.
    }
    \label{fig:fig1_workflow}
\end{figure}

\begin{figure}
    \centering
    \includegraphics[width=0.99\linewidth]{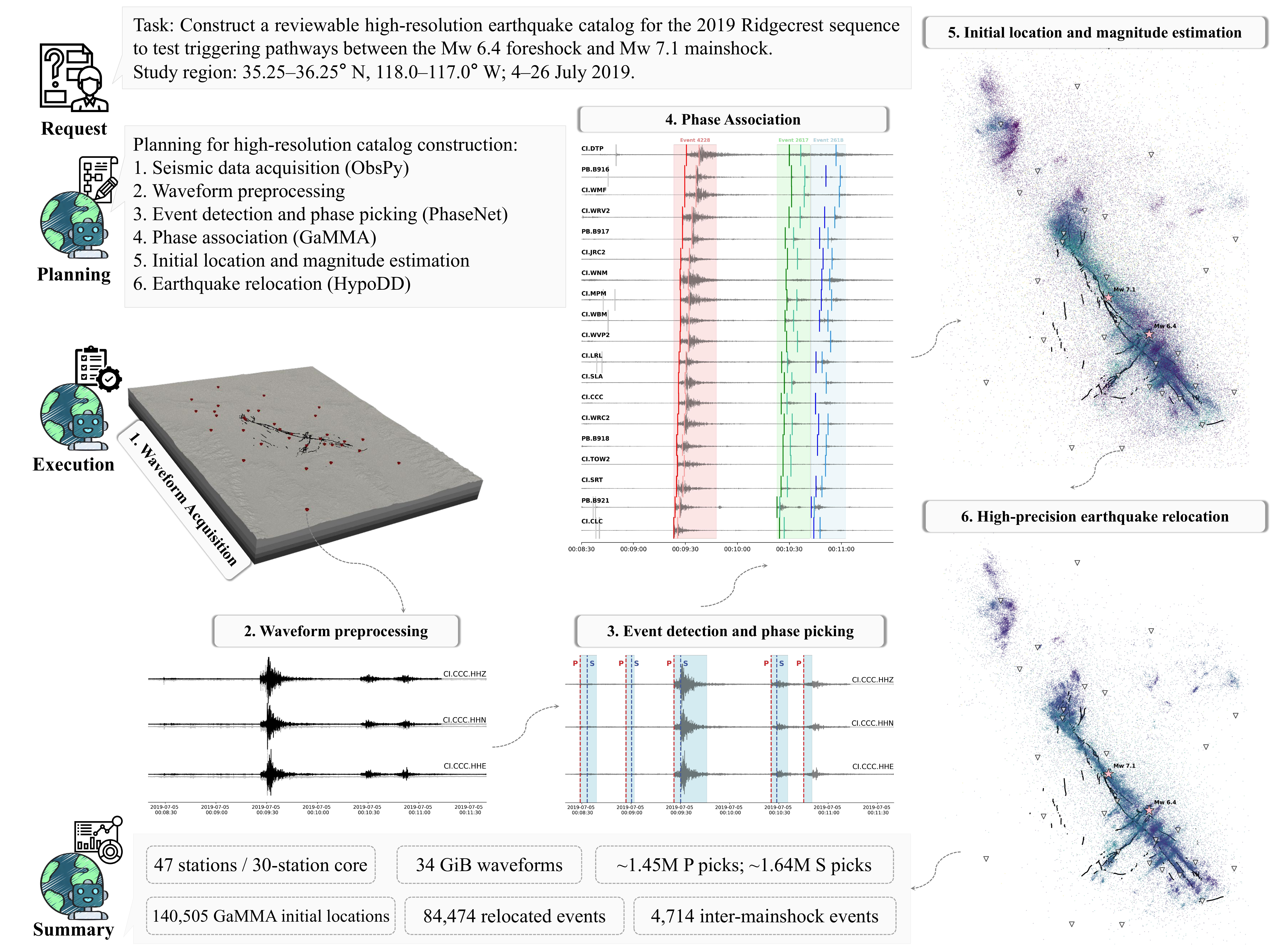}
    \caption{
    \textbf{TRACE construction of a reviewable Ridgecrest earthquake catalog workflow.}
    TRACE converts waveform records and task constraints into a documented catalog-building workflow whose data selections, processing parameters, phase picks, associated events, initial locations, local-magnitude estimates and relocation outputs can be inspected. The 2019 Ridgecrest sequence is shown as a worked example: TRACE plans a six-step route and executes waveform acquisition, preprocessing, PhaseNet phase picking, GaMMA association, initial location, and HypoDD-based relocation. Summary cards report 47 stations selected for waveform processing, of which the 30 nearest to the $M_w$ 6.4 epicenter were retained for association and relocation; 34 GiB of requested waveforms; approximately 1.45 million $P$ picks and 1.64 million $S$ picks; 140,505 associated events with initial locations; 84,474 relocated events; and 4,714 events between the $M_w$ 6.4 and $M_w$ 7.1 earthquakes. These catalog and provenance records provide the observational basis for evaluating triggering pathways from the $M_w$ 6.4 foreshock to the $M_w$ 7.1 mainshock.
    }
    \label{fig:fig2_ridgecrest_location}
\end{figure}

\begin{figure}
    \centering
    \includegraphics[width=0.99\linewidth]{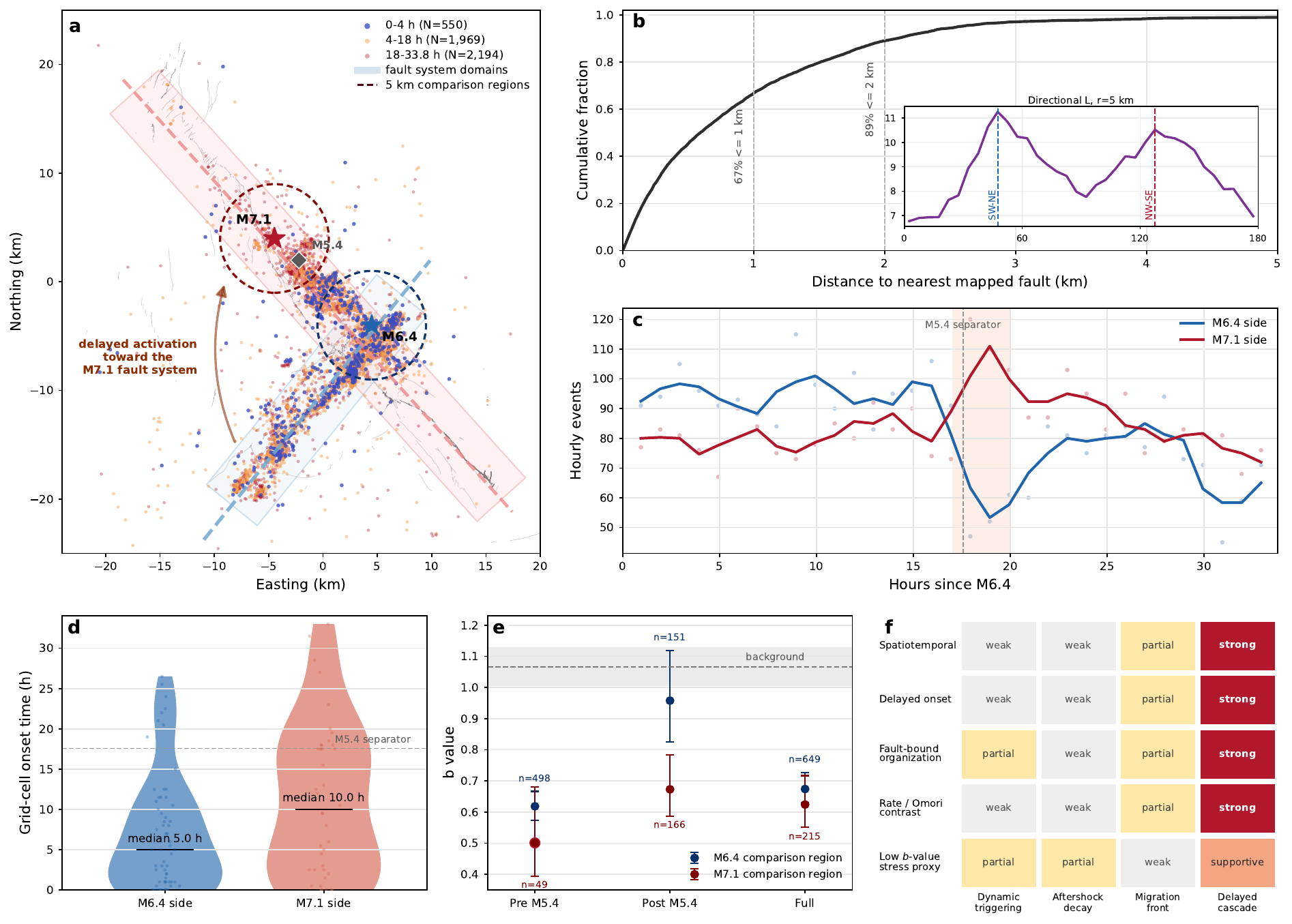}
    \caption{
    \textbf{Evidence for delayed cascading activation between the Ridgecrest $M_w$ 6.4 foreshock and $M_w$ 7.1 mainshock.}
    \textbf{a}, Seismicity over 34 h in three windows (0--4, 4--18 and 18--33.8 h) after the foreshock, with mapped SW-NE- and NW-SE-trending surface-rupture traces \cite{kendrick_2019_Geologic}, the $M$ 5.4 analytical separator and inferred activation direction toward the NW-SE fault family. Pale blue and red corridors define domains in \textbf{c,d}; darker dashed outlines mark the two 5 km comparison regions in \textbf{e}.
    \textbf{b}, Distances to the nearest mapped surface-rupture trace and directional spectrum; reference directions mark the dominant fault-family orientations, and vertical dashed lines mark 1 and 2 km.
    \textbf{c}, Hourly seismicity rates in the foreshock- and mainshock-side corridors; shading marks the separator interval.
    \textbf{d}, Grid-cell activation onset times in the same corridors; horizontal bars denote medians.
    \textbf{e}, Fixed-window $b$-values for the foreshock and mainshock comparison regions before and after the separator and over the full interval. Error bars denote standard uncertainties, $n$ event counts and the gray band the regional background range.
    \textbf{f}, Qualitative evidence-weighting matrix comparing delayed cascading activation with dynamic triggering, ordinary aftershock decay and a coherent migration front; cells summarize diagnostic consistency and do not represent an independent statistical test.
    }
    \label{fig:fig3_ridgecrest_analysis}
\end{figure}

\begin{figure}
    \centering
    \includegraphics[width=0.99\linewidth]{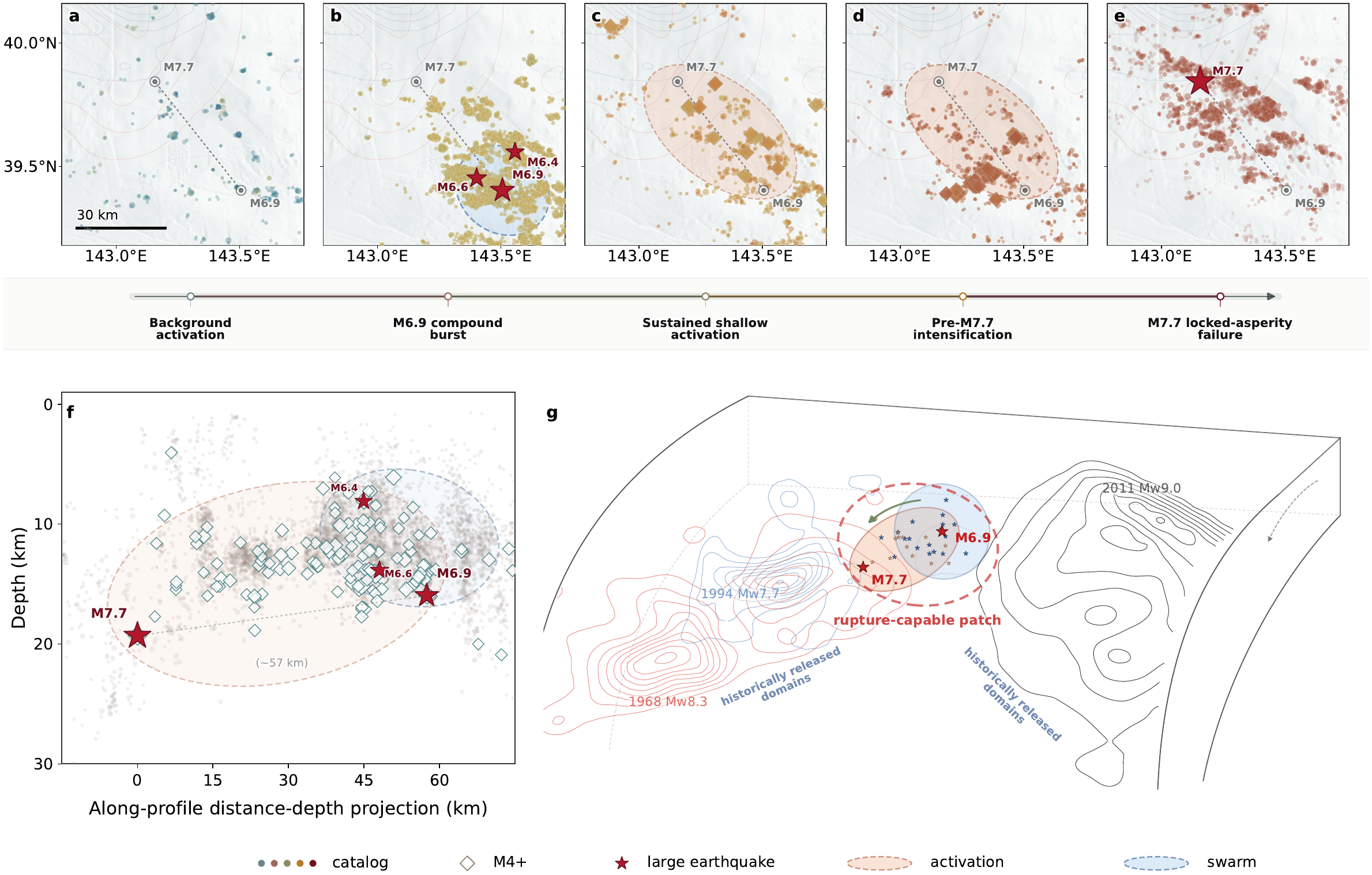}
    \caption{
    \textbf{Persistent shallow-interface activation before the 2026 Sanriku $M_J$ 7.7 earthquake.}
    The physical synthesis links stage-resolved seismicity, the focused shallow-interface geometry and neighboring historical large-slip regions.
    \textbf{a--e}, Five-stage evolution of relocated seismicity, from background activity through the November 2025 $M_J$ 6.9 compound burst, sustained shallow-interface activation, pre-$M_J$ 7.7 intensification and the $M_J$ 7.7 locked-asperity failure. Dashed outlines mark the $M_J$ 6.9 source region, the broader shallow activation zone and the historically bounded $M_J$ 7.7 patch. The central timeline summarizes the transition from distributed activity around the $M_J$ 6.9 source to the $M_J$ 7.7 rupture.
    \textbf{f}, Along-profile distance-depth projection of the focused shallow-interface region. The $M_J$ 6.9 and $M_J$ 7.7 reference locations are separated by $\sim$57 km, and $M \ge 4$ earthquakes persist in spatially separated patches across the focused profile rather than forming a continuous migration front.
    \textbf{g}, Three-dimensional synthesis of short-term activation and historical rupture geometry. Historical rupture contours mark neighboring regions of large coseismic slip, whereas the focused shallow-interface region occupies a patch bracketed by these regions, with the later $M_J$ 7.7 rupture occurring within the same bounded patch.
    }
    \label{fig:fig4_japan_mechanism}
\end{figure}

\begin{figure}
    \centering
    \includegraphics[width=0.99\linewidth]{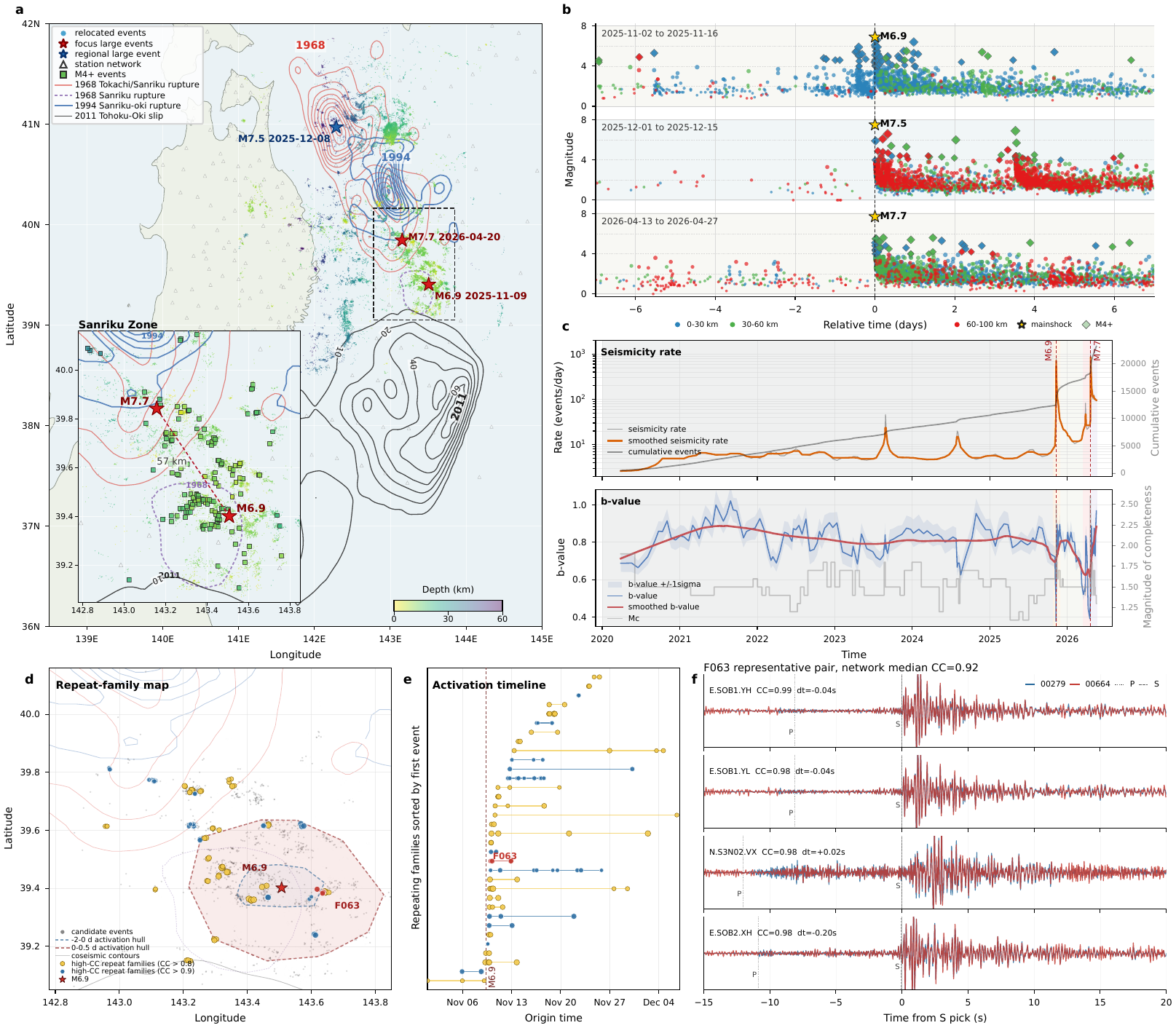}
    \tightcaptionspacing
    \caption{
    \textbf{Seismicity-rate, magnitude--frequency and repeating-earthquake evidence for Sanriku shallow-interface activation.}
    \textbf{a}, Relocated catalog colored by depth in the focused Sanriku region. Stars mark the $M_J$ 6.9 and $M_J$ 7.7 earthquakes (red) and the regional $M$ 7.5 earthquake (blue). Contours show the 2011 $M_w$ 9.0 Tohoku-Oki high-slip region \cite{iinuma_2012_Coseismic} and the 1994 $M_w$ 7.7 Sanriku-Oki and 1968 Tokachi-Oki (approximately $M_w$ 8.3) ruptures \cite{nagai_2001_Comparative}. The inset shows $M \ge 4$ events along the $\sim$57-km shallow-interface profile.
    \textbf{b}, Magnitude-time plots for the November 2025 $M_J$ 6.9, December 2025 regional $M$ 7.5 and April 2026 $M_J$ 7.7 sequences, each aligned to its largest-event origin time. Colors denote 0--30, 30--60 and 60--100 km epicentral-distance bins; diamonds mark $M \ge 4$ events.
    \textbf{c}, Long-term seismicity rate, cumulative event count, $b$-value and magnitude of completeness. Lines mark the $M_J$ 6.9 and $M_J$ 7.7 events; shading shows $\pm 1\sigma$ uncertainty in $b$-value estimates.
    \textbf{d}, Repeating-earthquake candidates and high-correlation families near the $M_J$ 6.9 source. Hulls enclose seismicity during the final 2 days before and first 0.5 days after the event; contours show historical coseismic slip. Colors indicate family thresholds; CC, cross-correlation.
    \textbf{e}, Repeating-family timeline ordered by first event; points denote events, segments family durations and the dashed line the $M_J$ 6.9 origin time.
    \textbf{f}, Representative waveforms from family F063. Colors denote the two events; vertical lines mark $P$ and $S$ picks. Labels give station-component cross-correlation and time shift; the title gives the network-median CC.
    }
    \label{fig:fig5_japan_evidence}
    \standardcaptionspacing
\end{figure}

\makeatletter
\renewcommand{\fnum@figure}{\textbf{Figure \thefigure}}
\renewcommand{\fnum@table}{\textbf{Table \thetable}}
\makeatother


\clearpage 

%
\bibliography{reference}

@Misc{methods,
  note = {Materials and methods are available as supplementary material},
}

@article{aoi_2020_MOWLAS,
  title = {{{MOWLAS}}: {{NIED}} Observation Network for Earthquake, Tsunami and Volcano},
  shorttitle = {Mowlas},
  author = {Aoi, Shin and Asano, Youichi and Kunugi, Takashi and Kimura, Takeshi and Uehira, Kenji and Takahashi, Narumi and Ueda, Hideki and Shiomi, Katsuhiko and Matsumoto, Takumi and Fujiwara, Hiroyuki},
  year = 2020,
  month = dec,
  journal = {Earth, Planets and Space},
  volume = {72},
  number = {1},
  pages = {126},
  issn = {1880-5981},
  doi = {10.1186/s40623-020-01250-x},
  urldate = {2026-06-21},
  language = {en}
}

@article{beyreuther_2010_ObsPy,
  title = {{{ObsPy}}: {{A Python Toolbox}} for {{Seismology}}},
  shorttitle = {{{ObsPy}}},
  author = {Beyreuther, M. and Barsch, R. and Krischer, L. and Megies, T. and Behr, Y. and Wassermann, J.},
  year = 2010,
  month = may,
  journal = {Seismological Research Letters},
  volume = {81},
  number = {3},
  pages = {530--533},
  issn = {0895-0695},
  doi = {10.1785/gssrl.81.3.530},
  urldate = {2024-05-04},
  language = {en}
}

@article{boiko_2023_Autonomous,
  title = {Autonomous Chemical Research with Large Language Models},
  author = {Boiko, Daniil A. and MacKnight, Robert and Kline, Ben and Gomes, Gabe},
  year = 2023,
  month = dec,
  journal = {Nature},
  volume = {624},
  number = {7992},
  pages = {570--578},
  issn = {0028-0836, 1476-4687},
  doi = {10.1038/s41586-023-06792-0},
  urldate = {2025-09-22},
  language = {en}
}

@article{chamberlain_2018_EQcorrscan,
  title = {{{EQcorrscan}}: {{Repeating}} and {{Near}}-{{Repeating Earthquake Detection}} and {{Analysis}} in {{Python}}},
  shorttitle = {{{EQcorrscan}}},
  author = {Chamberlain, Calum J. and Hopp, Chet J. and Boese, Carolin M. and Warren-Smith, Emily and Chambers, Derrick and Chu, Shanna X. and Michailos, Konstantinos and Townend, John},
  year = 2018,
  month = jan,
  journal = {Seismological Research Letters},
  volume = {89},
  number = {1},
  pages = {173--181},
  issn = {0895-0695, 1938-2057},
  doi = {10.1785/0220170151},
  urldate = {2026-03-18},
  language = {en}
}

@article{chen_2020_Cascading,
  title = {Cascading and Pulse-like Ruptures during the 2019 {{Ridgecrest}} Earthquakes in the {{Eastern California Shear Zone}}},
  author = {Chen, Kejie and Avouac, Jean-Philippe and Aati, Saif and Milliner, Chris and Zheng, Fu and Shi, Chuang},
  year = 2020,
  month = jan,
  journal = {Nature Communications},
  volume = {11},
  number = {1},
  pages = {22},
  issn = {2041-1723},
  doi = {10.1038/s41467-019-13750-w},
  urldate = {2026-07-11},
  language = {en}
}

@article{demets_2010_Geologically,
  title = {Geologically Current Plate Motions},
  author = {DeMets, Charles and Gordon, Richard G. and Argus, Donald F.},
  year = 2010,
  month = apr,
  journal = {Geophysical Journal International},
  volume = {181},
  number = {1},
  pages = {1--80},
  issn = {0956540X, 1365246X},
  doi = {10.1111/j.1365-246X.2009.04491.x},
  urldate = {2026-07-21},
  language = {en}
}

@article{dieterich_1994_Constitutive,
  title = {A Constitutive Law for Rate of Earthquake Production and Its Application to Earthquake Clustering},
  author = {Dieterich, James},
  year = 1994,
  month = feb,
  journal = {Journal of Geophysical Research: Solid Earth},
  volume = {99},
  number = {B2},
  pages = {2601--2618},
  issn = {0148-0227},
  doi = {10.1029/93JB02581},
  urldate = {2026-07-11},
  language = {en}
}

@article{ferber_2026_Autonomous,
  title = {Towards Autonomous Medical Artificial Intelligence Agents},
  author = {Ferber, Dyke and Hilgers, Lars and H{\"o}per, Christiane and {Kinny-K{\"o}ster}, Benedict and Eckardt, Jan-Niklas and {Egger-Heidrich}, Katharina and Bill, Marius and Schneider, Martin M. K. and Clusmann, Jan and Kadric, Lejla and Oehme, Marcel and {Mayrhofer-Schmid}, Maximilian and Oeser, Alexander and W{\"o}lflein, Georg and Wiest, Isabella C. and Middeke, Jan Moritz and Iafrate, A. John and Truhn, Daniel and J{\"a}ger, Dirk and Kather, Jakob Nikolas},
  year = 2026,
  month = jun,
  journal = {Nature},
  issn = {0028-0836, 1476-4687},
  doi = {10.1038/s41586-026-10675-5},
  urldate = {2026-06-20},
  language = {en}
}

@article{ghareeb_2026_Multiagent,
  title = {A Multi-Agent System for Automating Scientific Discovery},
  author = {Ghareeb, Ali E. and Chang, Benjamin and Mitchener, Ludovico and Yiu, Angela and Szostkiewicz, Caralyn J. and Shved, Dmytro and Gyimesi, Gavin J. and Laurent, Jon M. and Wright, Samantha M. and Razzak, Muhammed T. and White, Andrew D. and Finnemann, Silvia C. and Hinks, Michaela M. and Rodriques, Samuel G.},
  year = 2026,
  month = jul,
  journal = {Nature},
  volume = {655},
  number = {8122},
  pages = {497--505},
  issn = {0028-0836, 1476-4687},
  doi = {10.1038/s41586-026-10652-y},
  urldate = {2026-06-20},
  language = {en}
}

@article{gottweis_2026_Accelerating,
  title = {Accelerating Scientific Discovery with {{Co-Scientist}}},
  author = {Gottweis, Juraj and Weng, Wei-Hung and Daryin, Alexander and Tu, Tao and Sirkovic, Petar and Myaskovsky, Artiom and Glowaty, Grzegorz and Weissenberger, Felix and Orlandi, Alessio and Popovici, Dan and Palepu, Anil and Rong, Keran and Tanno, Ryutaro and Saab, Khaled and Zhang, Fan and Blum, Jacob and Carroll, Andrew and Kulkarni, Kavita and Toma{\v s}ev, Nenad and Zverinski, Dina and Rendulic, Ivor and Vedadi, Elahe and Hasler, Florian and Rimanic, Luka and Boia, Marina and Budiselic, Ivan and Feinstein, Ben and Bellaiche, Mathias and Sheffer, Tom and Freyberg, Jan and Ratcliff, Jeremy and Bertolli, Ottavia and Chou, Katherine and Hassidim, Avinatan and Gokturk, Burak and Vahdat, Amin and Guan, Yuan and Dhillon, Vikram and Vaishnav, Eeshit Dhaval and Lee, Byron and Costa, Tiago R. D. and Penad{\'e}s, Jos{\'e} R. and Peltz, Gary and Matias, Yossi and Manyika, James and Hassabis, Demis and Xu, Yunhan and Kohli, Pushmeet and Pawlosky, Annalisa and Karthikesalingam, Alan and Natarajan, Vivek},
  year = 2026,
  month = jul,
  journal = {Nature},
  volume = {655},
  number = {8122},
  pages = {487--496},
  issn = {0028-0836, 1476-4687},
  doi = {10.1038/s41586-026-10644-y},
  urldate = {2026-06-20},
  language = {en}
}

@article{gulia_2019_Realtime,
  title = {Real-Time Discrimination of Earthquake Foreshocks and Aftershocks},
  author = {Gulia, Laura and Wiemer, Stefan},
  year = 2019,
  month = oct,
  journal = {Nature},
  volume = {574},
  number = {7777},
  pages = {193--199},
  issn = {0028-0836, 1476-4687},
  doi = {10.1038/s41586-019-1606-4},
  urldate = {2026-07-18},
  language = {en}
}

@article{hu_2024_DASPy,
  title = {{{DASPy}}: {{A Python Toolbox}} for {{DAS Seismology}}},
  author = {Hu, Minzhe and Li, Zefeng},
  year = 2024,
  month = sep,
  journal = {Seismological Research Letters},
  volume = {95},
  number = {5},
  pages = {3055--3066},
  issn = {0895-0695, 1938-2057},
  doi = {10.1785/0220240124},
  language = {en}
}

@article{huang_2020_Spatiotemporal,
  title = {Spatio-Temporal Foreshock Evolution of the 2019 {{M}} 6.4 and {{M}} 7.1 Ridgecrest, California Earthquakes},
  author = {Huang, Hui and Meng, Lingsen and B{\"u}rgmann, Roland and Wang, Wei and Wang, Kang},
  year = 2020,
  month = dec,
  journal = {Earth and Planetary Science Letters},
  volume = {551},
  pages = {116582},
  issn = {0012821X},
  doi = {10.1016/j.epsl.2020.116582},
  urldate = {2026-08-19},
  language = {en}
}

@article{hutton_1987_ML,
  title = {The {{{\emph{ML}}}} Scale in {{Southern California}}},
  author = {Hutton, L. K. and Boore, David M.},
  year = 1987,
  month = dec,
  journal = {Bulletin of the Seismological Society of America},
  volume = {77},
  number = {6},
  pages = {2074--2094},
  issn = {1943-3573, 0037-1106},
  doi = {10.1785/BSSA0770062074},
  urldate = {2026-08-10},
  language = {en}
}

@article{iinuma_2012_Coseismic,
  title = {Coseismic Slip Distribution of the 2011 off the {{Pacific Coast}} of {{Tohoku Earthquake}} ({{M9}}.0) Refined by Means of Seafloor Geodetic Data},
  author = {Iinuma, T. and Hino, R. and Kido, M. and Inazu, D. and Osada, Y. and Ito, Y. and Ohzono, M. and Tsushima, H. and Suzuki, S. and Fujimoto, H. and Miura, S.},
  year = 2012,
  month = jul,
  journal = {Journal of Geophysical Research: Solid Earth},
  volume = {117},
  number = {B7},
  pages = {2012JB009186},
  issn = {0148-0227},
  doi = {10.1029/2012JB009186},
  urldate = {2026-07-21},
  language = {en}
}

@misc{jma_2026_SeismologicalBulletin,
  title = {{{JMA}} Unified Hypocenter Catalog and Arrival-Time Data},
  author = {{Japan Meteorological Agency}},
  year = 2026,
  publisher = {{Japan Meteorological Agency}},
  url = {https://www.data.jma.go.jp/eqev/data/bulletin/index_e.html},
  urldate = {2026-07-29}
}

@inproceedings{kendrick_2019_Geologic,
  title = {Geologic Observations of Surface Fault Rupture Associated with the Ridgecrest {{M6}}. 4 and {{M7}}. 1 Earthquake Sequence by the Ridgecrest Rupture Mapping Group},
  booktitle = {2019 {{SCEC Annual Meeting}}},
  author = {Kendrick, K. J. and Akciz, S. O. and Angster, S. J. and Avouac, J. and Bachhuber, J. L. and Bennett, S. E. and Blake, K. and Bork, S. and Brooks, B. A. and Burgess, P.},
  year = 2019,
  language = {en}
}

@article{king_1994_Static,
  title = {Static Stress Changes and the Triggering of Earthquakes},
  author = {King, Geoffrey C. P. and Stein, Ross S. and Lin, Jian},
  year = 1994,
  month = jun,
  journal = {Bulletin of the Seismological Society of America},
  volume = {84},
  number = {3},
  pages = {935--953},
  issn = {1943-3573, 0037-1106},
  doi = {10.1785/BSSA0840030935},
  urldate = {2026-06-21},
  language = {en}
}

@misc{mirwald_2025_SeismoStats,
  title = {{{SeismoStats}}: A Python Package for Statistical Seismology},
  shorttitle = {{{SeismoStats}}},
  author = {Mirwald, Aron and Schmid, Nicolas and Mizrahi, Leila and Han, Marta and Rohnacher, Alicia and Ritz, Vanille A. and Wiemer, Stefan},
  year = 2025,
  month = nov,
  number = {arXiv:2511.04521},
  eprint = {2511.04521},
  primaryclass = {physics},
  publisher = {arXiv},
  doi = {10.48550/arXiv.2511.04521},
  urldate = {2025-11-10},
  archiveprefix = {arXiv},
  language = {en}
}

@article{moreno_2010_2010,
  title = {2010 {{Maule}} Earthquake Slip Correlates with Pre-Seismic Locking of {{Andean}} Subduction Zone},
  author = {Moreno, Marcos and Rosenau, Matthias and Oncken, Onno},
  year = 2010,
  month = sep,
  journal = {Nature},
  volume = {467},
  number = {7312},
  pages = {198--202},
  issn = {0028-0836, 1476-4687},
  doi = {10.1038/nature09349},
  urldate = {2026-07-29},
  language = {en}
}

@article{mousavi_2020_Earthquake,
  title = {Earthquake Transformer---an Attentive Deep-Learning Model for Simultaneous Earthquake Detection and Phase Picking},
  author = {Mousavi, S. Mostafa and Ellsworth, William L. and Zhu, Weiqiang and Chuang, Lindsay Y. and Beroza, Gregory C.},
  year = 2020,
  month = aug,
  journal = {Nature Communications},
  volume = {11},
  number = {1},
  pages = {3952},
  issn = {2041-1723},
  doi = {10.1038/s41467-020-17591-w},
  urldate = {2023-07-18},
  language = {en},
  lccn = {1}
}

@article{nadeau_1999_Fault,
  title = {Fault {{Slip Rates}} at {{Depth}} from {{Recurrence Intervals}} of {{Repeating Microearthquakes}}},
  author = {Nadeau, Robert M. and McEvilly, Thomas V.},
  year = 1999,
  month = jul,
  journal = {Science},
  volume = {285},
  number = {5428},
  pages = {718--721},
  issn = {0036-8075, 1095-9203},
  doi = {10.1126/science.285.5428.718},
  urldate = {2026-06-21},
  language = {en}
}

@article{nagai_2001_Comparative,
  title = {Comparative {{Study}} on the {{Source Processes}} of {{Recurrent Large Earthquakes}} in {{Sanriku-oki Region}}: The 1968 {{Tokachi-oki Earthquake}} and the 1994 {{Sanriku-oki Earthquake}}},
  shorttitle = {Comparative Study on the Source Processes of Recurrent Large Earthquakes in Sanriku-Oki Region},
  author = {Nagai, Riko and Kikuchi, Masayuki and Yamanaka, Yoshiko},
  year = 2001,
  journal = {Zisin (Journal of the Seismological Society of Japan. 2nd ser.)},
  volume = {54},
  number = {2},
  pages = {267--280},
  issn = {0037-1114, 1883-9029, 2186-599X},
  doi = {10.4294/zisin1948.54.2_267},
  urldate = {2026-07-18},
  language = {en}
}

@article{nanjo_2020_Were,
  title = {Were Changes in Stress State Responsible for the 2019 {{Ridgecrest}}, {{California}}, Earthquakes?},
  author = {Nanjo, K. Z.},
  year = 2020,
  month = jun,
  journal = {Nature Communications},
  volume = {11},
  number = {1},
  pages = {3082},
  issn = {2041-1723},
  doi = {10.1038/s41467-020-16867-5},
  urldate = {2026-02-02},
  language = {en}
}

@misc{nied_2019_HiNet,
  title = {{{NIED Hi-net}}},
  author = {{National Research Institute for Earth Science {and} Disaster Resilience}},
  year = 2019,
  publisher = {{National research institute for earth science and disaster resilience}},
  doi = {10.17598/nied.0003},
  urldate = {2026-07-29},
  language = {en}
}

@misc{nied_2019_SNet,
  title = {{{NIED S-net}}},
  author = {{National Research Institute for Earth Science {and} Disaster Resilience}},
  year = 2019,
  publisher = {{National Research Institute for Earth Science and Disaster Resilience}},
  doi = {10.17598/nied.0007},
  urldate = {2026-08-19},
  language = {en}
}

@article{nishikawa_2023_Review,
  title = {A Review on Slow Earthquakes in the {{Japan Trench}}},
  author = {Nishikawa, Tomoaki and Ide, Satoshi and Nishimura, Takuya},
  year = 2023,
  month = jan,
  journal = {Progress in Earth and Planetary Science},
  volume = {10},
  number = {1},
  pages = {1},
  issn = {2197-4284},
  doi = {10.1186/s40645-022-00528-w},
  urldate = {2026-06-09},
  language = {en}
}

@article{obara_2016_Connecting,
  title = {Connecting Slow Earthquakes to Huge Earthquakes},
  author = {Obara, Kazushige and Kato, Aitaro},
  year = 2016,
  month = jul,
  journal = {Science},
  volume = {353},
  number = {6296},
  pages = {253--257},
  issn = {0036-8075, 1095-9203},
  doi = {10.1126/science.aaf1512},
  urldate = {2026-06-09},
  language = {en}
}

@article{ogata_1988_Statistical,
  title = {Statistical {{Models}} for {{Earthquake Occurrences}} and {{Residual Analysis}} for {{Point Processes}}},
  author = {Ogata, Yosihiko},
  year = 1988,
  month = mar,
  journal = {Journal of the American Statistical Association},
  volume = {83},
  number = {401},
  pages = {9--27},
  issn = {0162-1459, 1537-274X},
  doi = {10.1080/01621459.1988.10478560},
  urldate = {2026-02-21},
  language = {en}
}

@misc{omori_1895_Aftershocks,
  title = {On the After-Shocks of Earthquakes},
  author = {Omori, Fusakichi},
  year = 1895,
  number = {2},
  eprint = {2},
  publisher = {College of Science, Imperial University},
  doi = {10.15083/00037562},
  urldate = {2026-04-03},
  archiveprefix = {College of Science, Imperial University},
  language = {en}
}

@article{peng_2009_Migration,
  title = {Migration of Early Aftershocks Following the 2004 {{Parkfield}} Earthquake},
  author = {Peng, Zhigang and Zhao, Peng},
  year = 2009,
  month = dec,
  journal = {Nature Geoscience},
  volume = {2},
  number = {12},
  pages = {877--881},
  issn = {1752-0894, 1752-0908},
  doi = {10.1038/ngeo697},
  urldate = {2026-03-15},
  language = {en}
}

@article{peng_2010_Integrated,
  title = {An Integrated Perspective of the Continuum between Earthquakes and Slow-Slip Phenomena},
  author = {Peng, Zhigang and Gomberg, Joan},
  year = 2010,
  month = sep,
  journal = {Nature Geoscience},
  volume = {3},
  number = {9},
  pages = {599--607},
  issn = {1752-0894, 1752-0908},
  doi = {10.1038/ngeo940},
  urldate = {2026-06-09},
  language = {en}
}

@article{romera-paredes_2024_Mathematical,
  title = {Mathematical Discoveries from Program Search with Large Language Models},
  author = {{Romera-Paredes}, Bernardino and Barekatain, Mohammadamin and Novikov, Alexander and Balog, Matej and Kumar, M. Pawan and Dupont, Emilien and Ruiz, Francisco J. R. and Ellenberg, Jordan S. and Wang, Pengming and Fawzi, Omar and Kohli, Pushmeet and Fawzi, Alhussein},
  year = 2024,
  month = jan,
  journal = {Nature},
  volume = {625},
  number = {7995},
  pages = {468--475},
  issn = {0028-0836, 1476-4687},
  doi = {10.1038/s41586-023-06924-6},
  urldate = {2026-06-09},
  language = {en}
}

@article{ross_2019_Hierarchical,
  title = {Hierarchical Interlocked Orthogonal Faulting in the 2019 {{Ridgecrest}} Earthquake Sequence},
  author = {Ross, Zachary E. and Idini, Benjam{\'i}n and Jia, Zhe and Stephenson, Oliver L. and Zhong, Minyan and Wang, Xin and Zhan, Zhongwen and Simons, Mark and Fielding, Eric J. and Yun, Sang-Ho and Hauksson, Egill and Moore, Angelyn W. and Liu, Zhen and Jung, Jungkyo},
  year = 2019,
  month = oct,
  journal = {Science},
  volume = {366},
  number = {6463},
  pages = {346--351},
  issn = {0036-8075, 1095-9203},
  doi = {10.1126/science.aaz0109},
  urldate = {2025-10-28},
  language = {en}
}

@misc{scedc_2013_Southern,
  title = {Southern {{California Earthquake Center}}},
  author = {{SCEDC}},
  year = 2013,
  publisher = {Caltech},
  doi = {10.7909/C3WD3XH1},
  urldate = {2026-03-18},
  language = {en}
}

@article{scholz_2015_Stress,
  title = {On the Stress Dependence of the Earthquake {\emph{b}} Value},
  author = {Scholz, Christopher H.},
  year = 2015,
  month = mar,
  journal = {Geophysical Research Letters},
  volume = {42},
  number = {5},
  pages = {1399--1402},
  issn = {0094-8276, 1944-8007},
  doi = {10.1002/2014GL062863},
  urldate = {2026-06-21},
  language = {en}
}

@book{scholz_2019_Mechanics,
  title = {The Mechanics of Earthquakes and Faulting},
  shorttitle = {Earthquake {{Mechanics}}},
  author = {Scholz, Christopher H.},
  year = 2019,
  edition = {3rd},
  publisher = {Cambridge University Press},
  address = {Cambridge, UK},
  doi = {10.1017/9781316681473},
  urldate = {2026-04-03},
  isbn = {978-1-316-68147-3},
  language = {en}
}

@article{schorlemmer_2005_Variations,
  title = {Variations in Earthquake-Size Distribution across Different Stress Regimes},
  author = {Schorlemmer, Danijel and Wiemer, Stefan and Wyss, Max},
  year = 2005,
  month = sep,
  journal = {Nature},
  volume = {437},
  number = {7058},
  pages = {539--542},
  issn = {0028-0836, 1476-4687},
  doi = {10.1038/nature04094},
  urldate = {2026-06-21},
  language = {en}
}

@article{shapiro_1997_Estimating,
  title = {Estimating the Crust Permeability from Fluid-Injection-Induced Seismic Emission at the {{KTB}} Site},
  author = {Shapiro, Sergei A. and Huenges, Ernst and Borm, G{\"u}nter},
  year = 1997,
  month = nov,
  journal = {Geophysical Journal International},
  volume = {131},
  number = {2},
  pages = {F15-F18},
  issn = {0956540X, 1365246X},
  doi = {10.1111/j.1365-246X.1997.tb01215.x},
  urldate = {2026-02-21},
  language = {en}
}

@article{shelly_2007_Nonvolcanic,
  title = {Non-Volcanic Tremor and Low-Frequency Earthquake Swarms},
  author = {Shelly, David R. and Beroza, Gregory C. and Ide, Satoshi},
  year = 2007,
  month = mar,
  journal = {Nature},
  volume = {446},
  number = {7133},
  pages = {305--307},
  issn = {0028-0836, 1476-4687},
  doi = {10.1038/nature05666},
  urldate = {2026-06-09},
  language = {en}
}

@article{shi_1982_Standard,
  title = {The Standard Error of the Magnitude-Frequency {\emph{b}} Value},
  author = {Shi, Yaolin and Bolt, Bruce A.},
  year = 1982,
  month = oct,
  journal = {Bulletin of the Seismological Society of America},
  volume = {72},
  number = {5},
  pages = {1677--1687},
  issn = {1943-3573, 0037-1106},
  doi = {10.1785/BSSA0720051677},
  urldate = {2026-07-21},
  language = {en}
}

@article{stein_1999_Role,
  title = {The Role of Stress Transfer in Earthquake Occurrence},
  author = {Stein, Ross S.},
  year = 1999,
  month = dec,
  journal = {Nature},
  volume = {402},
  number = {6762},
  pages = {605--609},
  issn = {0028-0836, 1476-4687},
  doi = {10.1038/45144},
  urldate = {2026-02-21},
  language = {en}
}

@article{toda_2020_Long,
  title = {Long- and {{Short-Term Stress Interaction}} of the 2019 {{Ridgecrest Sequence}} and {{Coulomb-Based Earthquake Forecasts}}},
  author = {Toda, Shinji and Stein, Ross S.},
  year = 2020,
  month = aug,
  journal = {Bulletin of the Seismological Society of America},
  volume = {110},
  number = {4},
  pages = {1765--1780},
  issn = {0037-1106, 1943-3573},
  doi = {10.1785/0120200169},
  urldate = {2026-07-11},
  language = {en}
}

@article{uchida_2003_Interplate,
  title = {Interplate Quasi-static Slip off {{Sanriku}}, {{NE Japan}}, Estimated from Repeating Earthquakes},
  author = {Uchida, Naoki and Matsuzawa, Toru and Hasegawa, Akira and Igarashi, Toshihiro},
  year = 2003,
  month = aug,
  journal = {Geophysical Research Letters},
  volume = {30},
  number = {15},
  pages = {2003GL017452},
  issn = {0094-8276, 1944-8007},
  doi = {10.1029/2003GL017452},
  urldate = {2026-07-18},
  language = {en}
}

@article{uchida_2016_Periodic,
  title = {Periodic Slow Slip Triggers Megathrust Zone Earthquakes in Northeastern {{Japan}}},
  author = {Uchida, Naoki and Iinuma, Takeshi and Nadeau, Robert M. and B{\"u}rgmann, Roland and Hino, Ryota},
  year = 2016,
  month = jan,
  journal = {Science},
  volume = {351},
  number = {6272},
  pages = {488--492},
  issn = {0036-8075, 1095-9203},
  doi = {10.1126/science.aad3108},
  urldate = {2026-07-18},
  language = {en}
}

@article{uchida_2019_Repeating,
  title = {Repeating {{Earthquakes}}},
  author = {Uchida, Naoki and B{\"u}rgmann, Roland},
  year = 2019,
  month = may,
  journal = {Annual Review of Earth and Planetary Sciences},
  volume = {47},
  number = {1},
  pages = {305--332},
  issn = {0084-6597, 1545-4495},
  doi = {10.1146/annurev-earth-053018-060119},
  urldate = {2026-06-21},
  language = {en}
}

@misc{USGS_2017_Advanced,
  title = {Advanced {{National Seismic System}} ({{ANSS}}) {{Comprehensive Catalog}}},
  author = {{U. S. Geological Survey}},
  year = 2017,
  publisher = {U.S. Geological Survey},
  doi = {10.5066/F7MS3QZH},
  urldate = {2026-06-30}
}

@article{waldhauser_2000_DoubleDifference,
  title = {A {{Double-Difference Earthquake Location Algorithm}}: {{Method}} and {{Application}} to the {{Northern Hayward Fault}}, {{California}}},
  shorttitle = {A Double-Difference Earthquake Location Algorithm},
  author = {Waldhauser, F. and Ellsworth, W. L.},
  year = 2000,
  month = dec,
  journal = {Bulletin of the Seismological Society of America},
  volume = {90},
  number = {6},
  pages = {1353--1368},
  issn = {0037-1106},
  doi = {10.1785/0120000006},
  urldate = {2026-02-21},
  language = {en}
}

@article{wang_2023_Scientific,
  title = {Scientific Discovery in the Age of Artificial Intelligence},
  author = {Wang, Hanchen and Fu, Tianfan and Du, Yuanqi and Gao, Wenhao and Huang, Kexin and Liu, Ziming and Chandak, Payal and Liu, Shengchao and Van Katwyk, Peter and Deac, Andreea and Anandkumar, Anima and Bergen, Karianne and Gomes, Carla P. and Ho, Shirley and Kohli, Pushmeet and Lasenby, Joan and Leskovec, Jure and Liu, Tie-Yan and Manrai, Arjun and Marks, Debora and Ramsundar, Bharath and Song, Le and Sun, Jimeng and Tang, Jian and Veli{\v c}kovi{\'c}, Petar and Welling, Max and Zhang, Linfeng and Coley, Connor W. and Bengio, Yoshua and Zitnik, Marinka},
  year = 2023,
  month = aug,
  journal = {Nature},
  volume = {620},
  number = {7972},
  pages = {47--60},
  issn = {0028-0836, 1476-4687},
  doi = {10.1038/s41586-023-06221-2},
  urldate = {2024-11-11},
  language = {en}
}

@inproceedings{wei_2022_Chainofthought,
  title = {Chain-{{Of-Thought Prompting Elicits Reasoning}} in {{Large Language Models}}},
  booktitle = {Advances in {{Neural Information Processing Systems}} 35},
  author = {Wei, Jason and Wang, Xuezhi and Schuurmans, Dale and Bosma, Maarten and Ichter, Brian and Xia, Fei and Chi, Ed and Le, Quoc V and Zhou, Denny},
  year = 2022,
  pages = {24824--24837},
  publisher = {Curran Associates, Inc.},
  address = {New Orleans, Louisiana, USA},
  doi = {10.52202/068431-1800},
  urldate = {2026-02-21},
  language = {en}
}

@article{wirth_2022_Occurrence,
  title = {The Occurrence and Hazards of Great Subduction Zone Earthquakes},
  author = {Wirth, Erin A. and Sahakian, Valerie J. and Wallace, Laura M. and Melnick, Daniel},
  year = 2022,
  month = jan,
  journal = {Nature Reviews Earth \& Environment},
  volume = {3},
  number = {2},
  pages = {125--140},
  issn = {2662-138X},
  doi = {10.1038/s43017-021-00245-w},
  urldate = {2026-07-29},
  language = {en}
}

@article{woollam_2022_SeisBench,
  title = {{{SeisBench-A Toolbox}} for {{Machine Learning}} in {{Seismology}}},
  author = {Woollam, Jack and M{\"u}nchmeyer, Jannes and Tilmann, Frederik and Rietbrock, Andreas and Lange, Dietrich and Bornstein, Thomas and Diehl, Tobias and Giunchi, Carlo and Haslinger, Florian and Jozinovi{\'c}, Dario and Michelini, Alberto and Saul, Joachim and Soto, Hugo},
  year = 2022,
  month = may,
  journal = {Seismological Research Letters},
  volume = {93},
  number = {3},
  eprint = {2111.00786},
  primaryclass = {physics},
  pages = {1695--1709},
  issn = {0895-0695, 1938-2057},
  doi = {10.1785/0220210324},
  urldate = {2023-11-20},
  archiveprefix = {arXiv},
  language = {en}
}

@article{yamanaka_2004_Asperity,
  title = {Asperity Map along the Subduction Zone in Northeastern {{Japan}} Inferred from Regional Seismic Data},
  author = {Yamanaka, Yoshiko and Kikuchi, Masayuki},
  year = 2004,
  month = jul,
  journal = {Journal of Geophysical Research: Solid Earth},
  volume = {109},
  number = {B7},
  pages = {2003JB002683},
  issn = {0148-0227},
  doi = {10.1029/2003JB002683},
  urldate = {2026-07-18},
  language = {en}
}

@article{yoshida_2026_Evolution,
  title = {Evolution of {{Coupled Seismic}}-{{Aseismic Slip During}} the 2025 {{Mw}} 6.8 {{Sanriku}}-{{Oki}}, {{Japan}}, {{Megathrust Sequence}}},
  author = {Yoshida, Keisuke},
  year = 2026,
  month = jul,
  journal = {Geophysical Research Letters},
  volume = {53},
  number = {13},
  pages = {e2026GL121704},
  issn = {0094-8276, 1944-8007},
  doi = {10.1029/2026GL121704},
  urldate = {2026-06-30},
  language = {en}
}

@article{yue_2021_2019,
  title = {The 2019 {{Ridgecrest}}, {{California}} Earthquake Sequence: {{Evolution}} of Seismic and Aseismic Slip on an Orthogonal Fault System},
  shorttitle = {The 2019 Ridgecrest, California Earthquake Sequence},
  author = {Yue, Han and Sun, Jianbao and Wang, Min and Shen, Zhengkang and Li, Mingjia and Xue, Lian and Lu, Weifan and Zhou, Yijian and Ren, Chunmei and Lay, Thorne},
  year = 2021,
  month = sep,
  journal = {Earth and Planetary Science Letters},
  volume = {570},
  pages = {117066},
  issn = {0012821X},
  doi = {10.1016/j.epsl.2021.117066},
  urldate = {2026-07-11},
  language = {en}
}

@article{zhang_2022_LOCFLOW,
  title = {{{LOC-FLOW}}: {{An End-to-End Machine Learning-Based High-Precision Earthquake Location Workflow}}},
  shorttitle = {Loc-Flow},
  author = {Zhang, Miao and Liu, Min and Feng, Tian and Wang, Ruijia and Zhu, Weiqiang},
  year = 2022,
  month = sep,
  journal = {Seismological Research Letters},
  volume = {93},
  number = {5},
  pages = {2426--2438},
  issn = {0895-0695, 1938-2057},
  doi = {10.1785/0220220019},
  urldate = {2025-06-30},
  language = {en}
}

@article{zhu_2018_PhaseNet,
  title = {{{PhaseNet}}: {{A Deep-Neural-Network-Based Seismic Arrival Time Picking Method}}},
  shorttitle = {{{PhaseNet}}},
  author = {Zhu, Weiqiang and Beroza, Gregory C},
  year = 2019,
  journal = {Geophysical Journal International},
  volume = {216},
  pages = {261--273},
  issn = {0956-540X, 1365-246X},
  doi = {10.1093/gji/ggy423},
  urldate = {2022-12-10},
  language = {en},
  lccn = {2.934}
}

@article{zhu_2022_Earthquake,
  title = {Earthquake {{Phase Association Using}} a {{Bayesian Gaussian Mixture Model}}},
  author = {Zhu, Weiqiang and McBrearty, Ian W. and Mousavi, S. Mostafa and Ellsworth, William L. and Beroza, Gregory C.},
  year = 2022,
  month = may,
  journal = {Journal of Geophysical Research: Solid Earth},
  volume = {127},
  number = {5},
  eprint = {2109.09008},
  primaryclass = {physics},
  pages = {e2021JB023249},
  issn = {2169-9313, 2169-9356},
  doi = {10.1029/2021JB023249},
  urldate = {2023-11-20},
  archiveprefix = {arXiv},
  language = {en}
}
\bibliographystyle{sciencemag}






\paragraph*{Data and code availability:}
Data supporting the findings of this study are available from the sources described below and from the TRACE repository (\url{https://github.com/liufeng2317/TRACE-review-materials}). Continuous waveform data used for the 2019 Ridgecrest analysis were obtained from the Southern California Earthquake Data Center (SCEDC) \cite{scedc_2013_Southern}. For the Sanriku analysis, hypocenter, magnitude and phase-arrival records for 1 June 2025 to 2 May 2026 were obtained from the JMA unified hypocenter catalog and associated arrival-time products served through JMA/NIED data services \cite{jma_2026_SeismologicalBulletin}. Continuous S-net waveform records evaluated during catalog-entry quality control were obtained from NIED \cite{nied_2019_SNet}. Event waveform packages and supplemental daily continuous waveforms used for repeating-earthquake analysis were obtained from NIED Hi-net/JMA waveform services \cite{nied_2019_HiNet}. Access to the S-net and Hi-net/JMA waveform records requires standard NIED user registration; Hi-net and S-net are components of the MOWLAS observing system \cite{aoi_2020_MOWLAS}. Primary waveform and arrival-time data are available from the original providers under their stated access conditions and are not redistributed in the repository. Derived TRACE catalogs, relocation outputs, waveform-selection tables, repeating-earthquake pair and family tables, benchmark records and figure source data are provided in the TRACE repository. Historical rupture and slip models and convergence information used as contextual constraints are available from the cited literature. Independent S-net strain observations used as external validation evidence and source-mechanism, tremor and coupling products identified for further independent testing are available from the cited public Japanese data products and literature sources. These external data were not used as inputs to the earthquake-catalog relocation or repeating-earthquake analyses.



\subsection*{Supplementary materials}
Materials and Methods\\
Supplementary Text\\
Supplementary Figs. S1 to S13\\
Supplementary Tables S1 to S11\\
References \textit{(52--59)}


\newpage

\renewcommand{\thefigure}{S\arabic{figure}}
\renewcommand{\thetable}{S\arabic{table}}
\renewcommand{\theequation}{S\arabic{equation}}
\renewcommand{\thepage}{S\arabic{page}}
\setcounter{figure}{0}
\setcounter{table}{0}
\setcounter{equation}{0}
\setcounter{page}{1}

\clearpage
\setcounter{page}{1}
\renewcommand{\figurename}{Supplementary Fig.}
\renewcommand{\tablename}{Supplementary Table}

\begin{center}
{\large\bfseries Supplementary Materials for\\[2mm]}
{\large\bfseries \scititlesupp}
\end{center}

    This Supplementary Information is organized as a verification chain. It contains six Supplementary Notes, 13 Supplementary Figures and 11 Supplementary Tables. Supplementary Notes 1 and 2 define the general TRACE workflow, provenance records and evaluation protocol. Supplementary Notes 3 and 4 provide the Ridgecrest catalog-construction record and the seismicity evidence for delayed cascading activation. Supplementary Notes 5 and 6 provide the Sanriku analysis-region definition, stage-resolved activation, JMA magnitude $M_J$ 6.9 migration diagnostics, repeating-earthquake evidence, external validation evidence and proposed independent tests.

    The Supplementary Materials begin with Materials and Methods. Methodological figures and tables are presented in that section; the remaining Supplementary Text is organized into general workflow and benchmark materials, the Ridgecrest case study and the Sanriku case study. Figures and tables in the Supplementary Text are placed near the note to which they contribute, while their numbering follows the order of appearance in this Supplementary Information.

    \section*{Materials and Methods}

    \subsection*{Implementation of TRACE}

    \subsubsection*{System overview and input-output definition}

    TRACE is implemented as a constrained multi-agent workflow for earthquake-sequence analysis (Fig.~\ref{fig:fig1_workflow}a). It takes as input a seismological inference task, available observations or data-access instructions, optional literature or user-provided context, and a resource layer consisting of the Structured Knowledge Library and Validated Scientific Tools Library. The Structured Knowledge Library includes an Expert-Validated Case Library of reviewed workflows and mechanism summaries. Processing these inputs, TRACE returns a fully executable analytical pipeline, intermediate diagnostics, domain-specific evidence summaries and an evidence-supported physical interpretation, with viable alternatives and directions for independent testing. This structured end-to-end architecture allows any final interpretation to be transparently traced back to specific data products, algorithm choices, parameter settings, and diagnostic checks.

    The architecture of TRACE consists of three integrated modules designed to bridge the gap between an open-ended seismological inquiry and a rigorously traceable physical interpretation. First, the Planning and Task Decomposition Module formulates candidate physical hypotheses and executable analytical workflows. Next, the Automated Execution and Diagnostics Module deploys the required scientific tools, validating intermediate products against strict computational and geophysical criteria. Finally, the Scientific Integration and Inference Module synthesizes these validated outputs into structured evidence, forms a physical interpretation and evaluates viable alternatives against that evidence. The specific roles and information flow across this pipeline are detailed below and summarized in Supplementary Fig.~\ref{fig:ed_trace_architecture}.

    \subsubsection*{Core agent-based modules}

    \paragraph{Planning and Task Decomposition Module}

    The Planning Module translates a natural-language description of a seismological inference task into candidate physical interpretations and analysis pathways. Its purpose is to reduce an open-ended task to operations that can be executed by validated tools while retaining the physical alternatives under consideration.

    The \textbf{Planning Agent} first parses the task to identify the target sequence, relevant data streams, requested outputs, and candidate physical mechanisms. Constrained by data availability and optional user context, the agent narrows the analysis to feasible workflows. It then queries the Structured Knowledge Library, including the Expert-Validated Case Library, to retrieve seismological constraints, method applicability rules, and curated workflow templates. The agent also accesses the Validated Scientific Tools Library for callable functions, converting these resources into candidate analysis plans structured as ordered sequences of scientific operations and tool calls. Examples include waveform retrieval, event detection, phase picking, association, relocation, rate analysis, magnitude-frequency analysis, repeating-earthquake detection and source-mechanism comparison.

    A \textbf{Plan Aggregation Agent} removes redundant plans and filters the remaining candidates according to data availability, methodological consistency, physical plausibility and user-specified constraints. When essential assumptions remain ambiguous, an optional human-in-the-loop step can confirm the data window, accepted tools or interpretation scope. Any human-confirmed constraint is retained with the selected plan, which is then passed to the execution module.

    The planning-to-workflow transition is summarized as
    \begin{equation}
        \begin{aligned}
        \mathcal{P}&=\mathcal{G}(Q,D,L\mid\mathcal{K},\mathcal{T}),
        \qquad
        p^\star=\mathcal{A}(\mathcal{P}\mid\mathcal{C},\mathcal{H}),\\
        w^{(0)}&=\mathcal{W}(p^\star),
        \end{aligned}
    \end{equation}
    where $Q$, $D$ and $L$ denote the task, available data and optional context; $\mathcal{K}$ and $\mathcal{T}$ denote the knowledge and tool libraries; and $\mathcal{C}$ and $\mathcal{H}$ denote explicit constraints and optional human confirmation. The operators $\mathcal{G}$, $\mathcal{A}$ and $\mathcal{W}$ denote candidate-plan generation, plan aggregation and compilation of the selected plan into the initial executable workflow $w^{(0)}$, respectively.

    \paragraph{Automated Execution and Diagnostics Module}

    Before execution, a \textbf{Workflow Agent} converts the selected plan into a machine-executable workflow with explicit task dependencies, input-output schemas, parameter values and data-flow constraints. At the task level, the workflow contains operations such as waveform preprocessing, phase picking, event association, hypocentral location, relocation, catalog filtering, statistical analysis and visualization. At the operation level, each task is decomposed into specific function calls, for example, detrending, instrument-response removal, resampling, band-pass filtering or double-difference relocation.

    A \textbf{Coding Agent} instantiates the workflow as executable scripts by combining its operation definitions, callable functions from the Validated Scientific Tools Library, and task examples or documentation from the Structured Knowledge Library. Tasks with sequential dependencies are executed serially, whereas independent tasks can be parallelized. Each execution returns intermediate products, log files and diagnostic messages. These outputs are stored with their associated workflow step so that later interpretations can be traced to specific operations.

    A \textbf{Result Checking Agent} evaluates the intermediate and final products against predefined computational and physical criteria. These checks include file validity, numerical ranges, catalog completeness, spatial and temporal consistency, convergence of inversion or relocation routines, and plausibility of derived seismicity statistics. A visualization diagnostic step checks whether figures contain the expected data layers, labels and physically meaningful ranges. If a diagnostic fails, TRACE revises the workflow within allowed operations, for example, by adjusting a parameter, repairing code, changing a data subset or selecting an alternative tool. Persistent differences among valid alternatives are retained as separate evidence branches rather than forced into a single output.

    The diagnostic loop is expressed as
    \begin{equation}
        \begin{aligned}
        O^{(k)}&=\mathcal{E}\!\left[\mathcal{F}\!\left(w^{(k)}\mid\mathcal{T},\mathcal{K}\right)\right],\\
        w^{(k+1)}&=\mathcal{M}\!\left(w^{(k)},O^{(k)}\right)
        \quad \text{if } \mathcal{V}\!\left(O^{(k)}\right)=0,
        \end{aligned}
    \end{equation}
    where $k$ denotes the iteration, $\mathcal{F}$ code instantiation, $\mathcal{E}$ execution, $\mathcal{V}$ diagnostic validation and $\mathcal{M}$ a permitted workflow or code revision. Each iteration stores the executed code, inputs, outputs, diagnostic result and any permitted revision. The loop stops when $\mathcal{V}(O^{(k)})=1$ or when the remaining failures cannot be resolved without user intervention or additional data. Only outputs satisfying the applicable diagnostic criteria are denoted $O$ and passed to evidence synthesis; unresolved failures remain in the provenance record and are not used as supporting evidence.

    \paragraph{Scientific Integration and Inference Module}

    The \textit{Scientific Integration and Inference Module} converts validated workflow outputs into structured evidence summaries and classifies their relations to candidate physical interpretations as supporting, inconsistent or unresolved.

    Perspective-specific \textbf{Analysis Agents} extract evidence from predefined or dynamically selected analytical perspectives. For example, a spatiotemporal perspective evaluates migration, clustering, activation timing and spatial density; a statistical perspective evaluates seismicity rate, Omori-Utsu behavior and magnitude-frequency structure; and waveform and source perspectives evaluate waveform similarity, repeating-earthquake families and source characteristics. Each structured evidence summary records the source data product, method, parameters, diagnostic status and interpretation-relevant findings. This record links the final physical interpretation to the executed workflow and intermediate outputs.

    Evidence extraction and cross-perspective synthesis are summarized as
    \begin{equation}
        S_v=E_v(O),
        \qquad
        R=\mathcal{I}\!\left(\{S_v\}_{v\in\mathcal{R}}\mid\Theta,\mathcal{K}\right),
    \end{equation}
    where $E_v$ extracts the evidence summary for perspective $v$, $\mathcal{R}$ is the set of analytical perspectives, $\Theta$ is the set of candidate physical interpretations, and $R$ is the final physical interpretation report. The integration operator $\mathcal{I}$ applies the recorded seismological constraints to the evidence summaries and candidate interpretations.

    For problems involving heterogeneous observations or multiple viable physical interpretations, a \textbf{Cross-Perspective Analysis Agent} compares each evidence summary with each candidate interpretation and classifies the resulting relation as supporting, inconsistent or unresolved. The final report retains an evidence-supported interpretation, its evidence chain, viable alternatives and external validation targets. This process is a rule-constrained synthesis of recorded evidence rather than a statistical inversion or a learned geophysical model. The Ridgecrest and Sanriku applications used this report structure.

    \subsubsection*{Constrained reasoning engine and LLM configuration}

    The LLM backbone supports semantic parsing, workflow synthesis, code generation and evidence extraction. It also synthesizes candidate physical interpretations from checked workflow outputs and explicitly supplied contextual constraints retained in the provenance record. Numerical and graphical products are generated by the scientific tools. LLM-generated plans, code and summaries are constrained by the Structured Knowledge Library, tool-interface schemas, physical plausibility checks and diagnostic feedback from executed code.

    In the experiments reported here, GPT-5.1 was used as the primary LLM backbone for all agents. The cross-backbone comparison evaluated GPT-5.1, Claude Sonnet 4 and Gemini 3 Flash preview across the 104 Level 1 and Level 2 tasks, yielding 312 individual model--task runs. All backbones used the same task instructions, input data, system prompts, tool schemas, validation rules and retry settings. The reproducibility archive specifies provider model identifiers, access dates, decoding and randomization settings, system prompts, tool schemas and retry rules. These conditions define the cross-backbone comparison summarized in Supplementary Fig.~\ref{fig:ed_trace_benchmark}.

    \subsubsection*{Knowledge and tool libraries}
    The resource layer contains the scientific rules, reviewed examples and callable software used by the planning, execution and synthesis modules. It comprises a \textit{Structured Knowledge Library}, including an \textit{Expert-Validated Case Library}, and a \textit{Validated Scientific Tools Library}.

    \paragraph{Structured Knowledge Library}

    The Structured Knowledge Library encodes methodological rules, parameter ranges, workflow templates and documented failure modes derived from established seismological practice. Unlike the Validated Scientific Tools Library, which provides executable functions, the Structured Knowledge Library defines when a method is appropriate, what input data it requires, what assumptions it makes and what diagnostics should be checked.

    The library is organized by analysis class, including waveform processing, phase picking, association, earthquake location, relocation, magnitude estimation, catalog statistics, waveform similarity and source-mechanism analysis. Each entry contains a short method description, required inputs, tunable parameters, recommended ranges, expected outputs and failure conditions. The Expert-Validated Case Library stores reviewed workflow records and mechanism summaries from completed analyses. These materials provide reusable examples without replacing case-specific diagnostic checks. The library can be updated with new papers, code repositories and software documentation, but the analyses reported here used the curated resource set available at execution time.

    \paragraph{Validated Scientific Tools Library}

    The Validated Scientific Tools Library contains open-source software and expert-developed modules for seismic data acquisition, preprocessing and analysis. Here, validated means that each tool is either established in seismological practice or implements an established method, and that its implementation within TRACE has been tested against reference cases with outputs reviewed by domain experts. This validation concerns the scientific basis and operational reliability of the tool; case-specific applicability, parameter choices and outputs remain subject to workflow-level evaluation. Reference-case outputs and expert-review records for the included tools are retained in the reproducibility archive. The reported case studies used ObsPy \cite{beyreuther_2010_ObsPy}, PhaseNet \cite{zhu_2018_PhaseNet}, GaMMA \cite{zhu_2022_Earthquake}, HypoDD \cite{waldhauser_2000_DoubleDifference}, and case-specific catalog-statistics and waveform-similarity routines. The broader library also includes SeisBench \cite{woollam_2022_SeisBench}, DASPy \cite{hu_2024_DASPy}, EQcorrscan \cite{chamberlain_2018_EQcorrscan}, SeismoStats \cite{mirwald_2025_SeismoStats}, ETAS-type point-process modeling \cite{ogata_1988_Statistical}, and general scientific Python libraries. These additional tools supported benchmark tasks or remained available for alternative workflows.

    Each tool is represented by metadata describing its inputs, outputs, file formats, parameter ranges, operational limits and physical constraints. The Coding Agent uses this metadata to assemble executable workflows and, where needed, to implement procedures from heterogeneous software environments as standardized Python scripts. The implementation does not assume that a single workflow is globally optimal. When several tools or parameterizations are plausible, TRACE can execute and compare the alternatives or retain their differing outputs in the evidence record.

    \subsection*{Application workflows and input boundaries}

    The Ridgecrest and Sanriku applications used the same provenance scheme but began at different points in the observation-to-interpretation workflow. No case-specific target interpretation was supplied to the planning, execution or inference modules; contextual constraints and external evidence introduced during synthesis are described below. Both workflows retained their data sources and selections, tool choices, parameter settings, intermediate products, diagnostic checks, revisions and final evidence summaries (Supplementary Note 1).

    For Ridgecrest, TRACE operated in a continuous-waveform entry mode. It retrieved continuous three-component records and station metadata from the Southern California Earthquake Data Center for 4--26 July 2019 and preprocessed the waveforms \cite{scedc_2013_Southern}. TRACE utilized the PhaseNet model for seismic phase detection and arrival time picking, GaMMA for association and location estimation, and HypoDD for relocation \cite{zhu_2018_PhaseNet,zhu_2022_Earthquake,waldhauser_2000_DoubleDifference}. Mapped surface-rupture traces compiled by the Ridgecrest Rupture Mapping Group were supplied as an auxiliary geospatial dataset for quantitative distance and orientation analyses \cite{kendrick_2019_Geologic}. The resulting catalog was compared with independent operational and high-resolution catalogs for quality control and documented cross-catalog alignment before subsequent analyses of activation timing, fault organization, seismicity-rate evolution and magnitude-frequency distributions \cite{USGS_2017_Advanced,ross_2019_Hierarchical}. These catalog comparisons and alignments did not prescribe the subsequent physical interpretation. TRACE analyzed the data in this interval without a prescribed target interpretation; published Ridgecrest interpretations were used only for retrospective comparison after the analysis was complete. Supplementary Notes 3 and 4 document the processing parameters, intermediate products, catalog quality control and diagnostic evidence.

    For Sanriku, TRACE had access to both waveform records and official JMA catalog products. It operated in a trusted-catalog mode for relocation, beginning with the JMA event catalog and station-level $P$- and $S$-arrival records for 1 June 2025 to 2 May 2026 \cite{jma_2026_SeismologicalBulletin}. TRACE matched event identifiers to phase records, performed monthly double-difference relocation, and compared the $M_J$ 6.9, regional $M$ 7.5 and $M_J$ 7.7 clusters using distance, depth compatibility and shallow-domain support. No physical relation between these sequences was prescribed. Seismicity within the selected $M_J$ 6.9--$M_J$ 7.7 region was analyzed across fixed stages using seismicity-rate and magnitude-frequency diagnostics. Separate short-timescale diagnostics characterized migration around the $M_J$ 6.9 sequence, and a separately assembled waveform dataset was used for repeating-earthquake analysis. Historical rupture models and convergence estimates were used in the final fault-system synthesis. Independent S-net borehole strain observations were used during this synthesis as external validation evidence for coupled seismic and aseismic activity around the $M_J$ 6.9 sequence. Additional tremor, source-mechanism and coupling products were identified for further independent testing of the interpretation. Neither the strain observations nor these additional products entered catalog relocation or repeating-earthquake detection. Supplementary Notes 5 and 6 document the regional relocation, analysis-region definition, stage bookkeeping, migration diagnostics, repeating-earthquake workflow and independent evaluation of aseismic modulation.

    \subsection*{Evaluation framework}

    \subsubsection*{Scientific task design and benchmarking scenarios} \label{sec:task_design}

    The evaluation framework spans three tiers of complexity, scaling from individual tool execution to integrated, end-to-end analytical workflows. This structure comprised a 104-task benchmark for Levels 1 and 2, alongside two Level 3 case-study evaluations. Performance was assessed using task-specific reference solutions and expert controls for deterministic outputs, a five-point grading rubric for qualitative Level 1/2 tasks, and structured non-quantitative review criteria for Level 3 reports. To ensure cross-backbone comparisons, all models were tested under identical task definitions, input datasets, tool libraries, and evaluation criteria. This design isolates the agent's computational performance and workflow coherence from the geophysical interpretations detailed in the Results (Supplementary Table~\ref{tab:tab_seis_levels}).

    \noindent\textbf{Level 1: Atomic task capability.}
    Level 1 evaluates whether TRACE can correctly execute individual seismological operations with well-defined inputs and outputs. The 74 tasks cover data access and retrieval, data representation and formatting, signal processing, statistics and feature extraction, and scientific visualization. These tasks test computational consistency, parameter handling and basic tool invocation, rather than open-ended seismological inference.

    \noindent\textbf{Level 2: Multi-step analytical capability.}
    Level 2 evaluates whether TRACE can coordinate multiple operations into coherent analytical workflows. The 30 tasks represent common workflow patterns, including sequential chaining, heuristic branching, scalable batching and scanning, and constrained space exploration. These tasks test dependency handling, intermediate-state management, diagnostic correction and tool orchestration across linked processing steps.

    \noindent\textbf{Level 3: Integrated scientific analysis capability.}
    Level 3 evaluates whether TRACE can coordinate integrated observation-to-interpretation workflows from their respective case-specific inputs. The two case studies require TRACE to connect data streams, formulate candidate physical interpretations, execute multiple analyses and assemble evidence into a structured report. The resulting reports underwent structured expert review of workflow coherence and evidence organization without numerical scoring. The Ridgecrest and Sanriku Results sections evaluate the corresponding geophysical interpretations using their case-specific evidence.

    This hierarchy separates basic tool execution, multi-step workflow control and integrated scientific reporting, three capabilities that are often conflated in scientific-agent evaluation.

    \subsubsection*{Evaluation and assessment criteria}

    The framework used separate protocols for deterministic benchmark outputs, qualitatively rated benchmark outputs and Level 3 case reports. For Level 1 and Level 2 tasks with deterministic numerical outputs, TRACE results were compared with reference solutions generated through expert-designed control experiments. These comparisons used matched input data, task-specific success criteria and predefined numerical tolerances. Where appropriate, root mean square error (RMSE) and correlation coefficients quantified numerical agreement with the reference outputs.

    Level 1 and Level 2 outputs requiring qualitative assessment were evaluated on task-type-specific five-point scales. Each task was assessed only against the dimensions included in its predefined output requirements: quantitative accuracy, scientific correctness, and visualization and reporting quality, as applicable (Supplementary Table~\ref{tab:tab_result_eval}). The applicability of each dimension was fixed in the task inventory and was identical across backbones. Five expert raters independently scored these benchmark outputs, and the mean of their ratings was used as the final output-level score. Expert-generated reference outputs were assessed using the same task-specific criteria and provide the human-expert reference in Supplementary Fig.~\ref{fig:ed_trace_benchmark}c. Five experts reviewed the Level 3 case reports qualitatively across planning, implementation and synthesis using the criteria in Supplementary Table~\ref{tab:supp_l3_assessment_dimensions}; no numerical Level 3 score was assigned.

    Debugging efficiency was quantified at the benchmark-run level. Within each task-level round, an individual coding workflow could undergo at most 20 correction rounds, and the outer task loop was capped at three task-level rounds. Both loops occurred within a single model--task benchmark run and did not constitute independent evaluations. A correction round was counted when generated code, parameters, data handling or workflow structure was revised after an execution or diagnostic failure. Runs that did not satisfy the predefined computational or diagnostic criteria within these limits were recorded as failures. Automated workflow repairs were recorded separately from retrospective expert assessment. Level 1 and Level 2 scores were interpreted using predefined bands: $4.0 < s \le 5.0$ indicated high performance, $2.5 \le s \le 4.0$ intermediate performance, and $s < 2.5$ substantial methodological or technical deficiencies. General score definitions are provided in Supplementary Table~\ref{tab:tab_result_eval}; the non-quantitative Level 3 assessment dimensions are summarized in Supplementary Table~\ref{tab:supp_l3_assessment_dimensions}.

    \begin{figure}[!htbp]
        \centering
        \includegraphics[width=0.99\linewidth]{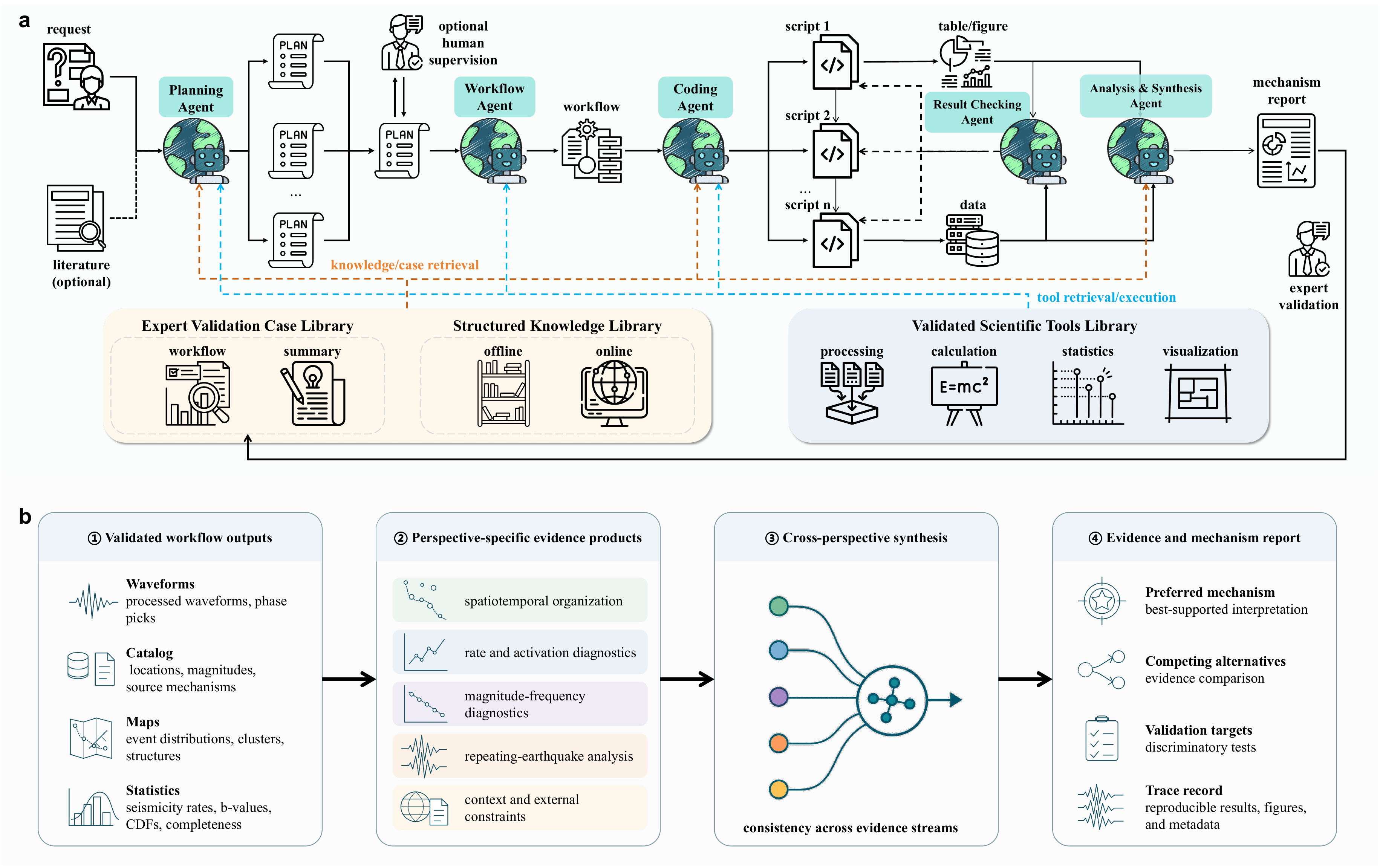}
        \caption{
        \textbf{Technical architecture and evidence-chain synthesis of TRACE.}
        \textbf{a}, Multi-agent architecture used to convert a seismological request and optional literature or contextual input into candidate plans, executable workflows, scripts, checked data products and a physical interpretation report. Planning, workflow, coding, result-checking, analysis and cross-perspective synthesis agents query the Expert-Validated Case Library, Structured Knowledge Library and Validated Scientific Tools Library. Dashed links distinguish knowledge and case retrieval from tool retrieval and execution. Optional human supervision and expert validation constrain ambiguous task boundaries and return reviewed workflows and summaries to the case library.
        \textbf{b}, Evidence-chain synthesis used by the Scientific Integration and Inference Module. Checked workflow outputs are converted into perspective-specific evidence summaries, integrated across evidence streams and assembled into a physical interpretation report containing an evidence-supported interpretation, viable alternatives, external validation targets and a trace record.
        }
        \label{fig:ed_trace_architecture}
    \end{figure}

    \clearpage

    \refstepcounter{figure}
    \label{fig:ed_trace_benchmark}
    \begin{center}
        \centering
        \includegraphics[width=0.84\linewidth]{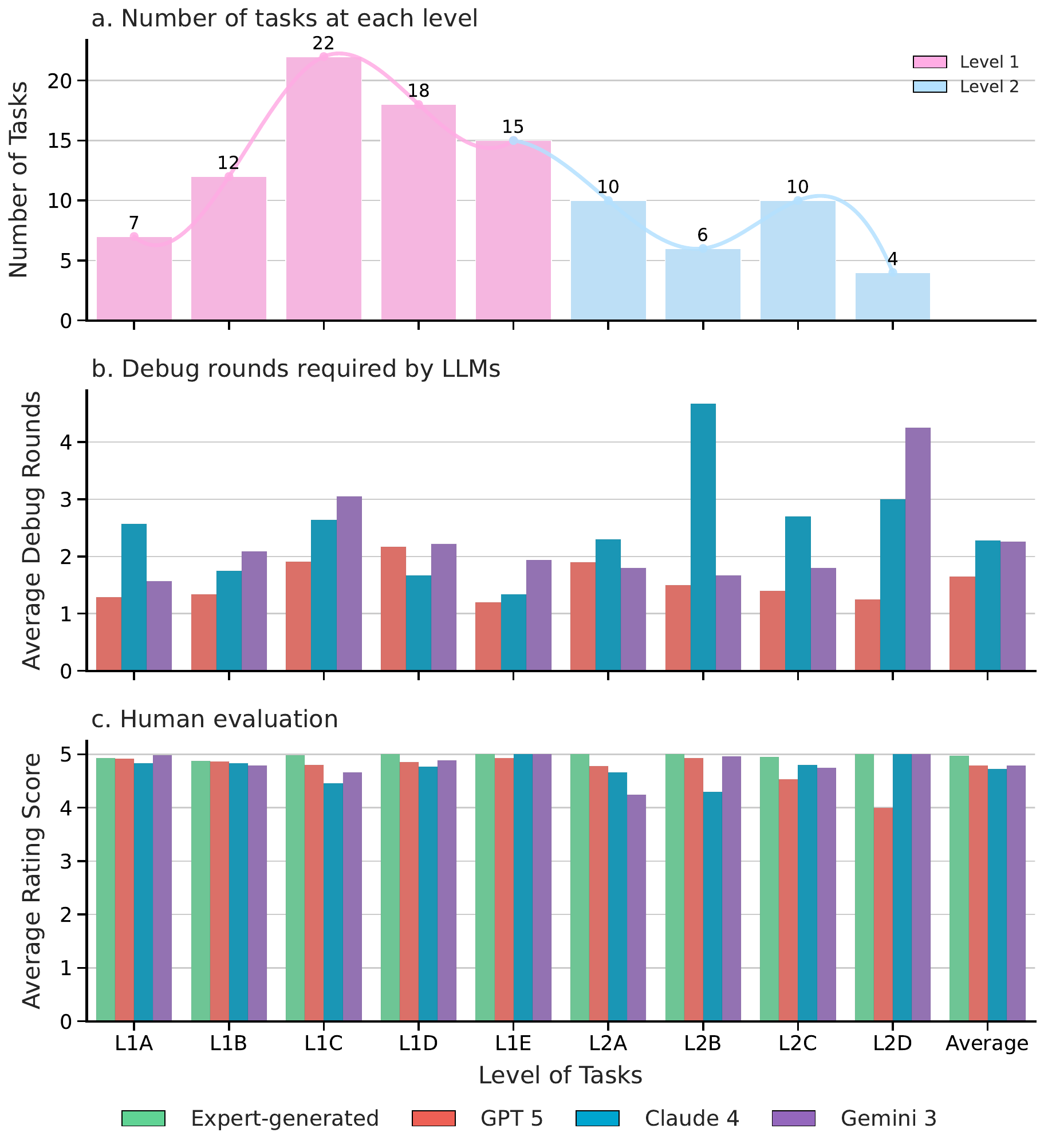}
    \end{center}
        \tightcaptionspacing
        \sciencecaptionof{figure}{
        \textbf{Benchmark evaluation of TRACE workflow execution and structured scientific reporting.}
        \textbf{a}, Distribution of benchmark tasks across atomic operations (Level 1, pink bars, $L1A$--$L1E$) and multi-step analytical workflows (Level 2, blue bars, $L2A$--$L2D$). Sub-levels correspond to representative task categories: data access and retrieval, data representation and formatting, signal processing, statistics and feature extraction, and scientific visualization for Level 1; and sequential chaining, heuristic branching, scalable batching and scanning, and constrained space exploration for Level 2. The vertical axis gives the number of tasks in each sub-level, with Level 1C ($n = 22$) forming the largest group.
        \textbf{b}, Debugging efficiency across Level 1 and Level 2 benchmark tasks and large language model (LLM) backbones. The vertical axis shows the average number of coding-workflow correction rounds. Each coding workflow allowed at most 20 correction rounds within a task-level round, and the outer task loop allowed at most three task-level rounds within the same model--task benchmark run.
        \textbf{c}, Structured scores for expert-generated reference outputs and outputs produced with GPT-5.1, Claude Sonnet 4 and Gemini 3 Flash preview across benchmark task categories. Scores are reported on a 1--5 scale using the task-type-specific criteria defined in Materials and Methods. The benchmark assesses workflow execution, diagnostic correction and structured reporting; case-specific geophysical interpretations are evaluated in the main Results.
        }
        \standardcaptionspacing

    \begin{table}[!ht]
        \centering
        \small
        \setlength{\belowcaptionskip}{8pt}
        \caption{\textbf{Task-complexity levels and representative TRACE capabilities in the seismological evaluation framework.}}
        \label{tab:tab_seis_levels}
        \renewcommand\arraystretch{1.25}
        \begin{tabularx}{\textwidth}{
            >{\centering\arraybackslash}m{0.2\textwidth}
            >{\centering\arraybackslash}m{0.62\textwidth}
            >{\centering\arraybackslash}m{0.1\textwidth}
            }
            \toprule
            \textbf{Levels} &
            \textbf{Description} &
            \textbf{Number of Tasks} \\ \midrule
            \rowcolor[HTML]{e9e9e9}
            Level 1: Atomic task capability &
            Performs fundamental seismological operations with procedural correctness and execution stability, including seismic data acquisition, format conversion, preprocessing, signal feature extraction, and scientific visualization. Emphasizes accurate tool invocation, parameter handling, and computational consistency in routine seismic data processing workflows. &
            74 \\

            Level 2: Multi-step analytical capability &
            Solves workflow-level seismological problems requiring coordinated multi-tool orchestration and dependency management. Integrates sequential toolchains, heuristic branching strategies, large-scale batch scanning, and parameter-space exploration under physical constraints. Evaluates adaptive pathway selection, intermediate state management, and long-chain analytical coherence. &
            30 \\

            \rowcolor[HTML]{e9e9e9}
            Level 3: Integrated scientific analysis capability &
            Conducts end-to-end scientific investigations under realistic research scenarios. Integrates cross-module data processing, multi-perspective analysis, hypothesis formulation, and evidence synthesis to assemble evidence-supported, testable interpretations. Assesses workflow coherence, cross-task integration, and interpretation boundaries in complex seismological studies. &
            2 \\

            \bottomrule
        \end{tabularx}
    \end{table}

    \clearpage
    
    \begin{table}[!ht]
        \centering
        \setlength{\belowcaptionskip}{8pt}
        \caption{\textbf{Five-point criteria for qualitative Level 1 and Level 2 benchmark outputs.}}
        \label{tab:tab_result_eval}
        \small
        \renewcommand\arraystretch{1.05}
        \begin{tabularx}{\textwidth}{
            >{\centering\arraybackslash}m{0.1\textwidth}
            >{\centering\arraybackslash}m{0.3\textwidth}
            >{\centering\arraybackslash}m{0.28\textwidth}
            >{\centering\arraybackslash}m{0.25\textwidth}
            }
            \toprule
            \textbf{Score} &
            \textbf{Quantitative Accuracy} &
            \textbf{Scientific Correctness} &
            \textbf{Visualization and Reporting Quality} \\ \midrule

            \rowcolor[HTML]{E9E9E9}
            5 points &
            Numerical outputs match reference solutions within predefined numerical tolerances. &
            All results are scientifically valid, logically consistent, and fully aligned with task objectives. No conceptual or methodological errors. &
            Figures are accurate, clearly labeled (axes, units, scales), and publication-ready. Visual design enhances interpretability. Text and figures are fully consistent. \\

            4 points &
            Minor numerical deviations from reference results but within acceptable tolerance. No impact on overall conclusions. &
            Results are scientifically correct with only minor interpretative omissions or limited analytical depth. &
            Figures are correct and interpretable, but minor issues exist (e.g., formatting inconsistencies, limited annotations). Overall presentation is clear. \\

            \rowcolor[HTML]{E9E9E9}
            3 points &
            Moderate deviation from reference values, but core trends or magnitudes remain correct. &
            Partial correctness: main scientific conclusion is identifiable, but secondary errors or incomplete interpretation exist. &
            Figures are understandable but contain noticeable labeling issues, limited clarity, or insufficient explanation. \\

            2 points &
            Significant numerical errors or unstable results that affect reliability. &
            Results contain conceptual misunderstandings or incorrect scientific interpretations. &
            Figures are poorly constructed or partially incorrect, limiting interpretability. \\

            \rowcolor[HTML]{E9E9E9}
            1 point &
            Numerical results are incorrect or fail to match reference benchmarks. &
            Scientific conclusions are invalid or inconsistent with task requirements. &
            No meaningful figures or coherent result presentation provided. \\

            \bottomrule
        \end{tabularx}
    \end{table}

    \clearpage

\section*{Supplementary Text}

\section*{Supplementary Note 1. TRACE workflow, architecture and recorded provenance}

    This note defines the structured provenance record that connects each TRACE interpretation to its permitted inputs, executed analyses, diagnostic checks and explicitly introduced contextual evidence. The benchmark structure and scoring criteria are described in Materials and Methods and Supplementary Note 2.

    \subsection*{Workflow scope and information boundaries}

    Each TRACE run begins with a seismological question, defined observational inputs, explicitly permitted contextual information and access to the Structured Knowledge Library and Validated Scientific Tools Library. The provenance record distinguishes these initial inputs from information introduced later for catalog quality control, retrospective comparison or physical synthesis. No case-specific target interpretation was supplied to the planning, execution or inference modules in either application.

    For the Ridgecrest application, mapped surface-rupture traces compiled by the Ridgecrest Rupture Mapping Group were supplied as an auxiliary geospatial dataset for quantitative distance and orientation analyses \cite{kendrick_2019_Geologic}, whereas reference catalogs supported catalog quality control and documented alignment. These inputs did not prescribe the subsequent physical interpretation. Published Ridgecrest interpretations were used only for retrospective comparison after the TRACE analysis was complete. For the Sanriku application, historical rupture models and convergence estimates entered the final fault-system synthesis. Independent S-net borehole strain observations were introduced as external validation evidence. Neither the strain observations nor additional tremor, source-mechanism or coupling products were used in catalog relocation or repeating-earthquake detection.

    \subsection*{Structured provenance record}

    The curated provenance record is organized by workflow layer (Supplementary Table~\ref{tab:supp_trace_provenance_schema}). It preserves the information needed to determine what TRACE received, what it executed, how outputs were checked and revised, and how checked observations entered the final interpretation.

    These structured records, executable scripts and linked output products provide the primary basis for scientific audit. Supplementary Notes 3--6 report the case-specific data selections, analytical settings, diagnostic products and evidence summaries needed to evaluate the manuscript claims. The versioned reproducibility release described in the Data and code availability statement contains the corresponding scripts and machine-readable records. Raw agent conversations and deployment-specific logs are retained separately and are not required to reconstruct the reported evidence chain.

    Within this common provenance scheme, the two applications served distinct scientific roles and began at different observational levels. Ridgecrest provided retrospective end-to-end validation of observation-to-interpretation reasoning. Beginning with continuous waveforms, TRACE constructed a double-difference-relocated earthquake catalog and evaluated the resulting seismicity evidence before retrospective comparison with published interpretations (Supplementary Notes 3 and 4). Sanriku served as a discovery application in an incompletely understood earthquake sequence. Beginning with the JMA event catalog and phase-arrival products, TRACE performed relocation, stage-resolved catalog analysis, repeating-earthquake analysis and final fault-system synthesis (Supplementary Notes 5 and 6). The two applications therefore instantiate continuous-waveform and trusted-catalog entry modes under the same provenance requirements.

\begin{table}[!ht]
    \centering
    \caption{\textbf{Core fields in the curated TRACE provenance record.} Each workflow layer retains the information needed to trace a physical interpretation back to its observational and computational basis.}
    \label{tab:supp_trace_provenance_schema}
    \begin{tabularx}{\textwidth}{>{\raggedright\arraybackslash}p{0.17\textwidth}>{\raggedright\arraybackslash}X>{\raggedright\arraybackslash}p{0.27\textwidth}}
    \toprule
    Provenance layer & Information retained & Audit function \\
    \midrule
    Input boundary & Scientific question; data sources; spatial and temporal selections; permitted contextual information; stage at which external evidence entered & Establishes what information was available at each stage \\
    Planning & Physical constraints; candidate physical interpretations; selected and retained analysis branches & Links the scientific question to the analyses without prescribing a conclusion \\
    Execution & Tool choices; executable scripts; parameter settings; input--output paths; intermediate products & Reconstructs the computational path from observations to analytical products \\
    Diagnostics and revision & Computational and physical checks; failed steps; diagnostic flags; permitted revisions; intervention status and source, when applicable & Distinguishes accepted outputs from unresolved failures and records how the workflow changed \\
    Evidence extraction & Source data product; method and parameters; diagnostic status; interpretation-relevant observation & Links each evidence statement to a checked analytical output \\
    Synthesis & Supporting, inconsistent or unresolved relations; evidence-supported interpretation; viable alternatives; external validation targets & Links the physical interpretation to the evidence while preserving its boundaries \\
    \bottomrule
    \end{tabularx}
\end{table}

\clearpage

\section*{Supplementary Note 2. Benchmark and case-study evaluation protocol}

    This note details the 104-task Level 1/Level 2 benchmark and the two Level 3 case-study evaluations used to assess TRACE. The benchmark used expert-generated reference outputs, expert-designed control experiments and structured scoring criteria to assess computational execution and workflow construction. Level 3 used structured, non-quantitative expert assessment to evaluate application-level workflow coherence and evidence organization. The complete machine-readable task inventory and benchmark run records are included in the reproducibility archive described in the Data and code availability statement.

    \subsection*{Evaluation hierarchy and task coverage}

    The evaluation framework distinguishes three levels of capability. Levels 1 and 2 form the cross-backbone benchmark, whereas Level 3 comprises two case-study evaluations of integrated observation-to-interpretation workflows. Level 1 tasks evaluate bounded operations with well-defined inputs and outputs, including data retrieval, format conversion, waveform processing, statistical feature extraction and scientific visualization. Level 2 tasks evaluate multi-step analytical workflow construction, including sequential chaining, heuristic branching, scalable scanning and constrained search. Level 3 evaluates integrated scientific analysis and evidence organization in the Ridgecrest and Sanriku applications. The two cases assess application-level workflow coherence and do not establish broad generalization across tectonic settings.

    The Level 1/Level 2 benchmark contains 104 core tasks, including 74 Level 1 tasks and 30 Level 2 tasks. Each task entry records the objective, input data, expected output, evaluation mode, scoring criteria and operation tags. Supplementary Tables~\ref{tab:supp_l1_l2_benchmark_coverage} and \ref{tab:supp_representative_l1_l2_tasks} summarize the category-level coverage and representative task content. The operation tags specify whether a task is single-step, multi-step, batched, branching or constrained-search, together with the primary skill and operation complexity. Across the 104 tasks and three evaluated backbones, the benchmark comprised 312 individual model--task runs. For each task, all backbones used the same task definition, input data, tool availability and evaluation criteria. Expert-generated reference outputs were retained for comparison.

    \subsection*{Benchmark scoring, aggregation and workflow revision}

    The Level 1/Level 2 benchmark followed two complementary evaluation tracks. Deterministic outputs were compared with task-specific reference solutions generated through expert-designed control experiments, using predefined success criteria and numerical tolerances. Outputs requiring qualitative assessment were evaluated on a five-point scale using the task-specific criteria for quantitative accuracy, scientific correctness, and visualization and reporting quality. Task-specific checks included schema validity, row counts, value tolerances, waveform alignment, metadata consistency, required visual elements and consistency between plotted and numerical products.

    Five expert raters independently scored each Level 1/Level 2 output requiring qualitative assessment. Their arithmetic mean defined the output-level rating,
    \begin{equation}
        \bar{s}_{o,d}=\frac{1}{5}\sum_{j=1}^{5}s_{o,d,j},
    \end{equation}
    where $s_{o,d,j}$ is the score assigned by expert $j$ to output $o$ under scoring dimension $d$. The expert ratings are repeated assessments of the same output rather than additional benchmark runs. For each backbone and dimension, the model-level rating was calculated from outputs whose predefined task requirements included that dimension. The applicable task sets were fixed before evaluation and were identical across backbones. Statistical-analysis and visualization ratings therefore summarize different task subsets defined by the corresponding output requirements. Expert-generated reference outputs were assessed using the same task-specific criteria and provide the human-expert reference in Supplementary Fig.~\ref{fig:ed_trace_benchmark}c.

    Each model--task benchmark run also recorded pass/fail status and the number of correction rounds, together with statistical-analysis or visualization ratings when required by the task. Within each task-level round, an individual coding workflow could undergo at most 20 correction rounds. The outer task loop was capped at three task-level rounds. These internal correction and task-level rounds belonged to the same benchmark run and were not counted as independent evaluations. A correction round was counted when code, parameters, data handling or workflow structure was revised after an execution or diagnostic failure. Workflow correction occurred during the benchmark run, whereas expert assessment was performed on its final output after the run had closed. A run was classified as a failure if its computational or diagnostic criteria remained unmet after the allowed internal rounds. The reported mean correction-round count summarizes coding-workflow corrections across successful and failed benchmark runs. Automated workflow repair was recorded separately from retrospective expert assessment.

    \subsection*{Level 1 and Level 2 benchmark results}

    Supplementary Table~\ref{tab:supp_l1_l2_model_results} reports descriptive model-level summaries. Each backbone contributed 104 model--task runs. Success rate is the fraction of these runs with pass status. Mean score and mean correction rounds are averaged over the corresponding run-level records. Statistical-analysis and visualization ratings summarize task-specific assessments when statistical analysis or figure generation formed part of the predefined output requirements. These summaries describe the evaluated runs and are not inferential comparisons of backbone performance.

    Of the 222 Level 1 runs, 221 satisfied the task-specific criteria, giving a pooled run-level success rate of 99.55\%. Of the 90 Level 2 runs, 89 passed, giving a pooled run-level success rate of 98.89\%. Both failed runs occurred with Claude Sonnet 4. One involved a DAS frequency--wavenumber (FK) spectrum task, and the other involved a branching task combining STA/LTA and autoregressive phase picking. The other two backbones completed both tasks. These backbone-specific failures identify difficult cases in signal processing and conditional workflow branching rather than tasks that were unsolved across all evaluated backbones. Together, the Level 1 and Level 2 results characterize bounded execution and multi-step analytical workflow performance across the evaluated tasks, providing the computational context for the integrated scientific applications.

    \subsection*{Level 3 case-study evaluations}

    Level 3 was evaluated separately from the Level 1/Level 2 benchmark. It assessed whether TRACE could organize each case-specific input set into a coherent observation-to-interpretation workflow, while preserving the information boundaries defined in Supplementary Note 1. Supplementary Table~\ref{tab:supp_l3_case_tasks} lists the two case-study evaluations and distinguishes initial observations from information introduced for quality control, synthesis or external validation.

    Five experts qualitatively reviewed the Level 3 reports across planning, implementation and synthesis. This structured review examined within-case workflow completeness, execution traceability, evidence organization and interpretation boundaries; no numerical Level 3 score was assigned. The review was not combined with the Level 1/Level 2 benchmark statistics or used to compare geophysical complexity across cases. The case-specific geophysical interpretations are evaluated by their underlying evidence in the main Results and Supplementary Notes 3--6.

    Supplementary Table~\ref{tab:supp_l3_assessment_dimensions} summarizes the criteria and records used for this non-quantitative assessment.

\begin{table}[!ht]
    \centering
    \caption{\textbf{Coverage of Level 1 and Level 2 benchmark tasks.} Counts are grouped by benchmark group, tested capability and operation pattern.}
    \label{tab:supp_l1_l2_benchmark_coverage}
    \begin{tabularx}{\textwidth}{>{\raggedright\arraybackslash}p{0.12\textwidth}>{\raggedright\arraybackslash}p{0.24\textwidth}>{\raggedright\arraybackslash}X>{\raggedright\arraybackslash}p{0.17\textwidth}>{\raggedleft\arraybackslash}p{0.10\textwidth}}
    \toprule
    Benchmark group & Benchmark category & Capability tested & Operation pattern & Tasks (n) \\
    \midrule
    L1-A & Data access and retrieval & Data retrieval & Single-step & 7 \\
    L1-B & Data representation and formatting & Data formatting & Single-step & 12 \\
    L1-C & Signal processing & Signal processing & Single-step & 22 \\
    L1-D & Statistics and feature extraction & Statistical feature extraction & Single-step & 18 \\
    L1-E & Scientific visualization & Scientific plotting & Single-step & 15 \\
    \addlinespace[0.2em]
    \multicolumn{4}{r}{Level 1 subtotal} & 74 \\
    \midrule
    L2-A & Sequential chaining & Sequential workflow construction & Multi-step & 10 \\
    L2-B & Heuristic branching & Conditional workflow branching & Branching & 6 \\
    L2-C & Scalable batching and scanning & Batched execution and scanning & Batch & 10 \\
    L2-D & Constrained space exploration & Constrained parameter exploration & Constrained search & 4 \\
    \addlinespace[0.2em]
    \multicolumn{4}{r}{Level 2 subtotal} & 30 \\
    \midrule
    \multicolumn{4}{r}{Total} & 104 \\
    \bottomrule
    \end{tabularx}
\end{table}

\clearpage

\begin{table}[!ht]
    \centering
    \small
    \renewcommand\arraystretch{0.9}
    \caption{\textbf{Representative Level 1 and Level 2 benchmark tasks.} One task is shown for each benchmark category.}
    \label{tab:supp_representative_l1_l2_tasks}
    \begin{tabularx}{\textwidth}{>{\raggedright\arraybackslash}p{0.11\textwidth}>{\raggedright\arraybackslash}p{0.24\textwidth}>{\raggedright\arraybackslash}X>{\raggedright\arraybackslash}p{0.15\textwidth}>{\raggedright\arraybackslash}p{0.12\textwidth}}
    \toprule
    Task ID & Benchmark category & Representative task & Operation pattern & Complexity \\
    \midrule
    L1-A-1 & Data access and retrieval & Download an event catalog from an FDSN server & Single-step & Low \\
    L1-B-1 & Data representation and formatting & Convert ASCII waveform data to MiniSEED & Single-step & Low \\
    L1-C-1 & Signal processing & Apply an AIC picker to a seismic trace & Single-step & Low \\
    L1-D-1 & Statistics and feature extraction & Compute an array response function & Single-step & Low \\
    L1-E-1 & Scientific visualization & Plot a focal-mechanism beachball from a QuakeML event & Single-step & Low \\
    \addlinespace[0.2em]
    L2-A-1 & Sequential chaining & Estimate back azimuth from STEAD waveforms using BazNet & Multi-step & Medium \\
    L2-B-1 & Heuristic branching & Check distributed acoustic sensing (DAS) channel quality and branch the workflow & Branching & High \\
    L2-C-1 & Scalable batching and scanning & Perform one-dimensional $b$-value significance testing & Batch & Medium-high \\
    L2-D-1 & Constrained space exploration & Convert DAS strain to velocity with curvelet constraints & Constrained search & High \\
    \bottomrule
    \end{tabularx}
\end{table}

\clearpage

\begin{table}[!ht]
    \centering
    \caption{\textbf{Model-level results for the Level 1 and Level 2 benchmark.} Each backbone contributed 104 model--task runs. Ratings and mean scores are reported on a five-point scale. Statistical-analysis and visualization ratings were calculated only when the corresponding component formed part of a task's predefined output requirements.}
    \label{tab:supp_l1_l2_model_results}
    \begin{tabularx}{\textwidth}{>{\raggedright\arraybackslash}X>{\centering\arraybackslash}p{0.07\textwidth}>{\centering\arraybackslash}p{0.10\textwidth}>{\centering\arraybackslash}p{0.10\textwidth}>{\centering\arraybackslash}p{0.11\textwidth}>{\centering\arraybackslash}p{0.13\textwidth}>{\centering\arraybackslash}p{0.09\textwidth}}
    \toprule
    Model & \makecell{Model--task\\runs} & Success rate & \makecell{Statistical\\analysis rating} & \makecell{Visualization\\rating} & \makecell{Mean\\correction\\rounds} & Mean score \\
    \midrule
    Claude Sonnet 4 & 104 & 98.08\% & 4.49 & 4.88 & 2.28 & 4.79 \\
    Gemini 3 Flash preview & 104 & 100.00\% & 4.64 & 4.86 & 2.26 & 4.82 \\
    GPT-5.1 & 104 & 100.00\% & 4.72 & 4.83 & 1.64 & 4.85 \\
    \bottomrule
    \end{tabularx}
\end{table}

\clearpage

\begin{table}[!ht]
    \centering
    \caption{\textbf{Integrated Level 3 scientific case-study evaluations.} The information boundary distinguishes initial observations from auxiliary, quality-control, synthesis and external-validation inputs.}
    \label{tab:supp_l3_case_tasks}
    \begin{tabularx}{\textwidth}{>{\raggedright\arraybackslash}p{0.08\textwidth}>{\raggedright\arraybackslash}p{0.10\textwidth}>{\raggedright\arraybackslash}p{0.20\textwidth}>{\raggedright\arraybackslash}X>{\raggedright\arraybackslash}p{0.22\textwidth}}
    \toprule
    Task ID & Case study & Scientific question & Workflow inputs and information boundary & Required report \\
    \midrule
    L3-001 & Ridgecrest & How was seismicity organized between the $M_w$ 6.4 foreshock and the $M_w$ 7.1 mainshock? & Initial observations: continuous waveforms and station metadata; auxiliary geospatial data: mapped surface-rupture traces; quality control and alignment: reference catalogs & Evidence-supported physical interpretation and viable alternatives, followed by retrospective comparison with published interpretations after the analysis was complete \\
    L3-002 & Sanriku & What evidence constrains shallow-interface activation before the April 2026 $M_J$ 7.7 rupture? & Initial observations: JMA event catalog and phase-arrival records; analysis-stage data: event waveforms for repeating-earthquake analysis; synthesis-stage constraints: historical rupture models and convergence estimates; external validation: independent S-net strain observations & Evidence-supported physical interpretation, viable alternatives and external validation targets \\
    \bottomrule
    \end{tabularx}
\end{table}

\clearpage

\begin{table}[!ht]
    \centering
    \caption{\textbf{Expert-assessment dimensions for Level 3 case reports.} The table defines the review criteria, common failure modes and the records used for assessment.}
    \label{tab:supp_l3_assessment_dimensions}
    \begin{tabularx}{\textwidth}{>{\raggedright\arraybackslash}p{0.15\textwidth}>{\raggedright\arraybackslash}X>{\raggedright\arraybackslash}p{0.26\textwidth}>{\raggedright\arraybackslash}p{0.23\textwidth}}
    \toprule
    Dimension & Assessment criterion & Common failure mode & Assessment evidence \\
    \midrule
    Planning & Scientific question decomposed into executable, physically relevant analyses & Missing analysis branch or unsupported assumption & Workflow plan and task decomposition \\
    Implementation & Executable scripts and tool calls linked to expected products & Broken execution, unstable parameters or missing quality control & Code, logs and output products \\
    Synthesis & Report separates observations from interpretation and records viable alternatives and external validation targets & Overclaiming, unsupported interpretation or unacknowledged uncertainty & Structured report and expert review \\
    \bottomrule
    \end{tabularx}
\end{table}

\clearpage

\section*{Supplementary Note 3. Ridgecrest waveform processing and catalog construction}

    This note documents how TRACE constructed and checked the high-resolution Ridgecrest earthquake catalog used for the delayed cascading analysis. It reports the data selection, processing counts (Supplementary Table~\ref{tab:supp_ridgecrest_processing_counts}), waveform-based local-magnitude estimation, relocation settings (Supplementary Table~\ref{tab:supp_ridgecrest_catalog_parameters}) and reference-catalog comparisons (Supplementary Table~\ref{tab:supp_ridgecrest_reference_catalogs}) that define the catalog as a documented observational input. The interpretation of delayed cascading activation is reserved for Supplementary Note 4, where the relocated catalog is evaluated together with timing, spatial, rate and magnitude-frequency diagnostics.

    \subsection*{Waveform data and station coverage}

    The Ridgecrest analysis required a catalog with fine-scale relative locations and a documented processing path. TRACE therefore constructed the catalog from continuous waveforms rather than relying only on a preselected event list. The study region spans $35.25^{\circ}$--$36.25^{\circ}$ N and $118.0^{\circ}$--$117.0^{\circ}$ W, covering the fault system activated between the $M_w$ 6.4 and $M_w$ 7.1 earthquakes. Continuous waveform records and StationXML metadata were retrieved from the Southern California Earthquake Data Center using FDSN-compatible ObsPy tools \cite{scedc_2013_Southern,beyreuther_2010_ObsPy}. The processing window covered 22 complete UTC days from 4 July to 26 July 2019.

    Station and channel selection used complete three-component high-gain triplets valid within the processing window. HH channels were selected where available, with EH channels used as fallback. The final waveform set contained 47 three-component stations from CI, GS, NN, PB and ZY networks. Of these, the 30 stations nearest to the $M_w$ 6.4 epicenter were retained for the final association and relocation route. Across the 22 daily windows, TRACE expected 1,034 station-day jobs and produced 853 quality-controlled station-day waveform inputs, corresponding to 82.5\% station-day availability. The raw waveform request contained 3,032 MiniSEED files with an approximate logical size of 34 GiB.

    Waveform preprocessing for phase picking was performed at the station-day level. Raw MiniSEED records were merged, trimmed to 24-hour UTC windows and checked for component completeness. Masked samples were filled to preserve continuous time axes for later sliding-window operations, and station days with insufficient component coverage were logged for review. Each trace was linearly detrended, demeaned and tapered using a Hann taper. Instrument responses were removed using the associated StationXML metadata, after which the data were resampled to 100 Hz and filtered within the seismic band used for phase picking. The processed output was written as three-component daily MiniSEED files and retained together with station metadata, file-level statistics and station-day completeness summaries.

    \subsection*{Phase picking, association and initial location}

    Phase picking was performed on the processed continuous waveforms using the pretrained PhaseNet model \cite{zhu_2018_PhaseNet}. TRACE applied the picker to daily three-component waveform windows and retained $P$ and $S$ arrivals with probability scores of at least 0.4 and finite phase times. This thresholded pick set contained 1,454,311 $P$ arrivals and 1,641,210 $S$ arrivals before applying the 30-station subset used for association. The final association route assigned 2,120,090 picks to associated events after station and association filtering, including 1,016,008 $P$ picks and 1,104,082 $S$ picks. Daily pick files were stored in tabular form and carried forward with their confidence scores, preserving the phase-level information behind the event catalog.

    Multi-station association and initial hypocentral location were then performed with GaMMA \cite{zhu_2022_Earthquake}. The final association run used a Bayesian Gaussian-mixture association model with DBSCAN pre-clustering, no amplitude term in association, an oversampling factor of 3 and a minimum of five picks per event. Initial locations were estimated using a one-dimensional velocity-model framework. The production GaMMA catalog contains 140,505 associated events, including origin time, latitude, longitude, depth and station-level $P$/$S$ arrival constraints.

    \subsection*{Local magnitude estimation}

    The Ridgecrest local magnitudes ($M_L$) used in subsequent analyses were estimated after event association from independently processed waveform records. Raw three-component records were demeaned, detrended and tapered with a 1\% cosine taper before instrument responses were removed to ground velocity using StationXML metadata. The traces were converted to simulated Wood--Anderson seismograms with ObsPy \cite{beyreuther_2010_ObsPy}, independently of the normalized waveforms used for PhaseNet picking.

    Associated $P$ and $S$ picks defined the measurement window for each station--event pair. The window extended from 0.5~s before to 3.0~s after the $S$ arrival, or from 0.2~s before to 1.0~s after the $P$ arrival when no $S$ pick was available. TRACE retained horizontal components with at least 80\% window coverage and measured the largest absolute zero-to-peak amplitude across them. Hypocentral distance $R_i$ was derived from the GaMMA location and station elevation. With amplitude $A_i$ in millimetres and $R_i$ in kilometres, the station local magnitude followed the southern California distance correction \cite{hutton_1987_ML}:
    \[
        M_{L,i}=\log_{10}(A_i)+1.11\log_{10}\left(\frac{R_i}{100}\right)
        +0.00189(R_i-100)+3.0.
    \]

    The event magnitude was the median of all valid station estimates. For events with an uncorrected median of at least 4.2, TRACE substituted the $S$-window median only when at least six $P$-window and six $S$-window estimates were available. The substitution also required the $S$-window median to exceed the uncorrected and $P$-window medians by at least 1.0 and 1.2 magnitude units, respectively. Events without a valid station estimate retained the provisional GaMMA value and were flagged as fallback cases. All magnitude-threshold and magnitude-frequency analyses used these final event magnitudes. No station corrections or Ridgecrest-specific recalibration of the attenuation relation were applied.

    \subsection*{Double-difference relocation}

    Initial locations from automated association remain affected by simplified velocity structure, phase-picking uncertainty and station-geometry effects. To improve relative source locations along the Ridgecrest fault system, TRACE applied double-difference relocation with HypoDD \cite{waldhauser_2000_DoubleDifference}. The relocation workflow converted the GaMMA phase and catalog outputs into HypoDD-compatible station, phase and initial-catalog files, constructed event-pair constraints and ran daily double-difference inversions. Events without sufficient differential-time constraints were not retained in the relocated catalog.

    Relocation used catalog-derived differential times generated by ph2dt from the associated $P$ and $S$ arrivals. The relocation was performed in 22 daily batches using the 30-station subset. Event pairs were formed with a maximum hypocentral separation of 8 km, up to 20 neighbors per event, a minimum of six links to define a neighboring event and a maximum of 40 observations per saved event pair. HypoDD was run with catalog differential times, both $P$ and $S$ phases, a maximum station-cluster distance of 80 km, an LSQR solver and two catalog-data weighting stages with damping values of 120 and 80. Waveform cross-correlation differential times were not used. The selected GaMMA and HypoDD production route used the same gradual one-dimensional velocity model with $V_{\mathrm{P}}/V_{\mathrm{S}}=1.73$.

    \subsection*{Final catalog statistics and quality control}

    The final relocated catalog used in the manuscript analyses contained 84,474 events. The retained events span 4 July 2019 00:56:37.590 UTC to 25 July 2019 23:59:29.320 UTC. After comparison with independent reference catalogs, the manuscript-analysis catalog retained events within the 0--30 km depth range and used small documented alignments in horizontal position and depth to maintain cross-catalog consistency. The unadjusted relocation outputs were retained for audit and reuse. Using the final event magnitudes described above, the catalog has a magnitude range of $M=-0.85$ to $M=7.10$, a median magnitude of 0.88 and an estimated completeness magnitude of $M_c=0.7$ based on the maximum-curvature method with 0.1-magnitude bins. It contains 54,551 events with $M \ge M_c$ and 35,462, 6,223 and 1,051 events with $M \ge 1$, $M \ge 2$ and $M \ge 3$, respectively. The corresponding unrelocated associated catalog, using the same event magnitudes, contains 52,260, 7,809 and 1,243 events above the same three magnitude thresholds.

    Relocation quality control focused on the interval between the $M_w$ 6.4 and $M_w$ 7.1 earthquakes, because this interval was used to evaluate delayed cascading activation. The GaMMA catalog contained 6,351 events in this window, of which 4,714 were retained after double-difference relocation. The relocated interval catalog contains 3,205, 734 and 151 events with $M \ge 1$, $M \ge 2$ and $M \ge 3$, respectively, with a median magnitude of 1.23. Relocation concentrated the seismicity between the two earthquakes into a clearer fault-aligned corridor while preserving the main along-profile activation pattern. This spatial sharpening supports use of the relocated catalog for diagnostic tests, while recognizing that relative relocation improves local event geometry rather than providing an absolute location-error bound.

\begin{figure}[!htbp]
    \centering
    \includegraphics[width=0.84\linewidth]{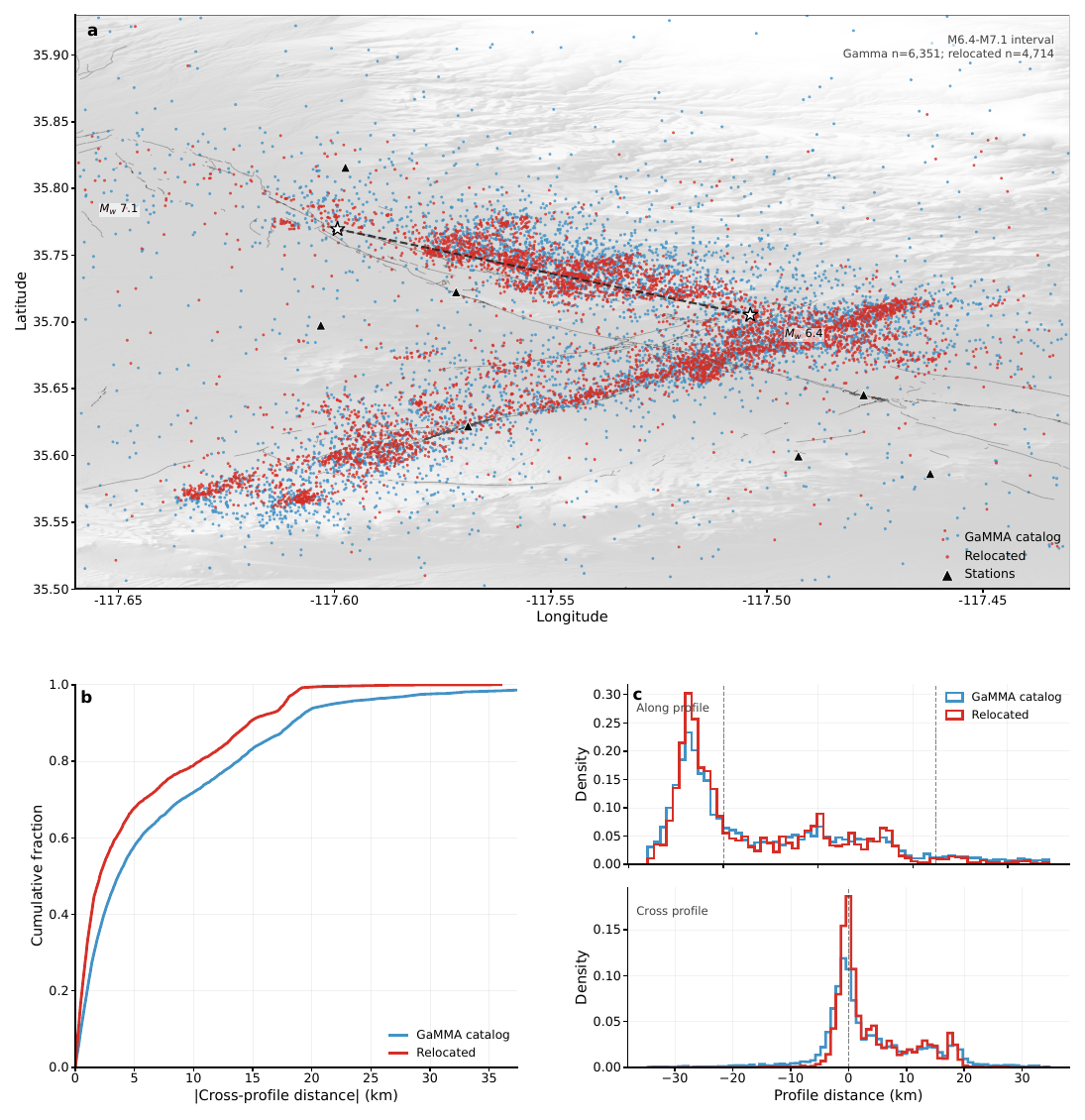}
    \caption{\textbf{Relocation sharpening for the Ridgecrest $M_w$ 6.4 to $M_w$ 7.1 interval.}
    \textbf{a}, Initial associated events and double-difference relocated events between the $M_w$ 6.4 and $M_w$ 7.1 earthquakes. Blue points show the GaMMA-associated catalog, red points show relocated events, and stars mark the two earthquakes. Relocation retains the principal activation corridor while reducing diffuse scatter away from the corridor. \textbf{b}, Cumulative distribution of absolute cross-profile distance from the activation corridor for the initial and relocated catalogs. The relocated catalog is more concentrated near the corridor. \textbf{c}, Along-profile and cross-profile density distributions for the same events. Relocation preserves the broad along-profile activation pattern while sharpening the cross-profile distribution. These diagnostics evaluate whether the relocated catalog is suitable for the spatial analyses in Supplementary Note 4. They are not used by themselves as evidence for delayed cascading activation.}
    \label{fig:supp_ridgecrest_relocation_qc}
    \end{figure}

    Supplementary Fig.~\ref{fig:supp_ridgecrest_relocation_qc} shows this relocation effect over the $M_w$ 6.4 to $M_w$ 7.1 interval. The figure supports the main-text use of a high-resolution relocated catalog for delayed cascading diagnostics, but does not by itself establish the mechanism.

    \subsection*{Comparison with independent reference catalogs}

\begin{figure}[!htbp]
    \centering
    \includegraphics[width=0.98\linewidth]{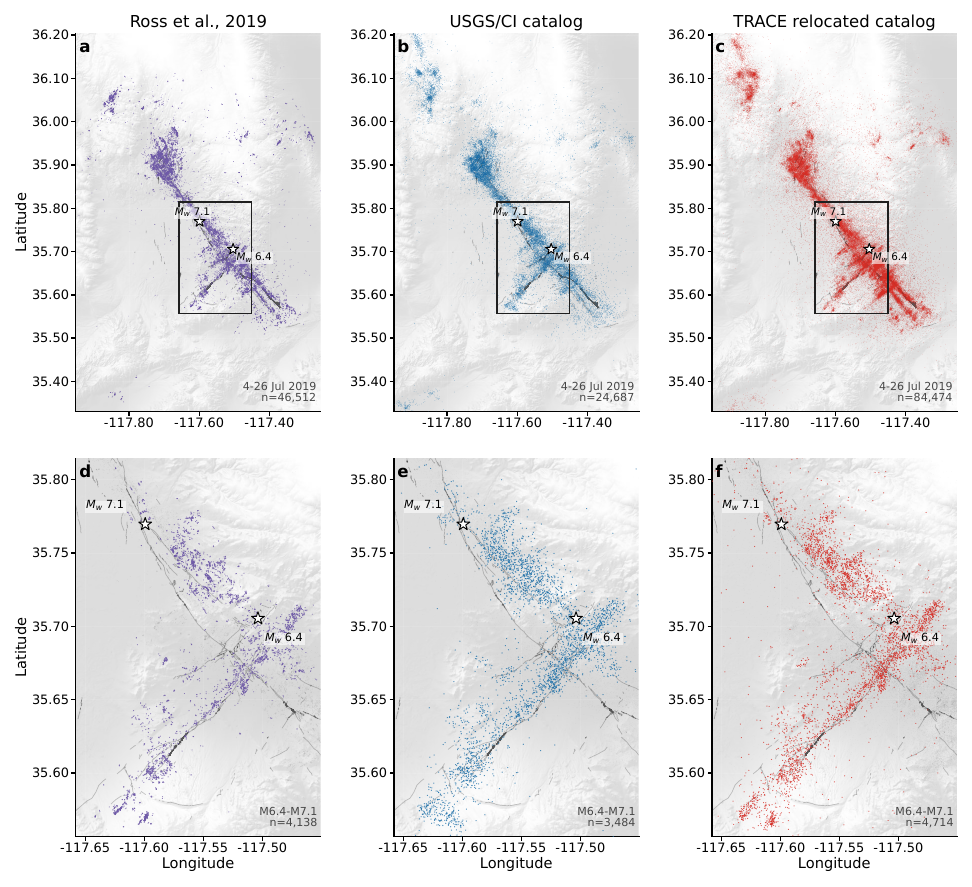}
    \caption{\textbf{Spatial coverage of Ridgecrest reference catalogs and the $M_w$ 6.4 to $M_w$ 7.1 analysis subset.}
    \textbf{a--c}, Full-window spatial distributions for the Ross et al. high-resolution catalog \cite{ross_2019_Hierarchical}, the USGS/ANSS ComCat operational catalog \cite{USGS_2017_Advanced} and the TRACE relocated catalog from 4 to 26 July 2019. The black rectangle marks the analysis region between the $M_w$ 6.4 and $M_w$ 7.1 earthquakes, and stars mark their epicenters. Event counts report the number of events shown under the corresponding catalog selection. \textbf{d--f}, Enlarged spatial distributions between the $M_w$ 6.4 and $M_w$ 7.1 origin times. The three catalogs recover the same fault-aligned activation corridor, supporting use of the TRACE relocated catalog for delayed cascading diagnostics.}
    \label{fig:supp_ridgecrest_catalog_spatial_coverage}
    \end{figure}

\begin{figure}[!htbp]
    \centering
    \includegraphics[width=0.98\linewidth]{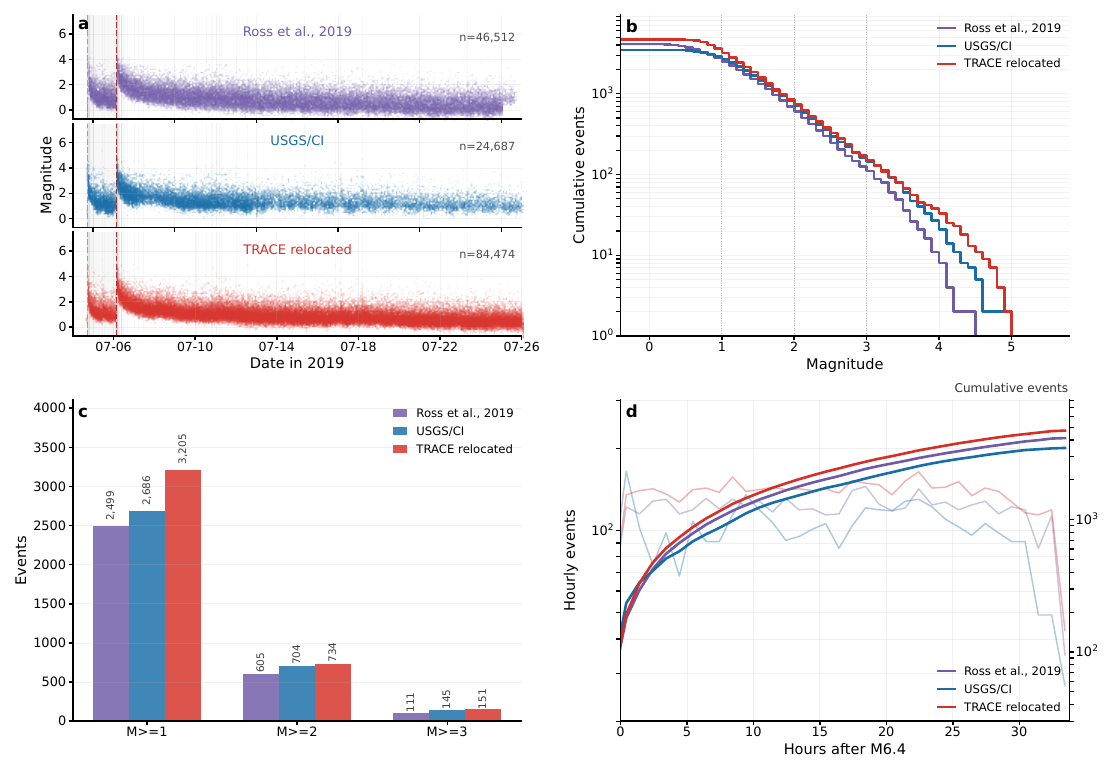}
    \caption{\textbf{Temporal and magnitude consistency of Ridgecrest reference catalogs.}
    \textbf{a}, Full-window time--magnitude distributions for the Ross et al. high-resolution catalog \cite{ross_2019_Hierarchical}, the USGS/ANSS ComCat operational catalog \cite{USGS_2017_Advanced} and the TRACE relocated catalog from 4 to 26 July 2019. The yellow and red dashed lines mark the $M_w$ 6.4 and $M_w$ 7.1 origin times, respectively, and the shaded interval marks the analysis window used for delayed cascading diagnostics. Event counts report the full-window catalog selections shown in this panel. \textbf{b}, Cumulative magnitude-frequency distributions for events in the analysis window. \textbf{c}, Event counts above $M \ge 1$, $M \ge 2$ and $M \ge 3$ in the same window, matching the magnitude thresholds used to compare catalog completeness for the diagnostic analyses. \textbf{d}, Hourly and cumulative event counts after the $M_w$ 6.4 earthquake. Thin lines show hourly counts and thick lines show cumulative counts. These comparisons show that the TRACE relocated catalog preserves the main temporal and magnitude structure of independent catalogs over the interval used for delayed cascading diagnostics.}
    \label{fig:supp_ridgecrest_catalog_temporal_magnitude}
    \end{figure}

    The final catalog was compared with independent Ridgecrest catalogs during the interval between the 4 July 2019 $M_w$ 6.4 earthquake and the 6 July 2019 $M_w$ 7.1 earthquake. This comparison was used as a boundary check on the TRACE catalog before the Ridgecrest diagnostic analysis. Supplementary Fig.~\ref{fig:supp_ridgecrest_catalog_spatial_coverage} compares full-window catalog coverage with the analysis subset used for delayed cascading diagnostics. Supplementary Fig.~\ref{fig:supp_ridgecrest_catalog_temporal_magnitude} compares the corresponding temporal and magnitude structure. The relocated TRACE catalog contains 4,714 events in the $M_w$ 6.4 to $M_w$ 7.1 interval, close to the event counts in the USGS/ANSS ComCat operational catalog \cite{USGS_2017_Advanced} and the Ross et al. high-resolution catalog (SCEDC QTM version) \cite{ross_2019_Hierarchical}. Its $M \ge 1$, $M \ge 2$ and $M \ge 3$ counts are also close to the ComCat counts. The unrelocated GaMMA catalog contains more events and a broader spatial footprint, whereas the HypoDD catalog is more conservative and better matched to the independent reference catalogs for the magnitude ranges used in the diagnostic analyses. Because the reference catalogs retain their own reported magnitude scales and procedures, these distribution-level comparisons were treated as boundary checks rather than as an event-by-event calibration of the TRACE magnitudes.

    This comparison shows that the TRACE catalog is suitable for diagnostic analysis over the interval between the $M_w$ 6.4 and $M_w$ 7.1 earthquakes while retaining a documented processing path. It does not by itself establish delayed cascading activation. The relocated catalog from this note is the input to Supplementary Note 4, where timing, fault alignment, seismicity rate evolution, Omori-Utsu behavior and magnitude-frequency structure are used to evaluate whether seismicity in this interval behaved as delayed cascading activation rather than as a nearly synchronous or ordinary aftershock response.

    \begin{table}[!htbp]
        \centering
        \caption{\textbf{Processing counts for the Ridgecrest TRACE catalog.}}
        \label{tab:supp_ridgecrest_processing_counts}
        \begin{tabularx}{\textwidth}{>{\raggedright\arraybackslash}p{0.38\textwidth}>{\raggedleft\arraybackslash}p{0.14\textwidth}>{\raggedright\arraybackslash}X}
        \toprule
        Processing output & Count (n) & Workflow use \\
        \midrule
        Selected three-component stations & 47 & Waveform processing and phase picking station set \\
        Stations used for association and relocation & 30 & Stations nearest to the $M_w$ 6.4 epicenter \\
        Quality-controlled station-day waveform inputs & 853 & Waveform inputs retained after station-day quality control \\
        $P$ arrivals retained after PhaseNet filtering & 1,454,311 & Thresholded PhaseNet $P$ picks before 30-station association \\
        $S$ arrivals retained after PhaseNet filtering & 1,641,210 & Thresholded PhaseNet $S$ picks before 30-station association \\
        Picks assigned to associated events & 2,120,090 & Event-level picks after station and association filtering \\
        GaMMA associated catalog & 140,505 & Associated-event catalog passed to subsequent processing \\
        TRACE relocated catalog & 84,474 & Relocated catalog used for Ridgecrest figures and diagnostic analyses \\
        \bottomrule
        \end{tabularx}
    \end{table}

    \clearpage
    
    \begin{table}[!htbp]
        \centering
        \caption{\textbf{Association and relocation parameters for the Ridgecrest catalog.}}
        \label{tab:supp_ridgecrest_catalog_parameters}
        \begin{tabularx}{\textwidth}{>{\raggedright\arraybackslash}p{0.20\textwidth}>{\raggedright\arraybackslash}p{0.34\textwidth}>{\raggedright\arraybackslash}X}
        \toprule
        Stage & Setting & Value \\
        \midrule
        Phase picking & PhaseNet probability threshold & $P \ge 0.4$; $S \ge 0.4$ \\
        Phase picking & Waveform sampling rate & 100 Hz \\
        \addlinespace[0.15em]
        Association & Association model & GaMMA Bayesian Gaussian-mixture model \\
        Association & DBSCAN pre-clustering & Enabled; $\epsilon=10$ km; minimum samples, 3 \\
        Association & Oversampling factor & 3 \\
        Association & Amplitude term & Not used \\
        Association & Station subset & 30 stations nearest to the $M_w$ 6.4 epicenter \\
        Association & Minimum event support & 5 picks per event \\
        Association & Search depth range & 0--30 km \\
        \addlinespace[0.15em]
        \texttt{ph2dt} & Maximum event-pair station distance & 80 km \\
        \texttt{ph2dt} & Maximum hypocentral separation & 8 km \\
        \texttt{ph2dt} & Maximum neighbors per event & 20 \\
        \texttt{ph2dt} & Minimum links per event pair & 6 \\
        \texttt{ph2dt} & Maximum observations per event pair & 40 \\
        \addlinespace[0.15em]
        HypoDD & Differential-time input & Catalog $P$ and $S$ differential times \\
        HypoDD & Cross-correlation differential times & Not used \\
        HypoDD & Relocation batching & 22 daily inversions \\
        HypoDD & Solver and initial model & LSQR; GaMMA initial locations \\
        HypoDD & Weighting and damping & Two catalog-data stages; damping 120 then 80 \\
        HypoDD & Relocation velocity model & Eleven $V_{\mathrm{P}}$ layers; $V_{\mathrm{P}}/V_{\mathrm{S}}=1.73$ \\
        \bottomrule
        \end{tabularx}
    \end{table}

    \clearpage

    \begin{table}[!htbp]
        \centering
        \caption{\textbf{Reference-catalog comparison for the Ridgecrest $M_w$ 6.4 to $M_w$ 7.1 interval.}}
        \label{tab:supp_ridgecrest_reference_catalogs}
        \begingroup
        \setlength{\tabcolsep}{3.5pt}
        \begin{tabularx}{\textwidth}{@{}>{\raggedright\arraybackslash}p{0.29\textwidth}>{\raggedleft\arraybackslash}p{0.08\textwidth}>{\raggedleft\arraybackslash}p{0.08\textwidth}>{\raggedleft\arraybackslash}p{0.07\textwidth}>{\raggedleft\arraybackslash}p{0.07\textwidth}>{\raggedleft\arraybackslash}p{0.08\textwidth}>{\raggedright\arraybackslash}X@{}}
        \toprule
        Catalog & Events & $N(M \ge 1)$ & $N(M \ge 2)$ & $N(M \ge 3)$ & Median $M$ & Use in TRACE audit \\
        \midrule
        Ross et al. high-resolution catalog, $n_{\mathrm{branch}}>1$ \cite{ross_2019_Hierarchical} & 4,138 & 2,499 & 605 & 111 & 1.15 & High-resolution reference \\
        USGS/ANSS ComCat (USGS/CI) \cite{USGS_2017_Advanced} & 3,484 & 2,686 & 704 & 145 & 1.37 & Operational reference \\
        TRACE associated catalog before relocation & 6,351 & 3,907 & 817 & 163 & 1.14 & Unrelocated TRACE reference \\
        TRACE HypoDD relocated catalog & 4,714 & 3,205 & 734 & 151 & 1.23 & Final catalog used for diagnostic tests \\
        \bottomrule
        \end{tabularx}
        \endgroup
    \end{table}

    \clearpage
    
    \clearpage

\section*{Supplementary Note 4. Seismicity evidence for delayed cascading activation in Ridgecrest}

    This note presents the seismicity diagnostics used to evaluate delayed cascading activation between the Ridgecrest $M_w$ 6.4 and $M_w$ 7.1 earthquakes. The analysis combines four complementary diagnostic groups rather than treating any single test as decisive. First, gridded onset and time window maps define when and where activation occurred. Second, fault distance and directional statistics test whether that activation followed mapped structures. Third, seismicity rate and Omori-Utsu diagnostics compare relaxation between the conjugate fault systems and within the $M_w$ 7.1 fault system. Finally, $b$-value estimates provide a complementary stress proxy view of the $M_w$ 7.1 comparison region. Together, these diagnostics evaluate whether seismicity in this interval was more consistent with delayed cascading activation than with immediate dynamic triggering, ordinary aftershock decay or a coherent migration front.

    \subsection*{Delayed spatiotemporal activation}

\begin{figure}[!htbp]
    \centering
    \includegraphics[width=0.98\linewidth]{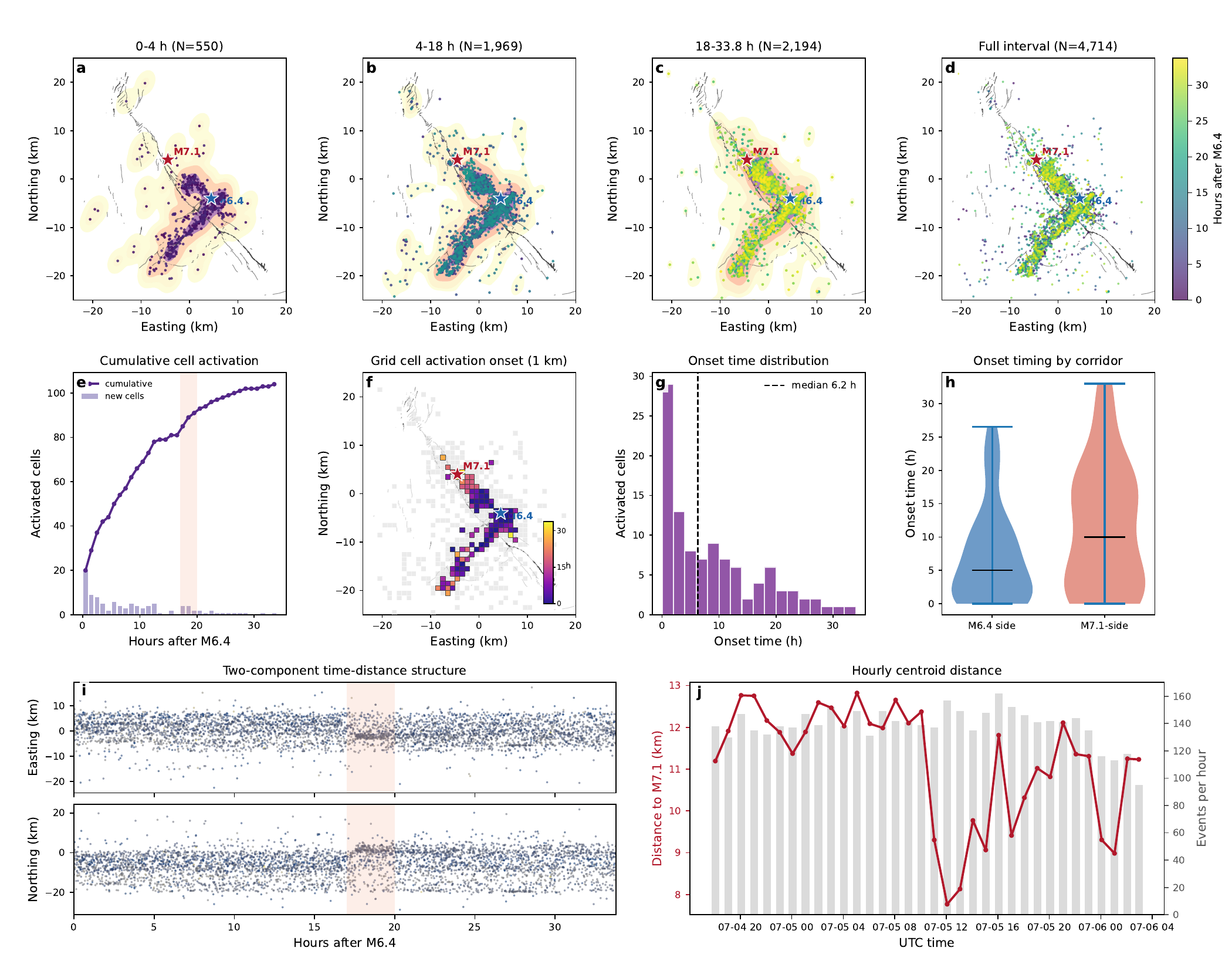}
    \caption{\textbf{Spatiotemporal activation during the Ridgecrest $M_w$ 6.4 to $M_w$ 7.1 interval.}
    \textbf{a--d}, Relocated seismicity in successive time windows after the $M_w$ 6.4 earthquake and over the full interval between the $M_w$ 6.4 and $M_w$ 7.1 earthquakes. The windows contain 550, 1,969, 2,194 and 4,714 events, respectively. Points are colored by time after the $M_w$ 6.4 earthquake; mapped surface-rupture traces \cite{kendrick_2019_Geologic} and earthquake locations provide spatial reference. \textbf{e}, Cumulative activation of 1 km grid cells and newly activated cells through time, with the shaded band marking the $M$ 5.4 analytical separator. \textbf{f}, Spatial distribution of grid cell activation onset times. \textbf{g}, Distribution of grid cell onset times, with a median of 6.2 h. \textbf{h}, Comparison of onset times between the $M_w$ 6.4 and $M_w$ 7.1 side corridors. \textbf{i}, Time-distance structure in easting and northing components. \textbf{j}, Hourly centroid distance to the $M_w$ 7.1 epicenter and event counts. The panels show delayed, spatially organized activation toward the $M_w$ 7.1 fault system. Timing alone is not used by itself to determine the triggering mechanism.}
    \label{fig:supp_ridgecrest_RA1_spatiotemporal}
    \end{figure}

    TRACE evaluated 4,714 relocated earthquakes between the 4 July 2019 $M_w$ 6.4 earthquake and the 6 July 2019 $M_w$ 7.1 earthquake. The $M_w$ 6.4 foreshock and $M_w$ 7.1 mainshock were separated by approximately 33.8 h, and the first 4 h after the $M_w$ 6.4 earthquake contained 550 relocated events. Time window maps and gridded activation diagnostics showed that activation did not occur everywhere immediately after the $M_w$ 6.4 event. Early seismicity remained concentrated around the $M_w$ 6.4 fault system, whereas later activation increasingly occupied the fault system that subsequently hosted the $M_w$ 7.1 rupture (Supplementary Fig.~\ref{fig:supp_ridgecrest_RA1_spatiotemporal}). Grid cell onset times had a median of 6.2 h, and corridor level onset distributions placed the $M_w$ 7.1 side later than the $M_w$ 6.4 side. These onset maps, corridor distributions and two-component time-distance projections define the timing evidence for delayed, spatially organized activation. They are not treated by themselves as evidence for the triggering mechanism.

    \subsection*{Organization near mapped surface ruptures}

\begin{figure}[!htbp]
    \centering
    \includegraphics[width=0.98\linewidth]{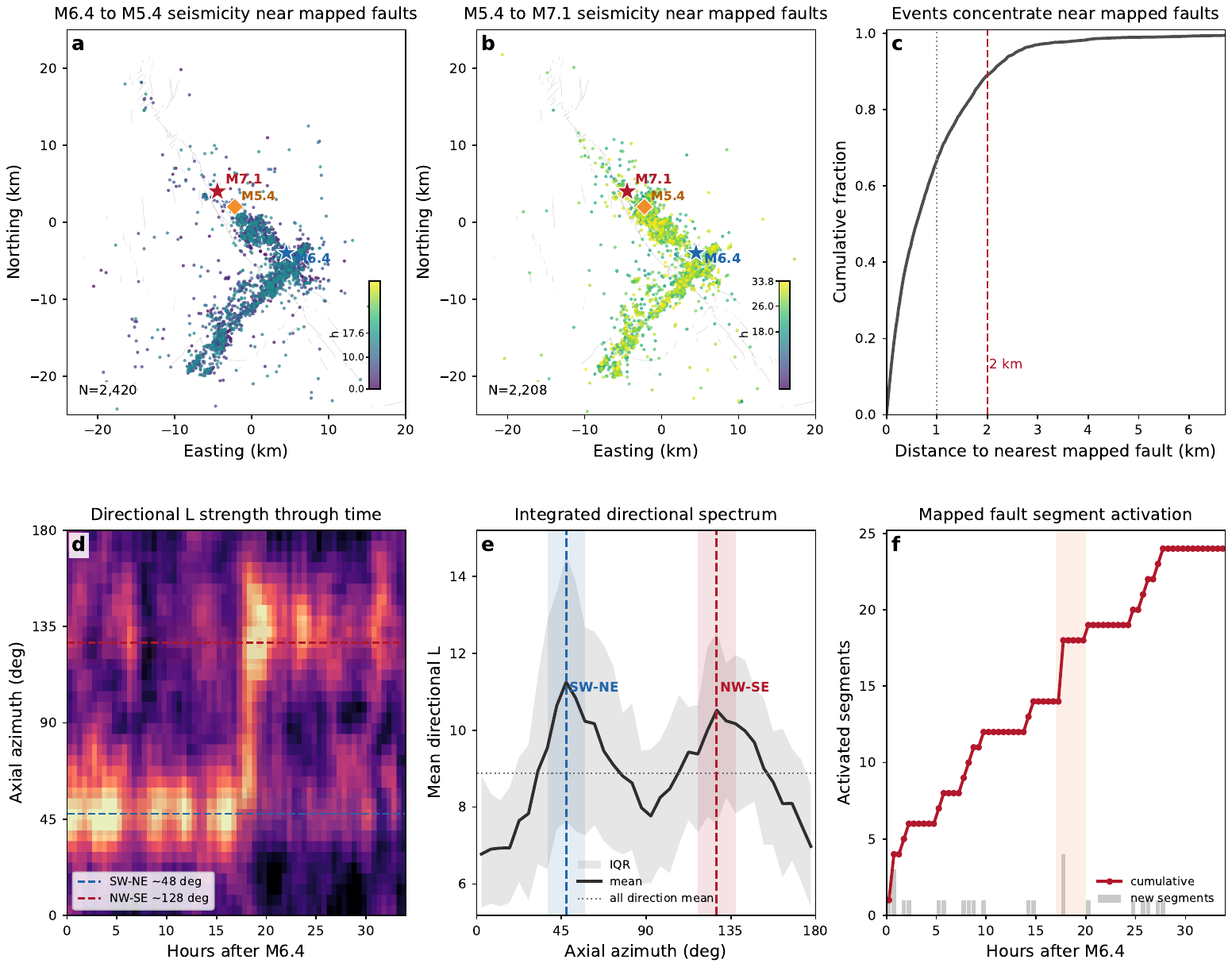}
    \caption{\textbf{Organization of Ridgecrest seismicity near mapped surface ruptures.}
    \textbf{a,b}, Relocated seismicity during the $M_w$ 6.4 to $M$ 5.4 and $M$ 5.4 to $M_w$ 7.1 intervals, plotted with mapped surface-rupture traces \cite{kendrick_2019_Geologic} and the $M_w$ 6.4, $M$ 5.4 and $M_w$ 7.1 locations. The windows contain 2,420 and 2,208 events, respectively. \textbf{c}, Cumulative distribution of distance to the nearest mapped surface-rupture trace, with 1 km and 2 km references. \textbf{d}, Directional $L$ strength through time, with the SW--NE and NW--SE fault-family directions marked. \textbf{e}, Integrated directional spectrum over the interval between the $M_w$ 6.4 and $M_w$ 7.1 earthquakes, showing enhanced directional strength near 48$^{\circ}$ and 128$^{\circ}$. \textbf{f}, Cumulative activation of segments defined from the mapped surface-rupture traces, with new segment activations shown as gray bars. The panels show fault-referenced and directionally organized activation rather than diffuse spatial expansion.}
    \label{fig:supp_ridgecrest_RA2_fault_bound}
    \end{figure}

    The delayed activation was then tested against the mapped surface-rupture geometry. Relocated events during both the early $M_w$ 6.4 to $M$ 5.4 interval and the later $M$ 5.4 to $M_w$ 7.1 interval were concentrated near mapped surface-rupture traces (Supplementary Fig.~\ref{fig:supp_ridgecrest_RA2_fault_bound}). The $M$ 5.4 event is used here as an analytical separator for comparing early and late activity between the $M_w$ 6.4 and $M_w$ 7.1 earthquakes, not as a prescribed causal boundary. Directional clustering analysis showed enhanced organization near the dominant SW--NE and NW--SE fault-family orientations, with integrated peaks near 48$^{\circ}$ and 128$^{\circ}$. Cumulative activation of segments defined from these traces continued through the $M_w$ 6.4 to $M_w$ 7.1 interval and reached 24 segments. These diagnostics showed that the evolving activity was fault-referenced and directionally organized, rather than an isotropic expansion of aftershock activity through the surrounding crust.

    \subsection*{Seismicity rate and Omori-Utsu diagnostics}

\begin{figure}[!htbp]
    \centering
    \includegraphics[width=0.98\linewidth]{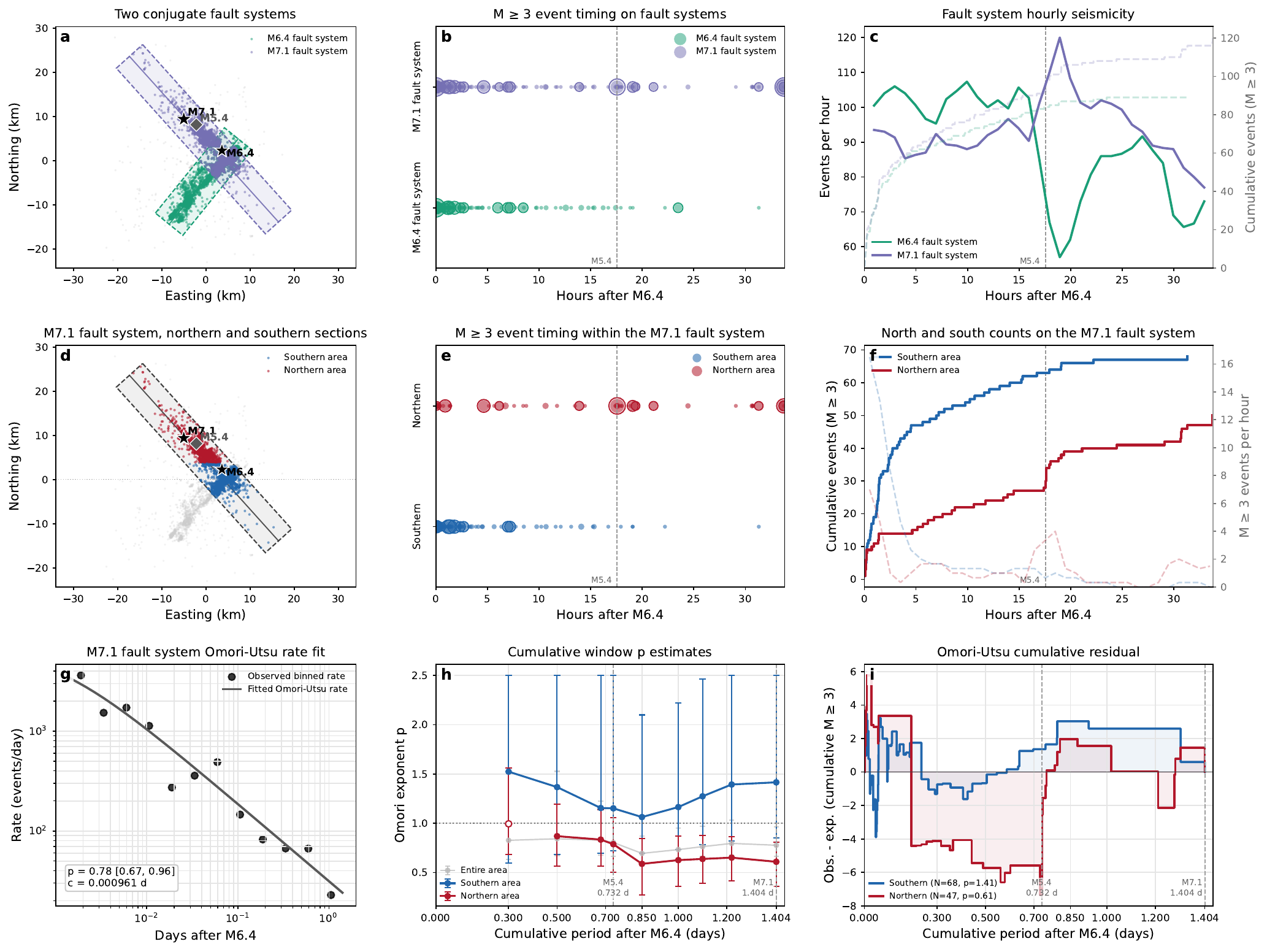}
    \caption{\textbf{Fault system seismicity rate and Omori-Utsu diagnostics.}
    \textbf{a--c}, Fault system scale comparison between the two conjugate fault systems associated with the $M_w$ 6.4 and $M_w$ 7.1 ruptures. Panel a maps event assignments to the two fault systems, panel b shows the timing of $M \ge 3$ events with symbol size scaled by magnitude, and panel c compares hourly and cumulative seismicity rates. \textbf{d--f}, North and south partitioning within the $M_w$ 7.1 fault system. Panel d maps the southern and northern sections, panel e shows the timing of $M \ge 3$ events within these sections, and panel f compares their cumulative and hourly $M \ge 3$ counts. \textbf{g--i}, Omori-Utsu diagnostics for the $M_w$ 7.1 fault system. Panel g shows a power-law rate fit for the full $M_w$ 7.1 fault system, with $p=0.78$ and a 95\% confidence interval of 0.67--0.96. Panel h shows cumulative window decay parameter estimates for the full, southern and northern sections. Panel i shows cumulative residuals between observed and expected $M \ge 3$ counts for the southern and northern subsets. These diagnostics indicate heterogeneous seismicity rate evolution within the $M_w$ 7.1 fault system rather than a single spatially uniform aftershock decay.}
    \label{fig:supp_ridgecrest_RA3_rate_omori}
    \end{figure}

    Seismicity rate diagnostics tested whether the two conjugate fault systems followed the same relaxation pattern after the $M_w$ 6.4 earthquake. Events were first assigned to the $M_w$ 6.4 and $M_w$ 7.1 fault systems, and the $M_w$ 7.1 fault system was then separated into southern and northern portions (Supplementary Fig.~\ref{fig:supp_ridgecrest_RA3_rate_omori}). The $M_w$ 6.4 fault system showed a stronger early response, whereas the $M_w$ 7.1 fault system retained later moderate-magnitude activity. For the $M_w$ 7.1 fault system, the fitted Omori-Utsu decay exponent was $p=0.78$ with a 95\% confidence interval of 0.67--0.96. Southern and northern subsets differed, with later activity retained more persistently in the northern portion closer to the later rupture. These contrasts indicate heterogeneous seismicity rate evolution along the $M_w$ 7.1 fault system rather than a single spatially uniform aftershock decay.

    \subsection*{$b$-value stress proxy diagnostics}

\begin{figure}[!htbp]
    \centering
    \includegraphics[width=0.98\linewidth]{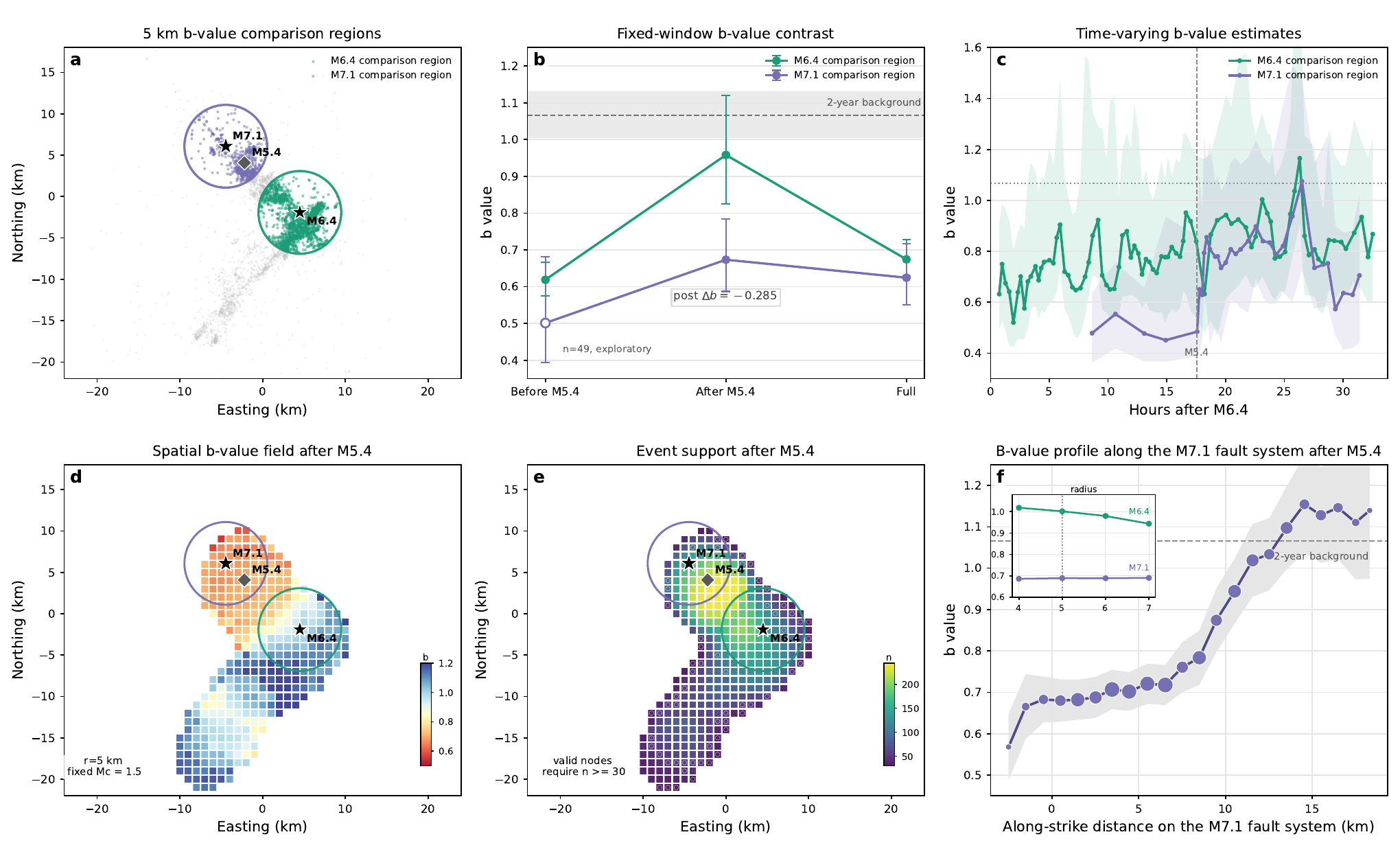}
    \caption{\textbf{$b$-value stress proxy diagnostics for the Ridgecrest $M_w$ 6.4 to $M_w$ 7.1 interval.}
    \textbf{a}, Definition of 5 km comparison regions centered on the $M_w$ 6.4 and $M_w$ 7.1 reference locations. \textbf{b}, Fixed window $b$-value estimates for the two comparison regions before and after the $M$ 5.4 separator and over the full interval between the $M_w$ 6.4 and $M_w$ 7.1 earthquakes. The $M_w$ 7.1 estimate before the separator is exploratory because of limited event support, whereas the estimate after the separator remains lower than the $M_w$ 6.4 comparison region ($\Delta b=-0.285$) and below the background level over the previous two years. \textbf{c}, Sliding window dynamic $M_c$ $b$-value estimates from 100-event windows. \textbf{d}, Spatial $b$-value field after $M$ 5.4, estimated with a 5 km radius and fixed $M_c=1.5$. \textbf{e}, Event support for the spatial $b$-value estimates, shown as the number of events above $M_c$; valid nodes require $n \ge 30$. \textbf{f}, Along-strike $b$-value profile on the $M_w$ 7.1 fault system, with an inset showing radius sensitivity of the comparison region estimates. The results document a relative low $b$ stress proxy contrast in the $M_w$ 7.1 comparison region after $M$ 5.4, while treating the $b$-value estimate as complementary evidence.}
    \label{fig:supp_ridgecrest_RA4_bvalue}
    \end{figure}

    Magnitude-frequency diagnostics provided a complementary stress proxy test. The analysis compared the 5 km $M_w$ 6.4 and $M_w$ 7.1 comparison regions before and after the $M$ 5.4 separator (Supplementary Fig.~\ref{fig:supp_ridgecrest_RA4_bvalue}). After the separator, the $M_w$ 7.1 comparison region had a lower $b$-value than the $M_w$ 6.4 comparison region, with a contrast of $\Delta b=-0.285$. It also remained below the background level over the previous two years. Spatial $b$-value mapping and along-strike profiles placed this lower $b$ structure on the $M_w$ 7.1 fault system. Because the $b$-value is an indirect stress proxy and time-varying estimates depend on event support, TRACE treated this result as a bounded magnitude-frequency constraint rather than a stand-alone diagnostic.

    The four diagnostic groups made three simpler explanations less consistent with the relocated seismicity. Immediate dynamic triggering alone would predict a nearly synchronous dominant response across the affected fault system, whereas the first hours after the $M_w$ 6.4 earthquake were dominated by the $M_w$ 6.4 fault system and later activation developed toward the $M_w$ 7.1 fault system. Ordinary aftershock decay alone would predict more similar relaxation across the fault network, whereas the $M_w$ 7.1 fault system showed heterogeneous seismicity rate behavior and late moderate-event activity. A coherent migration front would predict a more monotonic onset pattern, whereas the observed activation was organized around mapped surface ruptures, staged and spatially heterogeneous. The diagnostics were therefore more consistent with delayed cascading activation within a structurally complex fault network, with support coming from timing, fault organization, seismicity rate decay and $b$-value stress proxy evidence rather than from any single diagnostic.

    \clearpage

\section*{Supplementary Note 5. Sanriku catalog relocation and analysis-region definition}

    This note describes the JMA catalog, phase-arrival data and relocation workflow used to define the shallow Sanriku $M_J$ 6.9--$M_J$ 7.7 plate-interface region. It provides the data and window definitions used by the stage-resolved mechanism analysis and the higher-resolution $M_J$ 6.9 diagnostics in Supplementary Note 6. Supplementary Note 6 presents external validation evidence and proposed independent tests.

    \subsection*{JMA catalog and phase-arrival data}

    The Sanriku analysis used the official JMA earthquake catalog and phase-arrival records for 1 June 2025 to 2 May 2026 \cite{jma_2026_SeismologicalBulletin}. Magnitudes follow the JMA catalog unless otherwise stated; we denote the two primary JMA magnitudes as $M_J$ 6.9 and $M_J$ 7.7, whereas magnitude thresholds and secondary comparison events use the generic notation $M$. In this offshore Japanese setting, the JMA products provide dense, regionally calibrated event locations, magnitudes and station-level $P$ and $S$ arrivals, and recent Sanriku work has likewise used JMA hypocenter and phase data as the basis for double-difference relocation and sequence analysis \cite{yoshida_2026_Evolution}. TRACE first matched JMA event identifiers to station-level $P$- and $S$-arrival records, producing an arrival-constrained event dataset that contained event locations, magnitudes and phase-pick information. This matching step was required because the later relocation, stage-resolved rate analysis and repeating-earthquake screening all depend on a consistent event identifier across catalog, arrival and waveform products.

    TRACE also evaluated an alternative catalog-entry route based on deep-learning phase picking from continuous S-net waveforms \cite{nied_2019_SNet}. Human-supervised quality control revealed numerous implausible $S$ arrivals, and the resulting relocated catalog was judged less coherent than the JMA-based catalog. The alternative route was therefore not retained for subsequent scientific analysis. The intermediate products and the basis for this human-supervised decision were preserved in the provenance record. A separately assembled waveform dataset remained available for the repeating-earthquake cross-correlation described in Supplementary Note 6. This comparison illustrates why detections, associations and initial locations require local validation when no trusted phase product is available.

    The full arrival-matched dataset comprised 257,416 events, 4,033,429 event-station observation records and 1,625 stations. These records included 3,057,725 $P$ arrivals and 3,098,577 $S$ arrivals. The median pick coverage was 13 phase picks per event, with an interquartile range of 9--21 picks. For the Sanriku regional relocation, TRACE extracted events within $141.00^{\circ}$--$144.50^{\circ}$ E and $38.50^{\circ}$--$42.50^{\circ}$ N. This regional subset contained 33,702 events, 637,812 event-station records and 386 stations, with 522,050 $P$ arrivals, 423,024 $S$ arrivals and 307,262 station records containing both $P$ and $S$ arrivals. The median pick coverage increased to 17 picks per regional event, with an interquartile range of 11--26 picks.

    \subsection*{Regional extraction and double-difference relocation}

    TRACE applied a windowed HypoDD double-difference relocation workflow to improve relative hypocentral locations in the Sanriku regional dataset \cite{waldhauser_2000_DoubleDifference}. After excluding 220 events without reported magnitudes from the regional subset, the relocation input comprised 33,482 JMA catalog events. Of these, 29,036 entered the initial HypoDD solution, and double-difference locations were obtained for 25,646 events (Supplementary Fig.~\ref{fig:supp_sanriku_system_definition}a,b). Events that lacked sufficient phase and neighboring-event constraints were not retained in the relocated catalog. The relocated catalog therefore served as the main basis for spatial clustering, along-profile distance-depth projections and local analysis of the shallow plate-interface region, rather than as a one-to-one copy of the original catalog.

    Events were relocated independently within monthly time windows using catalog-derived differential times. Event pairs were formed with ph2dt from both $P$ and $S$ arrivals. Neighboring events were required to be within 20 km of each other (MAXSEP = 20 km), and stations or event pairs were required to lie within 180 km (MAXDIST = 180 km). Each event was linked to at most 12 neighbors (MAXNGH = 12), with at least 8 linked observations per event pair (MINLNK = 8) and at least 6 observations per event (MINOBS = 6). No more than 40 observations per event pair were retained (MAXOBS = 40). The HypoDD inversion used catalog $P+S$ differential times only (IDAT = 2, IPHA = 3), with cross-correlation terms disabled and the same 180 km maximum event-station distance. The inversion used two damping and weighting stages: 8 iterations with $P/S$ catalog weights of 1.0/0.7 and damping 120, followed by 12 iterations with $P/S$ weights of 0.7/0.3 and damping 80. A 1-D velocity model with $V_{\mathrm{P}}/V_{\mathrm{S}} = 1.73$ and seven $V_{\mathrm{P}}$ layers was used.

    The relocated regional catalog covered approximately $141.00^{\circ}$--$144.50^{\circ}$ E and $38.50^{\circ}$--$42.38^{\circ}$ N, with magnitudes from approximately $M=-0.5$ to $M=7.7$. The median relocated depth was approximately 16.5 km, with an interquartile range of 11.9--32.2 km, and approximately 72\% of relocated events lay between 0 and 30 km depth. TRACE then compared the $M_J$ 6.9, regional $M$ 7.5 and $M_J$ 7.7 earthquake clusters across this broader relocated catalog. The November 2025 $M_J$ 6.9 activity and the 20 April 2026 $M_J$ 7.7 earthquake were selected as the primary shallow candidate pair because they were spatially closer than the 8 December 2025 regional $M$ 7.5 earthquake, occupied comparable shallow depths and were linked by $M \ge 4$ seismicity in map view and along-profile projection (Supplementary Fig.~\ref{fig:supp_sanriku_system_definition}c--f). The focused analysis box used for quantitative stage statistics was $142.78^{\circ}$--$143.76^{\circ}$ E and $39.18^{\circ}$--$40.16^{\circ}$ N. Within this box, 7,845 relocated events were retained, including 7,821 events in the 0--30 km shallow range. This step defines the physical region analyzed below; it does not require the focused box to represent a single rupture plane.

    For the three-cluster association step, TRACE labeled the November 2025 $M_J$ 6.9 activity as M1, the 8 December 2025 regional $M$ 7.5 earthquake as M2 and the 20 April 2026 $M_J$ 7.7 earthquake as M3. It evaluated the candidate pairs using bounded spatial-support metrics rather than selecting the focused region by epicentral distance alone. Distance support was defined as $S_D=\mathrm{clip}(1-D/220,0,1)$, where $D$ is the pair separation in kilometers. Depth support was defined as $S_Z=\mathrm{clip}(1-|\Delta z|/40,0,1)$, where $\Delta z$ is the cluster-depth difference in kilometers. The shallow-domain support $S_S$ was set to 1 when both cluster depths were $\leq 30$ km and to 0 otherwise. The combined support was then $S_C=(S_D+S_Z+S_S)/3$. These scores provided a traceable screening criterion for selecting the correlation-analysis region. They were not treated as rupture probabilities.

    Supplementary Fig.~\ref{fig:supp_sanriku_system_definition} is the analysis-region definition figure for the Sanriku application. It shows that the $M_J$ 6.9--$M_J$ 7.7 analysis region was derived from relocation quality, three-cluster association and shallow-domain support, rather than imposed only by a rectangular map box. The figure supports the main-text statement that TRACE first isolated a physically motivated shallow-interface analysis region associated with the November 2025 $M_J$ 6.9 activation sequence and the April 2026 $M_J$ 7.7 earthquake. It defines the object of inference, rather than demonstrating the perturbed shallow-interface state itself.

    This defines the Sanriku scope of inference. In this trusted-catalog mode, TRACE used an authoritative event-level product as the starting point and made the subsequent matching, relocation, filtering, stage bookkeeping, waveform screening and mechanism synthesis traceable. In regions without comparable operational catalogs or phase-arrival products, the upstream detection, association and location steps would become part of the validation burden before the downstream TRACE reasoning could be interpreted in the same way.

    Having defined the shallow-interface analysis region, the next step was to keep the stage bookkeeping fixed across the rate, magnitude-threshold and waveform analyses. This prevents the later physical interpretation from shifting the analytical windows after the fact.

    \subsection*{Focused analysis region and stage bookkeeping}

    The focused shallow-interface analysis region was divided into five analytical stages for the rate, magnitude-threshold and shallow-interface analyses in main-text Fig. 4 and Supplementary Fig.~\ref{fig:supp_sanriku_stage_resolved}. These stages are fixed analytical windows, not claims of sharp physical transitions. They span background activity before the November 2025 $M_J$ 6.9 sequence, the two-week $M_J$ 6.9 activation sequence, the interval following that sequence, the final interval before the $M_J$ 7.7 rupture and the early $M_J$ 7.7 response. Here, this sequence denotes the fixed two-week analytical window around the November 2025 $M_J$ 6.9 event; activity involved multiple large events, including 61 $M \ge 4$ events, 23 $M \ge 5$ events and four $M \ge 6$ events, rather than a single dominant event. The same stage definitions were used to compute event counts and rates at the total, $M \ge 3$, $M \ge 4$, $M \ge 5$ and $M \ge 6$ levels.

    Across these five windows, the total event counts and rates were 731 events at 4.54 d$^{-1}$ during background activity, 2,273 events at 162.36 d$^{-1}$ during the $M_J$ 6.9 activation sequence, 1,169 events at 13.13 d$^{-1}$ during the following interval, 962 events at 16.31 d$^{-1}$ during the final pre-$M_J$ 7.7 interval and 2,710 events at 246.36 d$^{-1}$ during the early $M_J$ 7.7 response. At $M \ge 3$, the five stages contained 66, 262, 53, 67 and 226 events, respectively. At $M \ge 4$, they contained 17, 61, 15, 18 and 39 events, respectively. The $M_J$ 6.9 activation window also contained 23 $M \ge 5$ and four $M \ge 6$ events, whereas the final pre-$M_J$ 7.7 interval contained five $M \ge 5$ and two $M \ge 6$ events. These fixed windows provide the stage-level baseline for the higher-resolution $M_J$ 6.9 diagnostics in Supplementary Note 6.

    Supplementary Fig.~\ref{fig:supp_sanriku_stage_resolved} summarizes the stage-resolved activation evidence. Within the focused shallow-interface region, seismicity remained above the background level after the $M_J$ 6.9 activation sequence. To examine this evolution along the $M_J$ 6.9--$M_J$ 7.7 axis, TRACE separately calculated the stage metrics within an oriented profile domain rather than the rectangular analysis box defined above. In this domain, the total event rate rose from 5.4 d$^{-1}$ during background activity to 104.4 d$^{-1}$ during the $M_J$ 6.9 activation sequence. It then decreased to 18.2 d$^{-1}$ during the following interval, but remained above background. During the final pre-$M_J$ 7.7 interval, the total rate increased to 21.3 d$^{-1}$, and the $M \ge 3$ and $M \ge 4$ rates also increased relative to the preceding stage. A separate long-term background correction identified the local $M_J$ 6.9--$M_J$ 7.7 region as an anomaly across magnitude thresholds, with observed/expected count ratios of 7.70 for $M \ge 3$, 8.80 for $M \ge 4$ and 9.64 for $M \ge 5$ earthquakes.

    Magnitude--frequency diagnostics provided a second constraint. The $b$-value in the directed $M_J$ 6.9--$M_J$ 7.7 region was approximately 0.811 during the background window before the November 2025 $M_J$ 6.9 sequence. It decreased to 0.608 during the $M_J$ 6.9 activation window, partially recovered to 0.773 during the following interval and declined again to 0.653 during the final pre-$M_J$ 7.7 interval. Considered together with the stage-resolved rate changes, this pattern indicates that both activity level and magnitude--frequency structure changed during the sequence. The $b$-value changes were interpreted in light of established relationships between earthquake-size distribution and stress state \cite{schorlemmer_2005_Variations, scholz_2015_Stress}.

    \refstepcounter{figure}
    \label{fig:supp_sanriku_system_definition}
    \begin{center}
    \centering
    \includegraphics[width=0.84\linewidth]{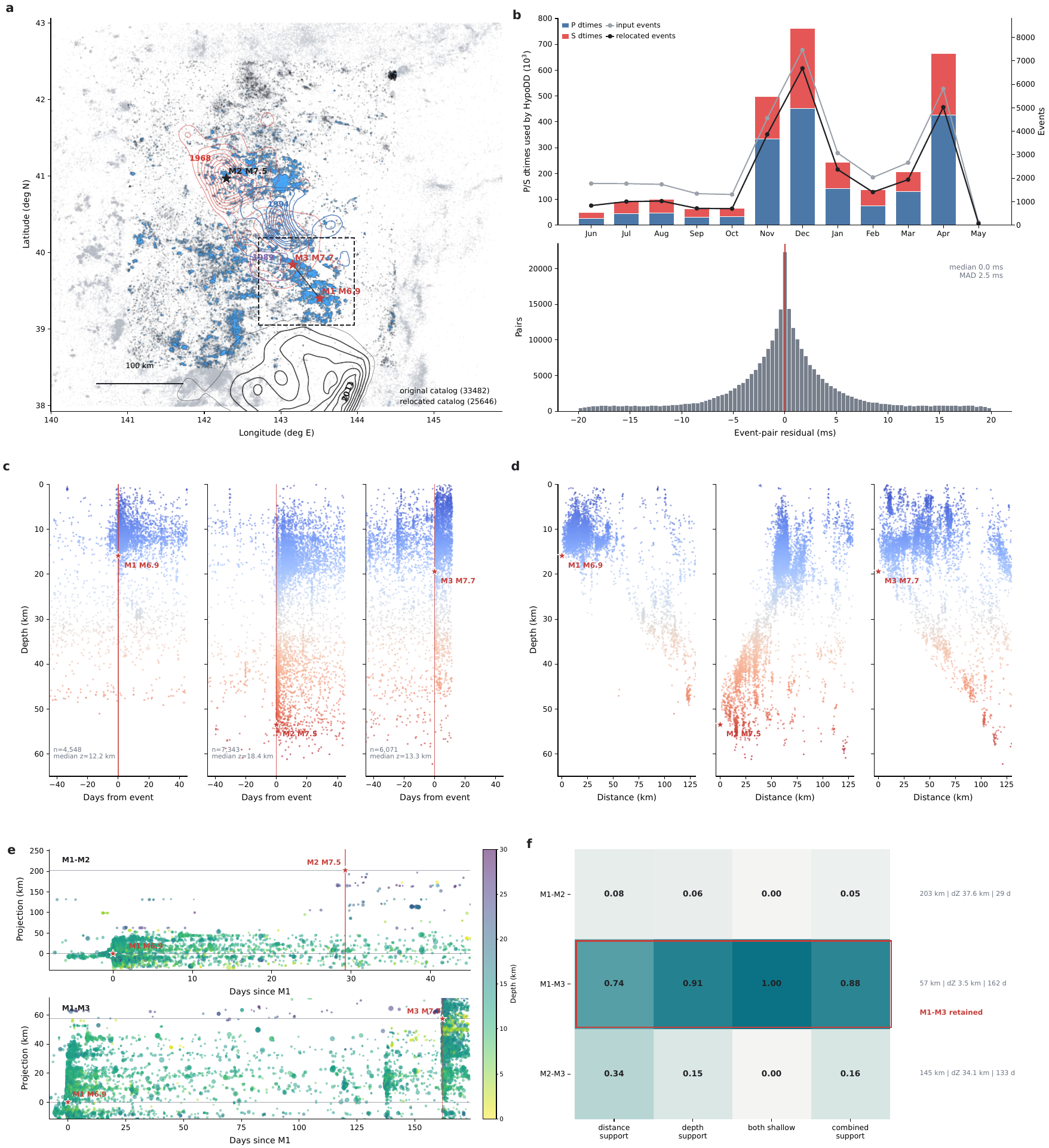}
    \end{center}
    \tightcaptionspacing
    \sciencecaptionof{figure}{\textbf{Relocation quality and data-driven definition of the Sanriku $M_J$ 6.9--$M_J$ 7.7 shallow-interface analysis region.}
    \textbf{a}, Regional map of the original JMA catalog (33,482 events) and the double-difference relocated catalog (25,646 events). The dashed box marks the focused Sanriku analysis area. M1 denotes the November 2025 $M_J$ 6.9 activity, M2 denotes the 8 December 2025 regional $M$ 7.5 earthquake, and M3 denotes the 20 April 2026 $M_J$ 7.7 earthquake. Historical rupture contours show the 2011 $M_w$ 9.0 Tohoku-Oki high-slip domain \cite{iinuma_2012_Coseismic}, the 1994 $M_w$ 7.7 Sanriku-Oki rupture and the 1968 Tokachi-Oki rupture (approximately $M_w$ 8.3) \cite{nagai_2001_Comparative}. \textbf{b}, Monthly $P$- and $S$-phase differential times used in HypoDD relocation, together with the numbers of input and relocated events in each monthly window (top), and the post-relocation event-pair residual distribution (bottom; median, 0.0 ms; median absolute deviation, 2.5 ms). \textbf{c,d}, Time-depth and distance-depth distributions for the M1, M2 and M3 clusters. M1 and M3 occupy comparable shallow depths, whereas the M2 event is deeper and more distant from M1. \textbf{e}, Along-axis projections for the M1--M2 and M1--M3 candidate pairings. The M1--M3 pairing defines the shorter and shallower analysis profile, with a 57 km axis and 6,153 events in the profile-domain selection. \textbf{f}, Pair-selection support scores based on distance, depth compatibility and shallow-domain consistency. M1--M3 had the highest combined support score (0.88) and was therefore retained as the focused shallow-interface region for the Sanriku rate, magnitude-frequency and repeating-earthquake analyses.}
    \standardcaptionspacing

    \refstepcounter{figure}
    \label{fig:supp_sanriku_stage_resolved}
    \begin{center}
    \centering
    \includegraphics[width=0.84\linewidth]{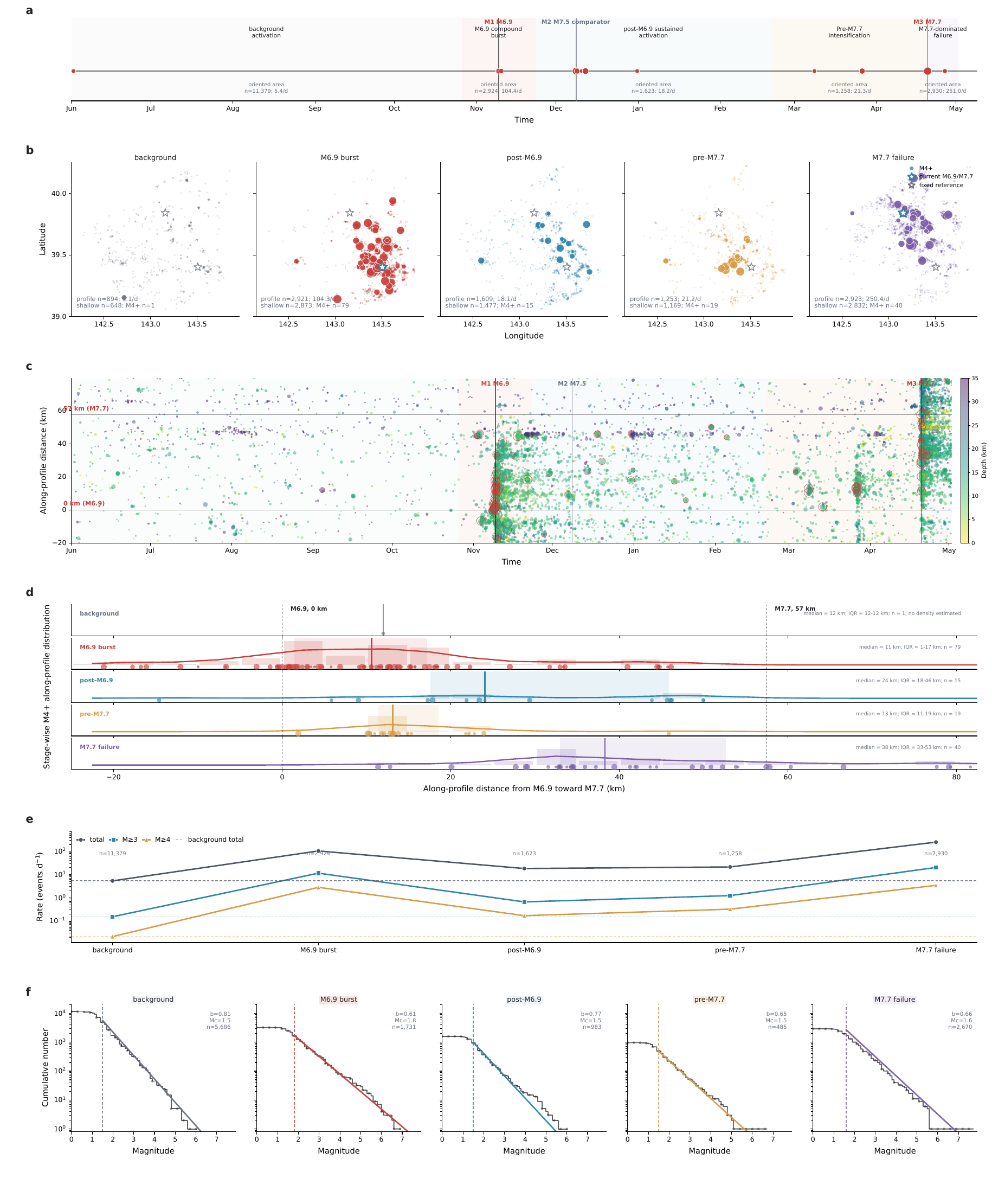}
    \end{center}
    \tightcaptionspacing
    \sciencecaptionof{figure}{\textbf{Stage-resolved seismicity evidence for sustained shallow activation in the Sanriku $M_J$ 6.9--$M_J$ 7.7 shallow-interface analysis region.}
    \textbf{a}, Fixed analytical windows for the Sanriku sequence. Shaded intervals mark the stage windows; M1, M2 and M3 denote the November 2025 $M_J$ 6.9 activation sequence, the 8 December 2025 regional $M$ 7.5 earthquake and the 20 April 2026 $M_J$ 7.7 earthquake, respectively. \textbf{b}, Stage-resolved maps of relocated seismicity in the focused Sanriku region. Small points show relocated events, larger circles mark $M \ge 4$ events, and stars mark fixed $M_J$ 6.9 and $M_J$ 7.7 reference locations. \textbf{c}, Time-distance distribution along the $M_J$ 6.9--$M_J$ 7.7 profile, colored by depth; the 0 km and 57 km lines mark the $M_J$ 6.9 and $M_J$ 7.7 reference locations. \textbf{d}, Stage-wise $M \ge 4$ along-profile distributions, showing activity near the $M_J$ 6.9 source and across the focused shallow-interface profile rather than monotonic migration toward $M_J$ 7.7. \textbf{e}, Stage-wise seismicity rates within the oriented profile domain for all events and the $M \ge 3$ and $M \ge 4$ subsets; dashed lines mark background rates. \textbf{f}, Cumulative magnitude--frequency distributions with completeness thresholds and Gutenberg--Richter fits; annotations report the event counts, completeness magnitudes and fitted $b$-values.}
    \standardcaptionspacing

    \clearpage

    \section*{Supplementary Note 6. $M_J$ 6.9 migration, repeating earthquakes and independent evaluation of aseismic modulation}

    This note uses the region definition and stage bookkeeping from Supplementary Note 5 to examine the short-timescale evolution around the November 2025 $M_J$ 6.9 sequence. It considers three analytical components: migration and magnitude-frequency diagnostics around the $M_J$ 6.9 event, waveform quality control for repeating earthquakes, and independent evidence and tests of aseismic-slip modulation. The earthquake-derived evidence is consistent with rapid shallow-interface activation around the $M_J$ 6.9 sequence. Independent strain observations provide additional support for an aseismic contribution, although the amount, geometry and stress-transfer role of aseismic slip remain unresolved.

    \subsection*{Migration and magnitude-frequency diagnostics around the $M_J$ 6.9 sequence}

    TRACE treated the Sanriku sequence as a recent discovery application. The reasoning began from the regional relocated catalog rather than from a prescribed physical relation between the $M_J$ 6.9 and $M_J$ 7.7 earthquakes. Supplementary Note 5 defines the $M_J$ 6.9--$M_J$ 7.7 shallow-interface region; here the analysis asks how seismicity evolved immediately before and after the $M_J$ 6.9 event within that region.

    The $M_J$ 6.9 reference event was the relocated catalog match at 2025-11-09 08:03:39.240 UTC, with longitude $143.50568^{\circ}$ E, latitude $39.403072^{\circ}$ N, depth 15.96 km and magnitude 6.9. TRACE compared three windows relative to this event: the broader context from 10 to 2 days before it, the final two days before it and the first 0.5 days after it. Using the 90th-percentile distance envelope as a bulk migration metric, the final two-day interval showed an advance of approximately $4.3~\mathrm{km~d^{-1}}$. In the first 0.5 days after the $M_J$ 6.9 event, the same metric had an apparent slope of approximately $16.2~\mathrm{km~d^{-1}}$. This value describes the bulk expansion of the activated seismicity envelope rather than a single continuous migration-front speed.

    The local activated-area analysis gave the same picture in map view. Within the 80 km neighbourhood of the $M_J$ 6.9 event, activity during the final two days before the event occupied a compact hull of approximately 307 km$^2$. During the first 0.5 days after the $M_J$ 6.9 event, the activated hull expanded to approximately 2,118 km$^2$ and included the early $M$ 6.6 aftershock. Event rates also increased sharply, from 396 events in the final two days (approximately 198 d$^{-1}$) to 415 events in the first 0.5 days after the $M_J$ 6.9 event (approximately 830 d$^{-1}$).

    Magnitude-frequency diagnostics were estimated for the same two key windows. Completeness magnitudes were $M_c=1.8$ for the final two-day interval before the $M_J$ 6.9 event and $M_c=2.0$ for the first 0.5 days after it. Classic maximum-likelihood $b$-value estimates gave $b=0.51 \pm 0.04$ for 201 events above completeness in the final two-day window and $b=0.43 \pm 0.02$ for 308 events above completeness in the first 0.5-day window. The uncertainties are the Shi and Bolt standard uncertainties for the classic maximum-likelihood $b$-value estimator \cite{shi_1982_Standard}. These low-$b$ short-window estimates indicate a comparatively large contribution from moderate events, but they are treated as short-window diagnostics rather than stable long-term regional $b$-value estimates.

    Supplementary Fig.~\ref{fig:supp_sanriku_swarm_migration} summarizes these diagnostics. The figure first places the November 2025 sequence near the northern edge of the 2011 Tohoku-Oki high-slip domain and within the historical Sanriku rupture zone. It then shows that the final two days before the $M_J$ 6.9 event were compact and directionally organized, whereas the first 0.5 days after the event activated a broader region at a much higher rate. This evidence supports rapid local shallow-interface activation around the $M_J$ 6.9 sequence, without requiring the later pattern to be a smooth one-dimensional migration front.

    \subsection*{Order-of-magnitude activated-area moment estimate}

    We converted the activated areas in Supplementary Fig.~\ref{fig:supp_sanriku_swarm_migration} into equivalent moment magnitudes to provide a dimensional size scale for possible aseismic slip. The calculation was performed separately for the final two-day interval before the $M_J$ 6.9 event and the first 0.5-day interval after it. This calculation does not invert for a slow-slip distribution; it gives an order-of-magnitude estimate of the moment implied if the activated area were affected by a representative average slip.

    Earthquake epicenters were first expressed in a local Cartesian coordinate system centered on the $M_J$ 6.9 reference event. The horizontal locations were then rotated into a sequence-aligned coordinate system using principal component analysis. The first principal component approximates the dominant along-subduction trend of the local event cloud, and the second component gives the across-axis direction. For each diagnostic window, the activated area was taken from the convex hull shown in Supplementary Fig.~\ref{fig:supp_sanriku_swarm_migration}c. This hull-based area follows the observed event cloud rather than imposing a rectangular or elliptical footprint.

    We converted the activated area into scalar moment using
    \[
    M_0=\mu A D,\qquad M_w=\frac{2}{3}\left(\log_{10}M_0-9.1\right),
    \]
    where $M_0$ is scalar moment, $\mu$ is shear modulus, $A$ is activated area and $D$ is average slip. Using $\mu=30~\mathrm{GPa}$ and a reference slip of $D=0.10~\mathrm{m}$, the 307 km$^2$ final two-day hull gives $M_0=9.21\times10^{17}~\mathrm{N\,m}$, equivalent to approximately $M_w$ 5.9. The 2,118 km$^2$ first 0.5-day hull gives $M_0=6.35\times10^{18}~\mathrm{N\,m}$, equivalent to approximately $M_w$ 6.5. Because the estimate scales linearly with the assumed average slip, these values are used only as order-of-magnitude moment-equivalent checks, not as measured aseismic moments.

    \subsection*{Repeating-earthquake detection near the $M_J$ 6.9 sequence}

    Repeating-earthquake analysis tested repeated activation of small rupture patches during the $M_J$ 6.9 activation sequence \cite{nadeau_1999_Fault, uchida_2019_Repeating}. The representative evidence in the main text focuses on the 1--15 November 2025 activation-sequence window, whereas the population-level quality-control analysis below uses an extended waveform-ready window from 1 November to 7 December 2025. Waveforms were obtained from JMA/Hi-net event waveform packages and converted to SAC format, with Hi-net operated as part of the NIED MOWLAS observing system \cite{aoi_2020_MOWLAS}. When an event waveform package lacked a station component that was available in the daily continuous data and had a matching phase arrival, TRACE cut a supplemental event window from the continuous waveform record. The unified waveform table contained 3,387 events, 329 stations, 1,008 station components and 182,886 event-station-component records. All listed SAC paths were checked to be accessible and non-empty. The quality-control figure displays the 1,839 waveform-ready events that fall within the focused $M_J$ 6.9 map window used for the population-level repeating-pair analysis; this plotted subset is therefore smaller than the full waveform table.

    Candidate event pairs were first screened by common observations, spatial distance and magnitude difference. For each retained candidate pair, TRACE computed phase-aligned cross-correlations in available $P$- and $S$-phase windows across station components. Stable evaluation required at least three stations and eight station components. This procedure produced 20,301 candidate event pairs, from which TRACE identified 231 high-confidence repeating pairs and 215 relaxed repeating candidates. The high-confidence repeating pairs formed 145 repeating-earthquake families, mostly small families with two to three events. The largest family contained six events. Representative pairs showed median network cross-correlation coefficients exceeding 0.9 across multiple stations and components.

    Supplementary Fig.~\ref{fig:supp_sanriku_repeating_qc} summarizes cross-correlation support for repeated small-asperity failure near the $M_J$ 6.9 source region and along the shallow interface toward the $M_J$ 7.7 reference location. It tests whether repeat-like pairs are spatially concentrated and supported by many station components across nested cross-correlation thresholds. The figure does not identify the loading source of the repeated failures; that question motivates the independent evaluation below.

    \subsection*{Independent evidence and tests of aseismic contributions to shallow-interface activation}

    The earthquake catalog, migration diagnostics, rate changes, magnitude-frequency diagnostics and repeating-earthquake families characterize earthquake-derived shallow-interface activation. Testing whether slow slip, afterslip or related aseismic processes modulated this activation requires independent tremor, deformation, source or coupling constraints. This validation strategy follows established links among slow slip, tremor, low-frequency earthquakes and large subduction earthquakes, including slow-earthquake processes along the Japan Trench \cite{peng_2010_Integrated, shelly_2007_Nonvolcanic, obara_2016_Connecting, nishikawa_2023_Review}. This Supplementary Information therefore distinguishes the earthquake-derived activation from the external observations needed to test its aseismic component.

    A recent S-net analysis of the 9 November 2025 Sanriku-Oki $M_w$ 6.8 sequence provides an independent example of such external validation. That study reported expansion before the largest event near locked-patch margins, rapid aftershock expansion to approximately 75 km, repeating-earthquake reactivation, strainmeter-consistent afterslip and a decrease in $b$-value before the largest event, and interpreted the sequence as coupled seismic and aseismic slip on a heterogeneous megathrust \cite{yoshida_2026_Evolution}. These observations support the physical plausibility of aseismic modulation for the $M_J$ 6.9-side activation. The core earthquake-derived activation pattern was obtained from catalog and waveform evidence, whereas the strain observations were introduced as external evidence when evaluating its aseismic component.

    Further independent tests could draw on tremor or low-frequency-earthquake observations, geodetic deformation, source-mechanism or slip models and regional plate-interface coupling models. Tremor or low-frequency-earthquake products can test whether slow-earthquake activity was temporally and spatially consistent with the $M_J$ 6.9--$M_J$ 7.7 shallow-interface region. GNSS, offshore pressure or seafloor geodetic constraints can test whether aseismic deformation or coupling change accompanied the staged activation. Source mechanisms and finite-fault models can test whether the contrast between the $M_J$ 6.9 activation sequence and the $M_J$ 7.7 rupture is compatible with plate-interface geometry. Regional interface and coupling models can test whether the focused region lies in a plausible weak-to-strong coupling transition zone. These potential tests are distinct from the evidence incorporated into the earthquake-catalog result.

    \clearpage
    \refstepcounter{figure}
    \label{fig:supp_sanriku_swarm_migration}
    \begin{center}
    \centering
    \includegraphics[width=0.9\linewidth]{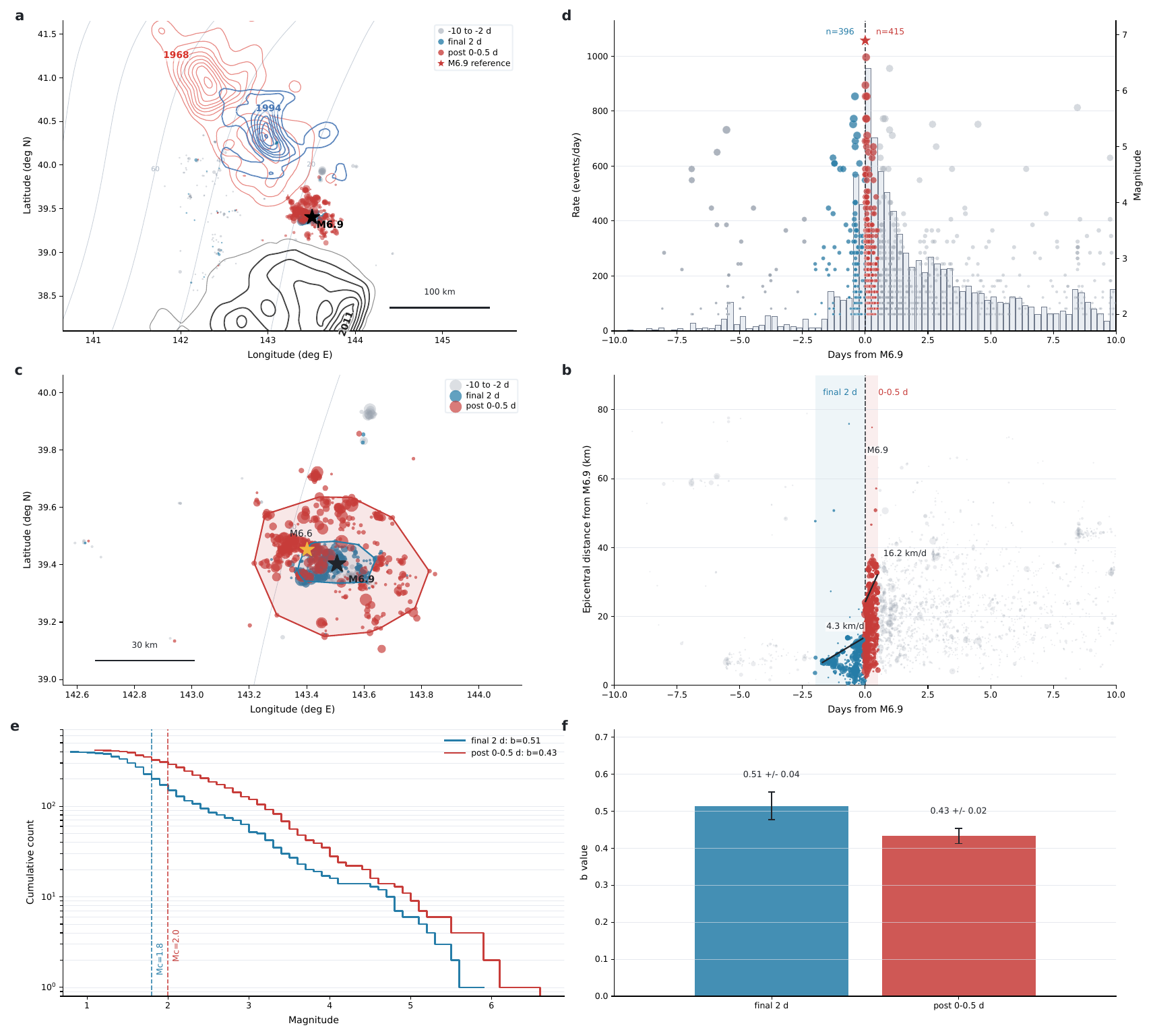}
    \end{center}
    \tightcaptionspacing
    \sciencecaptionof{figure}{\textbf{Migration, rate and magnitude-frequency diagnostics for the November 2025 Sanriku-Oki $M_J$ 6.9 sequence.}
    \textbf{a}, Regional map of seismicity from 10 days before to 0.5 days after the $M_J$ 6.9 event, shown with slab-depth contours and historical coseismic rupture or slip contours for the 1968 Tokachi/Sanriku event, the 1994 Sanriku-Oki event and the 2011 Tohoku-Oki earthquake. \textbf{b}, Time-distance view of seismicity within 80 km of the $M_J$ 6.9 event. Lines show fits to the 90th-percentile distance envelope during the final two days before the event and the first 0.5 days after it. The value for the first 0.5-day window is interpreted as an apparent slope of the activated seismicity envelope rather than a single-front propagation speed. \textbf{c}, Local map showing activated-area hulls for the final two-day interval before the $M_J$ 6.9 event and the first 0.5-day interval after it. The hull area expanded from approximately 307 km$^2$ to approximately 2,118 km$^2$. \textbf{d}, Event-rate and magnitude timeline for the 10-day interval before and 0.5-day interval after the $M_J$ 6.9 event. Within 80 km, the final two-day window contained 396 events, whereas the first 0.5 days after the $M_J$ 6.9 event contained 415 events and an early $M$ 6.6 aftershock. \textbf{e}, Cumulative magnitude-frequency distributions for the two diagnostic windows, with completeness thresholds marked at $M_c=1.8$ and $M_c=2.0$. \textbf{f}, Fixed-window $b$-value estimates for the same two windows. Error bars are standard uncertainties for the classic maximum-likelihood estimator.}
    \standardcaptionspacing

    \clearpage
    \refstepcounter{figure}
    \label{fig:supp_sanriku_repeating_qc}
    \begin{center}
    \centering
    \includegraphics[width=0.9\linewidth]{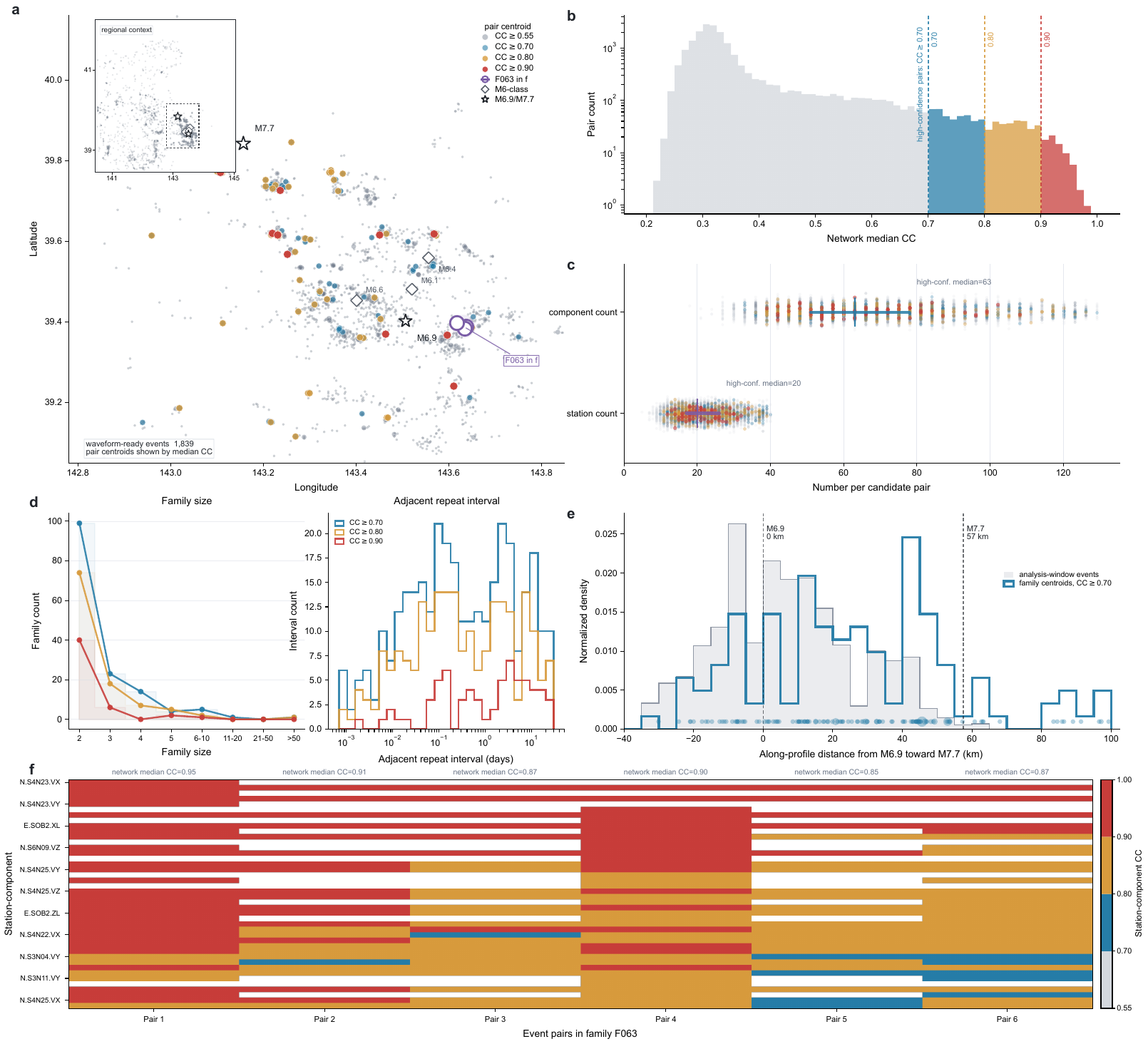}
    \end{center}
    \tightcaptionspacing
    \sciencecaptionof{figure}{\textbf{Population-level quality control for repeating-earthquake detections near the Sanriku $M_J$ 6.9 sequence.}
    \textbf{a}, Spatial distribution of the focused-window waveform-ready events and repeating-pair centroids in the extended analysis window from 1 November to 7 December 2025. Gray points show the waveform-ready subset used for the population-level QC map, colored circles show pair centroids for nested network-median-CC thresholds, diamonds mark $M$ 6-class events, and stars mark the fixed $M_J$ 6.9 and $M_J$ 7.7 reference locations. Family F063, used as the representative family in the main-text waveform panel, is highlighted. \textbf{b}, Distribution of network median cross-correlation coefficients for candidate event pairs. Dashed vertical lines mark the nested thresholds used for sensitivity analysis. Higher-threshold pairs are subsets of lower-threshold detections, and the colored high-CC tail shows the candidate population from which accepted repeat-like pairs were drawn. \textbf{c}, Station and component support for candidate pairs. Each point represents a candidate pair, colored by the same median-CC threshold levels. High-confidence pairs are supported by many stations and components, indicating that high similarity is not controlled by a single station or component. \textbf{d}, Family-size and adjacent-repeat-interval distributions for nested CC thresholds of 0.70, 0.80 and 0.90. The persistence of small repeating families under stricter thresholds shows that the family structure is not an artifact of one permissive correlation cutoff. \textbf{e}, Along-profile distribution of family centroids compared with all analysis-window events. The family centroids occupy the $M_J$ 6.9-side profile domain and the shallow profile segment toward the $M_J$ 7.7 reference location, with fewer centroids near the later $M_J$ 7.7 reference location. \textbf{f}, Station-component cross-correlation matrix for representative family F063. Columns show event pairs within the family and rows show station components. Top labels report the network median CC for each pair. White cells denote unavailable station-component observations rather than low correlations.}
    \standardcaptionspacing


\end{document}